\documentclass{article}
\usepackage{amssymb}
\usepackage{amsmath}

\newtheorem{theorem}{Theorem}
\newtheorem{lemma}[theorem]{Lemma}
\newtheorem{definition}{Definition}
\newtheorem{proposition}[theorem]{Proposition}
\newtheorem{example}{Example}
\def\proof{\noindent{\bfseries Proof. }}
\def\endproof{\mbox{\ \rule{.1in}{.1in}}}

\def\limfunc#1{\mathop{\rm #1}}

\begin{document}

\title{Exact form factors in integrable quantum field theories: 
the sine-Gordon model (II) }
\author{H. Babujian\thanks{%
Permanent address: Yerevan Physics Institute, Alikhanian Brothers 2,
Yerevan, 375036 Armenia.} \thanks{
e-mail: babujian@lx2.yerphi.am, babujian@physik.fu-berlin.de}~ and M.
Karowski\thanks{
e-mail: karowski@physik.fu-berlin.de} \\
Institut f\"ur Theoretische Physik\\
Freie Universit\"at Berlin,\\
Arnimallee 14, 14195 Berlin, Germany\\
}
\date{\today}
\maketitle

\begin{abstract}
A general model independent approach using the `off-shell Bethe An\-satz' is
presented to obtain an integral representation of generalized form factors.
The general techniques are applied to the quantum sine-Gordon model alias
the massive Thirring model. Exact expressions of all matrix elements are
obtained for several local operators. In particular soliton form factors of
charge-less operators as for example all higher currents are investigated.
It turns out that the various local operators correspond to specific scalar
functions called p-functions. The identification of the local operators is
performed. In particular the exact results are checked with Feynman graph
expansion and full agreement is found. Furthermore all eigenvalues of the
infinitely many conserved charges are calculated and the results agree with
what is expected from the classical case. Within the frame work of
integrable quantum field theories a general model independent `crossing'
formula is derived. Furthermore the `bound state intertwiners' are
introduced and the bound state form factors are investigated. The general
results are again applied to the sine-Gordon model. The integrations are
performed and in particular for the lowest breathers a simple formula for
generalized form factors is obtained. \\[8pt]
\end{abstract}

\section{Introduction}

Form factors for integrable model in 1+1 dimensions were first investigated
by Vergeles and Gryanik \cite{VG} for the sinh-Gordon model and by Weisz 
\cite{W} for the sine-Gordon model. The `form factor program' was formulated
in \cite{KW} where the concept of generalized form factors was introduced.
In that article consistency equations were formulated which are expected to
be satisfied by these objects. Thereafter this approach was developed
further and studied in the context of several explicit models by Smirnov 
\cite{Sm} who proposed the form factor equations $(i)-(v)$ (see below) as
extensions of similar formulae in the original article \cite{KW}. The
formulae were proven in \cite{BFKZ}. In the last decade a large number of
articles were published on form factors (see e.g. references in \cite{BFKZ}%
). More recent papers on solitonic matrix elements in the sine-Gordon model
are \cite{Lu,LZ} and for the $SU(2)$-Thirring model \cite{KLP,NPT,NT}. Also
there is a nice application \cite{GNT,CET} of form factors in condensed
matter physics. The one dimensional Mott insulators can be described in
terms of the quantum sine-Gordon model.

In the present article the new approach to the `form factor program'
presented in \cite{BFKZ} is developed further. It uses the `off-shell Bethe
Ansatz' to obtain an integral representation of generalized form factors.
The approach applies also to general integrable models in 1+1 dimensions
where the nested version \cite{BKZ} of the `off-shell Bethe Ansatz has to be
used. Applications of the general case will be published elsewhere \cite
{BFKZ1,BK3}, here we restrict ourselves essentially to the simple no-nested
version. That means the general techniques are applied to the quantum
sine-Gordon alias massive Thirring model. The article is a continuation of
the previous one \cite{BFKZ} where the soliton field (an operator with
nonvanishing charge) has been investigated. Here exact expressions of all
matrix elements are obtained for several charge-less local operators.

We repeat the investigation of the current and the energy momentum tensor
which have been discussed before by Smirnov\cite{Sm}.\footnote{%
In this work similar integral representations of form factors were proposed.
We have checked that for small coupling the results for the four-particle
matrix element of the current and the energy momentum tensor agree with
ours. We could not prove that both representations agree in general.} Our
main new results are: 1) We propose the form factors of the local operators $%
\overline{\psi }\psi (x),\,\overline{\psi }\gamma ^{5}\psi (x)$ and of the
infinitely many conserved currents. 2) In order to identify the operators we
perform Feynman graph expansions and compare the results with expansions of
the exact expressions. 3) We calculate all eigenvalues of the infinitely
many conserved charges. 4) Within the frame work of integrable quantum field
theories we derive a general model independent `crossing' formula
(correcting for a sign mistake in a similar formula proposed by Smirnov\cite
{Sm}). 5) We develop the concept of the `bound state intertwiner' in the
context of bound state form factors and prove several relations. 6) Using
these techniques we derive a formula for breather form factors.

The `form factor program' is part of `bootstrap program' for integrable
quantum field theories in 1+1-dimensions. This program \emph{classifies}
integrable quantum field theoretic models and in addition it provides their
explicit exact solutions in term of all Wightman functions. These results
are obtained in three steps:

\begin{enumerate}
\item  The S-matrix is calculated by means of general properties as
unitarity and crossing, the Yang-Baxter equations (which are a consequence
of integrability) and the additional assumption of `maximal analyticity'.
This means that the two-particle S-matrix is an analytic function in the
physical plane (of the Mandelstam variable $(p_{1}+p_{2})^{2}$) and
possesses there only those poles which are of physical origin.

\item  Generalized form factors which are matrix elements of local operators 
\[
^{out}\left\langle \,p_{m}^{\prime },\ldots ,p_{1}^{\prime }\left| \mathcal{O%
}(x)\right| p_{1},\ldots ,p_{n}\,\right\rangle ^{in}\,
\]
are calculated using the S-matrix obtained in 1. More precisely, the
equations $(i)-(v)$ given below on page \pageref{pf} are used as solved.
These equations follow from LSZ-assumptions and again the additional
assumption of `maximal analyticity' (see also \cite{BFKZ}).

\item  The Wightman functions are obtained by inserting a complete set of
intermediate states. In particular the two point function for an hermitian
operator $\mathcal{O}(x)$ reads 
\begin{multline}
\langle \,0\left| \mathcal{O}(x)\,\mathcal{O}(0)\right| \,0\,\rangle
=\sum_{n=0}^{\infty }\frac{1}{n!}\int \dots \int \frac{dp_{1}\ldots dp_{n}}{%
\,(2\pi )^{n}2\omega _{1}\dots 2\omega _{n}}  \label{0.4} \\
\times \left| \left\langle \,0\left| \mathcal{O}(0)\right| p_{1},\ldots
,p_{n}\,\right\rangle ^{in}\right| ^{2}e^{-ix\sum p_{i}}.
\end{multline}
\end{enumerate}

The on-shell program i.e. the exact determination of the scattering matrix
using the Yang-Baxter equation was formulated in \cite{KTTW,KT} (for reviews
see also \cite{K,ZZ}). Off-shell considerations were carried out in \cite
{VG,W} and in \cite{KW,BKW}, where the concept of a generalized form factor
was introduced. The explicit evaluation of all the integrals and the sum in (%
\ref{0.4}) remains an open challenge. A progress towards a solution of this
problem has recently been achieved by Korepin et al. \cite{Korepin}. Up to
now it has even not been proven that the sum over the intermediate states
converges.\footnote{%
However, it is known \cite{CM} that the higher particle contributions are
very small compared to the leading ones.} We expect that our new
representations of form factors will help to solve these problems. The
`bootstrap program' does not use classical Lagrangians and any quantization
procedure to construct the quantum models. We have contact with the
classical models only, when at the end we compare our exact results with
Feynman graph expansions which are based on these Lagrangians.

In the previous paper \cite{BFKZ} an integral representation for general
soliton matrix elements of the fundamental fermi-field of the massive
Thirring model has been proposed. In the present paper we generalize this
formula and investigate in particular charge-less local operators. The
strategy is as follows:

For a state of $n$ particles of kind $\alpha _{i}$ with momenta $%
p_{i}=m\sinh \theta _{i}$ and a local operator $\mathcal{O}(x)$ the
generalized form factor is defined by 
\[
\langle \,0\,|\,\mathcal{O}(x)\,|\,\alpha _{1}(p_{1}),\dots ,\alpha
_{n}(p_{n})\,\rangle ^{in}=e^{-ix(p_{1}+\cdots +p_{n})}\mathcal{O}_{%
\underline{\alpha }}(\underline{\theta })~,~~\mathrm{for}~\theta _{1}>\dots
>\theta _{n}. 
\]
where the short notation $\underline{\alpha }=(\alpha _{1},\dots ,\alpha
_{n})$ and $\underline{\theta }=(\theta _{1},\dots ,\theta _{n})$ has been
used. We make the Ansatz 
\[
\mathcal{O}_{\underline{\alpha }}(\underline{\theta })=\int_{\mathcal{C}_{%
\underline{\theta }}}dz_{1}\cdots \int_{\mathcal{C}_{\underline{\theta }%
}}dz_{m}\,h(\underline{\theta },{\underline{z}})\,p^{\mathcal{O}}(\underline{%
\theta },{\underline{z}})\,\Psi _{\underline{\alpha }}(\underline{\theta },{%
\underline{z}}) 
\]
with the Bethe state $\Psi _{\underline{\alpha }}(\underline{\theta },{%
\underline{z}})$ defined by eq.~(\ref{2.4}) and the integration contours $%
\mathcal{C}_{\underline{\theta }}$ of figure~\ref{f5.1}. The scalar function 
$h(\underline{\theta },{\underline{z}})$ (see eqs. (\ref{1.3})-(\ref{2.12}))
is uniquely determined by the S-matrix and the `p-function' $p^{\mathcal{O}}(%
\underline{\theta },{\underline{z})}$ depends on the operator $\mathcal{O}%
(x) $. By means of the Ansatz we transform the properties $(i)-(v)$ of the
co-vector valued function $\mathcal{O}_{\underline{\alpha }}(\underline{%
\theta })$ to properties $(i^{\prime })-(v^{\prime })$ of the scalar
function $p^{\mathcal{O}}(\underline{\theta },{\underline{z})}$ which are
easily solved. In particular we obtain the p-functions for the local
operators\footnote{%
The symbol $\mathcal{N}$ refers to normal products of local quantum fields.
In perturbation theory they are defined by Zimmerman's \cite{Zi} subtraction
method, for example.} $\mathcal{N}\left[ \overline{\psi }\psi \right] (x)$, $%
\mathcal{N}\left[ \overline{\psi }\gamma ^{5}\psi \right] (x)$, the current $%
j^{\mu }(x)=\mathcal{N}\left[ \overline{\psi }\gamma ^{\mu }\psi \right] (x)$%
, the energy momentum tensor $T^{\mu \nu }(x)=\tfrac{i}{2}\mathcal{N}\left[ 
\overline{\psi }\gamma ^{\mu }\overleftrightarrow{\partial ^{\nu }\rule%
{0in}{0.14in}}\psi \right] (x)-g^{\mu \nu }\mathcal{L}^{MT}$ and the
infinitely many higher conserved currents $J_{L}^{\mu }(x)$ 
\begin{equation}
\begin{array}{llrcl}
1) &  & p^{\overline{\psi }\psi }(\underline{\theta },\underline{z}) & = & 
N_{n}^{\overline{\psi }\psi }\,q_{-}(\underline{\theta },\underline{z}) \\%
[3mm] 
2) &  & p^{\overline{\psi }\gamma ^{5}\psi }(\underline{\theta },\underline{z%
}) & = & N_{n}^{\overline{\psi }\gamma ^{5}\psi }\,q_{+}(\underline{\theta },%
\underline{z}) \\[3mm] 
3) &  & p^{j^{\pm }}(\underline{\theta },\underline{z}) & = & \pm
N_{n}^{j}\left( \sum\limits_{i=1}^{n}e^{\mp \theta _{i}}\right) ^{-1}\,q_{+}(%
\underline{\theta },\underline{z}) \\[3mm] 
4) &  & p^{T^{\pm \pm }}(\underline{\theta },\underline{z}) & = & 
N_{n}^{T}\sum\limits_{i=1}^{n}e^{\pm \theta _{i}}\left(
\sum\limits_{i=1}^{n}e^{\mp \theta _{i}}\right) ^{-1}\,q_{-}(\underline{%
\theta },\underline{z}) \\[3mm] 
&  & p^{T^{+-}}(\underline{\theta },\underline{z}) & = & -N_{n}^{T}\,q_{-}(%
\underline{\theta },\underline{z}) \\[3mm] 
5) &  & p^{J_{L}^{\pm }}(\underline{\theta },\underline{z}) & = & \pm
N_{n}^{J_{L}}\sum\limits_{i=1}^{n}e^{\pm \theta
_{i}}\sum\limits_{i=1}^{m}e^{Lz_{i}}\,,\quad (L=\pm 1,\pm 3,\dots )\,.
\end{array}
\label{4.3}
\end{equation}
\[
\text{where}\quad q_{\pm }(\underline{\theta },\underline{z}%
)=\sum\limits_{i=1}^{n}e^{-\theta _{i}}\sum\limits_{i=1}^{m}e^{z_{i}}\pm
\sum\limits_{i=1}^{n}e^{\theta _{i}}\sum\limits_{i=1}^{m}e^{-z_{i}}\,. 
\]
The identification with the operators is made by comparing the exact results
with Feynman graph expansions. Properties as charge, behavior under Lorentz
transformations etc. will also become obvious.

The article is organized as follows: In section \ref{s2} we recall some
formulae of \cite{BFKZ} which we need in the following. In section \ref{s3}
we present a general formula for solitonic form factors. In section \ref{s4}
we discuss several explicit examples and perform perturbative checks for
two- and four-particle form factors. As an example we also investigate the
asymptotic behavior when one soliton momentum goes to infinity and compare
the result with typical bosonic behavior. Section \ref{s5} contains the
derivation of a general crossing formula. Again LSZ-assumptions are used.
Also the charges of the infinitely many higher local conservation laws on
all states are calculated. In sections \ref{s6} we formulate the `bootstrap'
principle and introduce the `bound state intertwiners' (see also \cite{Q}).
Using these techniques we derive in section \ref{s7} from the pure soliton
anti-soliton form factors the mixed breather soliton and the pure breather
form factors. Section \ref{s8} contains conclusions and an outlook. Several
proofs and explicit calculations are delegated to appendices.

\section{Recall of formulae}

\label{s2}

In this section we recall some formulae which we shall need in the following
sections to present our results. All this material can be found in \cite
{BFKZ} including the original references. Coleman \cite{Co} had shown that
the sine-Gordon and the massive Thirring model are equivalent on the quantum
level. The corresponding classical models are defined by their Lagrangian's 
\begin{eqnarray}
\mathcal{L}^{SG} &=&\tfrac{1}{2}\partial _{\mu }\varphi \partial ^{\mu
}\varphi +\frac{\alpha }{\beta ^{2}}\left( \cos \beta \varphi -1\right) 
\nonumber \\
\mathcal{L}^{MT} &=&\overline{\psi }(i\gamma \partial -M)\psi -\tfrac{1}{2}%
g\,j^{\mu }j_{\mu }\,\,,\quad \left( j^{\mu }=\overline{\psi }\gamma ^{\mu
}\psi \right) .  \label{0.2}
\end{eqnarray}

\subsection{The S-matrix}

The sine-Gordon model alias massive Thirring model describes the interaction
of several types of particles: solitons, anti-solitons alias fermions and
anti-fermions and a finite number of charge-less breathers, which may be
considered as bound states of solitons and anti-solitons. In this work we
will concentrate on states consisting of solitons and anti-solitons.
Integrability of the model implies that the n-particle S-matrix factorizes
into two particle S-matrices. In particular scattering conserves the number
of particles and even their momenta. The two particle S-matrix contains the
following scattering amplitudes: the two-soliton amplitude $a(\theta )$, the
forward and backward soliton anti-soliton amplitudes $b(\theta )$ and $%
c(\theta )$:\footnote{%
This S-matrix has been obtained first by Zamolodchikov \cite{Za}
extrapolating a semiclassical result and by means of the `bootstrap program'
using the Yang-Baxter relations in \cite{KTTW}.} 
\begin{gather}
b(\theta )=\frac{\sinh \theta /\nu }{\sinh (i\pi -\theta )/\nu }\,a(\theta
)\,,~~~~c(\theta )=\frac{\sinh i\pi /\nu }{\sinh (i\pi -\theta )/\nu }%
\,a(\theta )\,,  \nonumber \\
a(\theta )=\exp \int_{0}^{\infty }\frac{dt}{t}\,\frac{\sinh \frac{1}{2}%
(1-\nu )t}{\sinh \frac{1}{2}\nu t\,\cosh \frac{1}{2}t}\,\sinh t\frac{\theta 
}{i\pi }\,.  \label{s}
\end{gather}
The parameter $\theta $ is the absolute value of the rapidity difference $%
\theta =|\theta _{1}-\theta _{2}|$ where $\theta _{i}$ are the rapidities of
the particles given by the momenta $p_{i}=M\sinh \theta _{i}$. The parameter 
$\nu $ is related to the sine-Gordon and the massive Thirring model coupling
constant by 
\[
\nu =\frac{\beta ^{2}}{8\pi -\beta ^{2}}=\frac{\pi }{\pi +2g} 
\]
where the second equality is due to Coleman \cite{Co}.

We list some general properties of the two-particle S-matrix. As usual in
this context we use in the notation 
\[
v^{1\dots n}\in V^{1\dots n}=V_{1}\otimes \cdots \otimes V_{n} 
\]
for a vector in a tensor product space. The vector components are denoted by 
$v^{\alpha _{1}\dots \alpha _{n}}$. A linear operator connecting two such
spaces with matrix elements $A_{\alpha _{1}\dots \alpha _{n}}^{\alpha
_{1}^{\prime }\dots \alpha _{n^{\prime }}^{\prime }}$ is denoted by 
\[
A_{1\dots n}^{1^{\prime }\dots n^{\prime }}:~V^{1\dots n}\to V^{1^{\prime
}\dots n^{\prime }} 
\]
where we omit the upper indices if they are obvious. All vector spaces $%
V_{i} $ are isomorphic to a space $V$ whose basis vectors label all kinds of
particles (here solitons and anti-solitons, i.e. $V\cong \Bbb{C}^{2}$). An
S-matrix such as $S_{ij}$ acts nontrivial only on the factors $V_{i}\otimes
V_{j}$.

The physical S-matrix in the formulas above is given for positive values of
the rapidity parameter $\theta $. For later convenience we will also
consider an auxiliary matrix $\dot{S}(\theta _{1},\theta _{2})$ regarded as
a function depending on the individual rapidities of both particles $\theta
_{1},\theta _{2}$ or some times also on the difference $\theta _{1}-\theta
_{2}$%
\[
\dot{S}_{12}(\theta _{1},\theta _{2})=\dot{S}_{12}(\theta _{1}-\theta
_{2})=\left\{ 
\begin{array}{lll}
(\sigma S)_{12}(|\theta _{1}-\theta _{2}|) & \text{for} & \theta _{1}>\theta
_{2} \\ 
(S\sigma )_{21}^{-1}(|\theta _{1}-\theta _{2}|) & \text{for} & \theta
_{1}<\theta _{2}
\end{array}
\right. 
\]
with $\sigma $ taking into account the statistics of the particles. It is a
diagonal matrix $\sigma _{12}$ with entries $-1$ if both particles are
fermions and $+1$ otherwise (see \cite{K1}). The matrix $\dot{S}(\theta
_{1},\theta _{2})$ is an analytic function in terms of both variables $%
\theta _{1}$ and $\theta _{2}$. This follows from unitarity $S^{\dagger }S=1$
and the fact that the physical S-matrix is the boundary value of a real
analytic function $S(s+i\epsilon )$ as a function of the Mandelstam variable 
$s=(p_{1}+p_{2})^{2}$ such that $S^{\dagger }(s+i\epsilon )=S(s-i\epsilon )$
or 
\begin{equation}
S^{\dagger }(\theta )=S^{-1}(\theta )=S(-\theta )  \label{1.65}
\end{equation}
The auxiliary matrix $\dot{S}_{12}$ acts on the factors $V_{1}\otimes V_{2}$
and in addition exchanges these factors, e.g. 
\[
\dot{S}_{12}(\theta )\,:\,V_{1}\otimes V_{2}\to V_{2}\otimes V_{1}\,. 
\]
It may be depicted as 
\[
\dot{S}_{12}(\theta _{1},\theta _{2})~~=~~ 
\begin{array}{c}
\unitlength3mm\begin{picture}(5,4) \put(1,0){\line(1,1){4}}
\put(5,0){\line(-1,1){4}} \put(0,.6){$\theta_1$} \put(4.8,.6){$\theta_2$}
\end{picture}
\end{array}
\]
Here and in the following we associate a rapidity variable $\theta _{i}\in 
\Bbb{C}$ to each space $V_{i}$ which is graphically represented by a line
labeled by $\theta _{i}$ or simply by $i$. In terms of the auxiliary
S-matrix the Yang-Baxter equation has the general form 
\begin{equation}
\dot{S}_{12}(\theta _{12})\,\dot{S}_{13}(\theta _{13})\,\dot{S}_{23}(\theta
_{23})=\dot{S}_{23}(\theta _{23})\,\dot{S}_{13}(\theta _{13})\,\dot{S}%
_{12}(\theta _{12})  \label{1.68}
\end{equation}
which graphically simply reads 
\[
\begin{array}{c}
\unitlength6mm\begin{picture}(9,4) \put(0,1){\line(1,1){3}}
\put(0,3){\line(1,-1){3}} \put(2,0){\line(0,1){4}} \put(4.3,2){$=$}
\put(6,0){\line(1,1){3}} \put(6,4){\line(1,-1){3}} \put(7,0){\line(0,1){4}}
\put(.2,.5){$1$} \put(1.3,0){$2$} \put(3,.2){$3$} \put(5.5,.2){$1$}
\put(7.3,0){$2$} \put(8.4,.4){$3$} \end{picture}~~~.
\end{array}
\]
Unitarity and crossing may be written and depicted as 
\begin{equation}
\dot{S}_{21}(\theta _{21})\dot{S}_{12}(\theta _{12})=1~:~~~~~ 
\begin{array}{c}
\unitlength3mm\begin{picture}(8,5) \put(0,1){\line(1,1){2}}
\put(2,1){\line(-1,1){2}} \put(0,3){\line(1,1){2}} \put(2,3){\line(-1,1){2}}
\put(6,1){\line(0,1){4}} \put(8,1){\line(0,1){4}} \put(3.7,2.7){$=$}
\put(0,-.5){$1$} \put(1.5,-.5){$2$} \put(6,-.5){$1$} \put(7.5,-.5){$2$}
\end{picture}
\end{array}
\label{1.8}
\end{equation}
\begin{gather}
\dot{S}_{12}(\theta _{1}-\theta _{2})=\mathbf{C}^{2\bar{2}}\,\dot{S}_{\bar{2}%
1}(\theta _{2}+i\pi -\theta _{1})\,\mathbf{C}_{\bar{2}2}=\mathbf{C}^{1\bar{1}%
}\,\dot{S}_{2\bar{1}}(\theta _{2}-(\theta _{1}-i\pi ))\,\mathbf{C}^{\bar{1}1}
\label{1.9} \\[0.12in]
\begin{array}{c}
\unitlength3mm\begin{picture}(4,5) \put(0,1){\line(1,1){4}}
\put(4,1){\line(-1,1){4}} \put(0,-.5){$1$} \put(3.7,-.5){$2$} \end{picture}
\end{array}
\quad =\quad 
\begin{array}{c}
\unitlength3mm\begin{picture}(6,5) \put(1,1){\line(1,1){4}}
\put(4,1){\line(-1,2){2}} \put(1,5){\oval(2,8)[lb]}
\put(5,1){\oval(2,8)[tr]} \put(3.5,-.5){$1$} \put(5.7,-.5){$2$} \end{picture}
\end{array}
\quad =\quad 
\begin{array}{c}
\unitlength3mm\begin{picture}(6,5) \put(2,1){\line(1,2){2}}
\put(5,1){\line(-1,1){4}} \put(1,1){\oval(2,8)[lt]}
\put(5,5){\oval(2,8)[br]} \put(0,-.5){$1$} \put(2,-.5){$2$} \end{picture}
\end{array}
~~~~~~~~  \nonumber
\end{gather}
where $\mathbf{C}^{1\bar{1}}$ and $\mathbf{C}_{1\bar{1}}$ are charge
conjugation matrices. For the sine-Gordon model the matrix elements are $%
\mathbf{C}^{\alpha \bar{\beta}}=\mathbf{C}_{\alpha \bar{\beta}}=\delta
_{\alpha \beta }$ where $\bar{\beta}$ denotes the anti-particle of $\beta $.
We have introduced the graphical rule that a line changing the ``time
direction'' also interchanges particles and anti-particles and changes the
rapidity as $\theta \to \theta \pm i\pi $. We depict this as 
\[
\mathbf{C}_{\alpha \bar{\beta}}= 
\begin{array}{c}
\unitlength4mm\begin{picture}(6,3) \put(2,1){\oval(2,4)[t]}
\put(0,1){$\theta$} \put(3.3,1){$\theta-i\pi$} \put(.7,0){$\alpha$}
\put(2.7,-.1){$\bar\beta$} \end{picture}
\end{array}
,~~~~\mathbf{C}^{\alpha \bar{\beta}}= 
\begin{array}{c}
\unitlength4mm\begin{picture}(6,3) \put(2,2){\oval(2,4)[b]}
\put(0,1){$\theta$} \put(3.3,1){$\theta+i\pi$} \put(.7,2.2){$\alpha$}
\put(2.7,2.2){$\bar\beta$} \end{picture}
\end{array}
. 
\]
Similar crossing relations will be used below to formulate the properties of
form factors.

Finally we note a property of the two-particle S-matrix 
\begin{equation}
\dot{S}_{\alpha \beta }^{\delta \gamma }(0)=-\delta _{\alpha }^{\delta
}\delta _{\beta }^{\gamma }  \label{1.10}
\end{equation}
which turns out to be true for all examples. This means that $\dot{S}$ for
zero momentum difference is equal to minus the permutation operator.

\subsection{Form factors}

For a state of $n$ particles of kind $\alpha _{i}$ with momenta $p_{i}$ and
a local operator $\mathcal{O}(x)$ we define the form factor functions $%
\mathcal{O}_{\alpha _{1},\dots ,\alpha _{n}}(\theta _{1},\dots ,\theta _{n})$
by 
\begin{equation}
\langle \,0\,|\,\mathcal{O}(x)\,|\,\alpha _{1}(p_{1}),\dots ,\alpha
_{n}(p_{n})\,\rangle ^{in}=e^{-ix(p_{1}+\cdots +p_{n})}\mathcal{O}_{%
\underline{\alpha }}(\underline{\theta })\,,\;\mathrm{for}\;\theta
_{1}>\dots >\theta _{n}.  \label{f}
\end{equation}
For all other arrangements of the rapidities the functions $\mathcal{O}_{%
\underline{\alpha }}({\underline{\theta }})$ are given by analytic
continuation. Note that in general this analytic continuation does 
\underline{not} provide the physical values of the form factor. These are
given for ordered rapidities as indicated above and for other orders of
course by the statistics of the particles. The $\mathcal{O}_{\underline{%
\alpha }}({\underline{\theta }})$ are considered as the components of a
co-vector valued function $\mathcal{O}_{1\dots n}({\underline{\theta }})\in
V_{1\dots n}=\left( V^{1\dots n}\right) ^{\dagger }$which may be depicted as 
\[
\mathcal{O}_{1\dots n}({\underline{\theta }})= 
\begin{array}{c}
\unitlength4mm\begin{picture}(6,4) \put(3,2){\oval(6,2)}
\put(3,2){\makebox(0,0){${\cal O}$}} \put(1,0){\line(0,1){1}}
\put(5,0){\line(0,1){1}} \put(0,0){$\theta_1$} \put(5.3,0){$\theta_n$}
\put(2.5,.5){$\dots$} \end{picture}
\end{array}
. 
\]

Now we formulate the main properties of form factors in terms of the
functions $\mathcal{O}_{1\dots n}({\underline{\theta }})$ which follow from
general LSZ-assumptions and ``maximal analyticity''. The later condition
means that $\mathcal{O}_{1\dots n}({\underline{\theta }})$ is a meromorphic
function with respect to all $\theta $'s and all poles in the `physical'
strips $0<\mathop{\rm Im}\theta _{ij}<\pi $~$(\theta _{ij}=\theta
_{i}-\theta _{j}\,i<j)$ are of physical origin, as for example bound state
poles as discussed in section \ref{s6}.

\paragraph{\textbf{Properties:\label{pf}}}

The co-vector valued function $\mathcal{O}_{1\dots n}({\underline{\theta }})$
is meromorphic with respect to all variables $\theta _{1},\dots ,\theta _{n}$
and

\begin{itemize}
\item[$(i)$]  it satisfies the symmetry property under the permutation of
both the variables $\theta _{i},\theta _{j}$ and the spaces $i,j$ at the
same time 
\[
\mathcal{O}_{\dots ij\dots }(\dots ,\theta _{i},\theta _{j},\dots )=\mathcal{%
O}_{\dots ji\dots }(\dots ,\theta _{j},\theta _{i},\dots )\,\dot{S}%
_{ij}(\theta _{i}-\theta _{j}) 
\]
for all possible arrangements of the $\theta $'s,

\item[$(ii)$]  it satisfies the periodicity property under cyclic
permutation of the rapidity variables and spaces 
\[
\mathcal{O}_{1\dots n}(\theta _{1},\theta _{2},\dots ,\theta _{n},)=\mathcal{%
O}_{2\dots n1}(\theta _{2},\dots ,\theta _{n},\theta _{1}-2\pi i)\sigma _{%
\mathcal{O}1} 
\]

\item[$(iii)$]  and it has poles determined by one-particle states in each
sub-channel. In particular the function $\mathcal{O}_{\underline{\alpha }}({%
\underline{\theta }})$ has a pole at $\theta _{12}=i\pi $ such that 
\[
\mathop{\rm Res}_{\theta _{12}=i\pi }\mathcal{O}_{1\dots n}(\theta
_{1},\dots ,\theta _{n})=2i\,\mathbf{C}_{12}\,\mathcal{O}_{3\dots n}(\theta
_{3},\dots ,\theta _{n})\left( \mathbf{1}-S_{2n}\dots S_{23}\right) 
\]
where $\mathbf{C}_{12}$ is the charge conjugation matrix.

\item[$(iv)$]  \label{iv}If the model also possesses bound states, the
function $\mathcal{O}_{\underline{\alpha }}({\underline{\theta }})$ has
additional poles. If for instance the particles 1 and 2 form a bound state
(12), there is a pole at $\theta _{12}=iu_{12}^{(12)}~(0<u_{12}^{(12)}<\pi )$
such that 
\[
\mathop{\rm Res}_{\theta _{12}=iu_{12}^{(12)}}\mathcal{O}_{12\dots n}(\theta
_{1},\theta _{2},\dots ,\theta _{n})\,=\mathcal{O}_{(12)\dots n}(\theta
_{(12)},\dots ,\theta _{n})\,\sqrt{2}\Gamma _{12}^{(12)}\, 
\]
where the bound state intertwiner $\Gamma _{12}^{(12)}$ and the relations of
the rapidities $\theta _{1},\theta _{2},\theta _{(12)}$ and the fusion angle 
$u_{12}^{(12)}$ will be discussed in section \ref{s6} below.

\item[$(v)$]  Since we are dealing with relativistic quantum field theories
Lorentz covariance in the form 
\[
\mathcal{O}_{1\dots n}(\theta _{1}+\mu ,\dots ,\theta _{n}+\mu )=e^{s\mu }\,%
\mathcal{O}_{1\dots n}(\theta _{1},\dots ,\theta _{n}) 
\]
holds if the local operator transforms as $\mathcal{O}\to e^{s\mu }\mathcal{O%
}$ where $s$ is the ``spin'' of $\mathcal{O}$.
\end{itemize}

In the formulae $(i)$ the statistics of the particles is taken into account
by $\dot{S}$ which means that $\dot{S}_{12}=-S_{12}$ if both particles are
fermions and $\dot{S}_{12}=S_{12}$ otherwise. In $(ii)$ the statistics of
the operator $\mathcal{O}$ is taken into account by $\sigma _{\mathcal{O}%
1}=-1$ if both the operator $\mathcal{O}$ and particle 1 are fermionic and $%
\sigma _{\mathcal{O}1}=1$ otherwise.

The properties $(i)-(iv)$ may be depicted as 
\[
\begin{array}{rrcl}
(i) & 
\begin{array}{c}
\unitlength3.2mm\begin{picture}(7,3) \put(3.5,2){\oval(7,2)}
\put(3.5,2){\makebox(0,0){${\cal O}$}} \put(1,0){\line(0,1){1}}
\put(3,0){\line(0,1){1}} \put(4,0){\line(0,1){1}} \put(6,0){\line(0,1){1}}
\put(1.4,.5){$\dots$} \put(4.4,.5){$\dots$} \end{picture}
\end{array}
& = & 
\begin{array}{c}
\unitlength3.2mm\begin{picture}(7,4) \put(3.5,3){\oval(7,2)}
\put(3.5,3){\makebox(0,0){${\cal O}$}} \put(1,0){\line(0,1){2}}
\put(3,0){\line(1,2){1}} \put(4,0){\line(-1,2){1}} \put(6,0){\line(0,1){2}}
\put(1.4,1){$\dots$} \put(4.4,1){$\dots$} \end{picture}
\end{array}
\\ 
(ii) & 
\begin{array}{c}
\unitlength4mm\begin{picture}(5,4) \put(2.5,2){\oval(5,2)}
\put(2.5,2){\makebox(0,0){${\cal O}$}} \put(1,0){\line(0,1){1}}
\put(2,0){\line(0,1){1}} \put(4,0){\line(0,1){1}} \put(2.4,.5){$\dots$}
\end{picture}
\end{array}
& = & 
\begin{array}{c}
\unitlength3.2mm\begin{picture}(7,4) \put(0,0){\line(0,1){1}}
\put(6,1){\oval(2,2)[b]} \put(3.5,1){\oval(7,6)[t]} \put(3.5,2){\oval(5,2)}
\put(3.5,2){\makebox(0,0){${\cal O}$}} \put(2,0){\line(0,1){1}}
\put(4,0){\line(0,1){1}} \put(2.4,.5){$\dots$} \end{picture}
\end{array}
\sigma _{\mathcal{O}1} \\ 
(iii) & \dfrac{1}{2i}\,\mathop{\rm Res}\limits_{\theta _{12}=i\pi }~~ 
\begin{array}{c}
\unitlength3.2mm\begin{picture}(6,4) \put(3,2){\oval(6,2)}
\put(3,2){\makebox(0,0){${\cal O}$}} \put(1,0){\line(0,1){1}}
\put(2,0){\line(0,1){1}} \put(3,0){\line(0,1){1}} \put(5,0){\line(0,1){1}}
\put(3.4,.5){$\dots$} \end{picture}
\end{array}
& = & 
\begin{array}{c}
\unitlength3.2mm\begin{picture}(5,4) \put(.5,0){\oval(1,2)[t]}
\put(3,2){\oval(4,2)} \put(3,2){\makebox(0,0){${\cal O}$}}
\put(2,0){\line(0,1){1}} \put(4,0){\line(0,1){1}} \put(2.4,.5){$\dots$}
\end{picture}
\end{array}
- 
\begin{array}{c}
\unitlength3.2mm\begin{picture}(6,5) \put(0,0){\line(0,1){3}}
\put(3,3){\oval(6,4)[t]} \put(3,3){\oval(6,4)[br]} \put(3,0){\oval(4,2)[tl]}
\put(3,3){\oval(4,2)} \put(3,3){\makebox(0,0){${\cal O}$}}
\put(2,0){\line(0,1){2}} \put(4,0){\line(0,1){2}} \put(2.4,1.5){$\dots$}
\end{picture}
\end{array}
\sigma _{\mathcal{O}1} \\ 
(iv) & \dfrac{1}{\sqrt{2}}\,\mathop{\rm Res}\limits_{\theta
_{12}=iu_{12}^{(12)}} 
\begin{array}{c}
\unitlength3.2mm\begin{picture}(5,3) \put(2.5,2){\oval(5,2)}
\put(2.5,2){\makebox(0,0){${\cal O}$}} \put(1,0){\line(0,1){1}}
\put(2,0){\line(0,1){1}} \put(4,0){\line(0,1){1}} \put(2.4,.5){$\dots$}
\end{picture}
\end{array}
& = & 
\begin{array}{c}
\unitlength3.2mm%
\begin{picture}(5,4) \put(2.5,3){\oval(5,2)} 
\put(2.5,3){\makebox(0,0){${\cal O}$}}
\put(1.5,0){\oval(1,2)[t]}
\put(1.5,1){\line(0,1){1}} \put(4,0){\line(0,1){2}} \put(2.4,1){$\dots$} 
\end{picture}
\end{array}
\end{array}
\]

As was shown in \cite{BFKZ} the properties $(i)-(iii)$ follow from general
LSZ-assump\-tions and ``maximal analyticity''. The bound state form factor
given by $(iv)$ was discussed in \cite{BFKZ} for special cases. In section 
\ref{s6} we investigate the general case and show that the bound state form
factor is consistent with the `bootstrap principle' which means that it also
satisfies $(i)-(iii)$ if the constituents do. In section \ref{s5} we derive
from the same assumptions a general crossing relation which implies $(ii)$
and $(iii)$. Conversely, it has been shown \cite{Sm,La,Q} that functions
satisfying the properties $(i)-(v)$ and the general crossing relation
represent \emph{local} operators i.e. they are form factors of $x$-dependent
operators $\mathcal{O}(x)$ which commute (anti-commute) for space like
differences of the arguments.

We will now provide a constructive and systematic way of how to solve the
properties $(i)-(v)$ for a co-vector valued function $f_{1\dots n}({%
\underline{\theta }})$ once the scattering matrix is given. These solutions
are candidates of form factors. To capture the vectorial structure of the
form factors we will employ the techniques of the algebraic Bethe Ansatz
which we briefly explain now.

\subsection{The `off-shell' Bethe Ansatz co-vectors}

As usual in the context of algebraic Bethe Ansatz we define the monodromy
matrix as 
\begin{eqnarray}
T_{1\dots n,0}({\underline{\theta }},\theta _{0}) &=&\dot{S}_{10}(\theta
_{1}-\theta _{0})\,\dot{S}_{20}(\theta _{2}-\theta _{0})\cdots \dot{S}%
_{n0}(\theta _{n}-\theta _{0})  \label{2.1} \\
&=& 
\begin{array}{c}
\unitlength3mm\begin{picture}(10,4) \put(0,2){\line(1,0){10}}
\put(2,0){\line(0,1){4}} \put(4,0){\line(0,1){4}} \put(8,0){\line(0,1){4}}
\put(1,0){$1$} \put(3,0){$ 2$} \put(7,0){$ n$} \put(9,.8){$ 0$}
\put(5,1){$\dots$} \end{picture}
\end{array}
.  \nonumber
\end{eqnarray}
It is a matrix acting in the tensor product of the ``quantum space'' $%
V^{1\dots n}=V_{1}\otimes \cdots \otimes V_{n}$ and the ``auxiliary space'' $%
V_{0}$ (all $V_{i}\cong \Bbb{C}^{2}$ = soliton-anti-soliton space). The
sub-matrices $A,B,C,D$ with respect to the auxiliary space are defined by 
\[
T_{1\dots n,0}({\underline{\theta }},z)\equiv \left( 
\begin{array}{cc}
A_{1\dots n}({\underline{\theta }},z) & B_{1\dots n}({\underline{\theta }},z)
\\ 
C_{1\dots n}({\underline{\theta }},z) & D_{1\dots n}({\underline{\theta }},z)
\end{array}
\right) \,. 
\]
A Bethe Ansatz co-vector in $V_{1\dots n}$ is given by 
\begin{equation}
\begin{array}{rcl}
\Psi _{1\dots n}({\underline{\theta }},\underline{z}) & = & \Omega _{1\dots
n}C_{1\dots n}({\underline{\theta }},z_{1})\cdots C_{1\dots n}({\underline{%
\theta }},z_{m}) \\ 
\begin{array}{c}
\unitlength5mm\begin{picture}(6,4) \put(3,2){\oval(6,2)}
\put(3,2){\makebox(0,0){$\Psi$}} \put(1,0){\line(0,1){1}}
\put(5,0){\line(0,1){1}} \put(0,0){$\theta_1$} \put(5.3,0){$\theta_n$}
\put(2.5,.5){$\dots$} \end{picture}
\end{array}
& = & 
\begin{array}{c}
\unitlength5mm\begin{picture}(6,4.5) \put(0,1){\line(1,0){6}}
\put(0,1){\vector(1,0){.5}} \put(6,1){\vector(-1,0){.5}}
\put(0,3){\line(1,0){6}} \put(0,3){\vector(1,0){.5}}
\put(6,3){\vector(-1,0){.5}} \put(1,0){\vector(0,1){4}}
\put(5,0){\vector(0,1){4}} \put(0,0){$\theta_1$} \put(4,0){$\theta_n$}
\put(5.4,.3){$z_m$} \put(5.4,3.3){$z_1$} \put(2.5,2){$\dots$}
\put(.3,1.7){$\vdots$} \put(5.3,1.7){$\vdots$} \end{picture}
\end{array}
\end{array}
\label{2.4}
\end{equation}
where $\underline{z}=(z_{1},\dots ,z_{m})$. Usually one has the restriction $%
2m\leq n$ and the charge of the state is $q=n-2m=$ number of solitons minus
number of anti-solitons. The solitons are depicted by $\uparrow $ or $%
\leftarrow $ and anti-solitons by $\downarrow $ or $\rightarrow $. The
co-vector $\Omega _{1\dots n}$ is the ``pseudo-vacuum'' consisting only of
solitons (highest weight states) 
\[
\Omega _{1\dots n}=\uparrow \otimes \cdots \otimes \uparrow \,. 
\]
It satisfies 
\[
\begin{array}{rcl}
\Omega _{1\dots n}\,B_{1\dots n}({\underline{\theta }},z) & = & 0 \\ 
\Omega _{1\dots n}\,A_{1\dots n}({\underline{\theta }},z) & = & 
\prod\limits_{i=1}^{n}\dot{a}(\theta _{i}-z)\Omega _{1\dots n} \\ 
\Omega _{1\dots n}\,D_{1\dots n}({\underline{\theta }},z) & = & 
\prod\limits_{i=1}^{n}\dot{b}(\theta _{i}-z)\Omega _{1\dots n}\,.
\end{array}
\]
The eigenvalues of the matrices $A$ and $D$, i.e. the functions $\dot{a}=-a$
and $\dot{b}=-b$ are given by the amplitudes of the scattering matrix (\ref
{s}). In the following we use the co-vector $\Psi _{1\dots n}({\underline{%
\theta }},\underline{z})$ in its `off-shell' version which means that we do
not fix the parameters $\underline{z}$ by means of Bethe Ansatz equations
but we integrate over the $z$'s \cite{B,B1,BF}.

\section{The general form factor formula}

\label{s3}

In this section we present our main result. We derive a general formula in
terms of an integral representation which allows to construct form factors
i.e. matrix elements of local fields given by eq.~(\ref{f}). More precisely,
we construct co-vector valued functions which satisfy the properties $%
(i)-(v) $ on page \pageref{pf}.

\begin{figure}[tbh]
\[
\unitlength4.2mm%
\begin{picture}(27,13)
\thicklines
\put(1,0){
\put(0,0){$\bullet~\theta_n-2\pi i$}
\put(0,2){$\bullet$}\put(.5,1.6){$\theta_n-i\pi\nu$}
\put(.19,3.2){\circle{.3}~$\theta_n-i\pi$}
\put(0,6){$\bullet~~\theta_n$}
\put(.2,6.2){\oval(1,1)}\put(-.1,5.71){\vector(-1,0){0}}
\put(.19,7.2){\circle{.3}~$\theta_n+i\pi(\nu-1)$}
\put(0,9){$\bullet~\theta_n+i\pi$}
\put(.19,11.2){\circle{.3}~$\theta_n+i\pi(2\nu-1)$}
}
\put(8,6){\dots}
\put(12,0){
\put(0,0){$\bullet~\theta_2-2\pi i$}
\put(0,2){$\bullet$}\put(.5,1.6){$\theta_2-i\pi\nu$}
\put(.19,3.2){\circle{.3}~$\theta_2-i\pi$}
\put(0,6){$\bullet~~\theta_2$}
\put(.2,6.2){\oval(1,1)}\put(-.1,5.71){\vector(-1,0){0}}
\put(.19,7.2){\circle{.3}~$\theta_2+i\pi(\nu-1)$}
\put(0,9){$\bullet~\theta_2+i\pi$}
\put(.19,11.2){\circle{.3}~$\theta_2+i\pi(2\nu-1)$}
}
\put(20,1){
\put(0,0){$\bullet~\theta_1-2\pi i$}
\put(0,2){$\bullet$}\put(.5,1.6){$\theta_1-i\pi\nu$}
\put(.19,3.2){\circle{.3}~$\theta_1-i\pi$}
\put(0,6){$\bullet~~\theta_1$}
\put(.2,6.2){\oval(1,1)}\put(-.1,5.71){\vector(-1,0){0}}
\put(.19,7.2){\circle{.3}~$\theta_1+i\pi(\nu-1)$}
\put(0,9){$\bullet~\theta_1+i\pi$}
\put(.19,11.2){\circle{.3}~$\theta_1+i\pi(2\nu-1)$}
}
\put(9,2.7){\vector(1,0){0}}
\put(0,3.2){\oval(34,1)[br]}
\put(27,3.2){\oval(20,1)[tl]}
\end{picture}
\]
\caption{\textit{The integration contour $\mathcal{C_{\protect\underline{%
\theta }}}$ (for the repulsive case $\nu >1$). The bullets belong to poles
of the integrand resulting from $u(\theta _{i}-u_{j})\,\phi (\theta
_{i}-u_{j})$ and the small open circles belong to poles originating from $%
t(\theta _{i}-u_{j})$ and $r(\theta _{i}-u_{j})$. }}
\label{f5.1}
\end{figure}
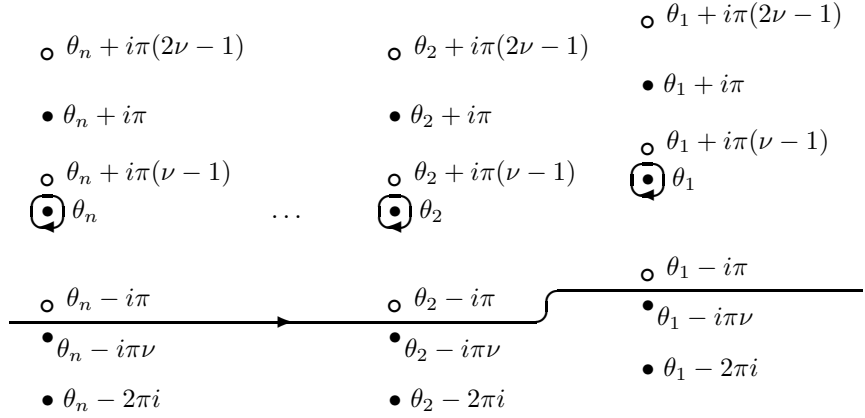

As a candidate of a generalized form factor of a local operator $\mathcal{O}%
(x)$ we make the following Ansatz for the co-vector valued function 
\begin{equation}
\fbox{$\rule{0in}{0.17in}~\mathcal{O}_{1\dots n}(\underline{\theta })=\int_{%
\mathcal{C}_{\underline{\theta }}}dz_{1}\cdots \int_{\mathcal{C}_{\underline{%
\theta }}}dz_{m}\,h(\underline{\theta },{\underline{z}})\,p^{\mathcal{O}}(%
\underline{\theta },{\underline{z}})\,\Psi _{1\dots n}(\underline{\theta },{%
\underline{z}})~$}  \label{1.2}
\end{equation}
with the Bethe Ansatz state $\Psi _{1\dots n}(\underline{\theta },{%
\underline{z}})$ defined by eq.~(\ref{2.4}). For all integration variables $%
z_{j}$ $(j=1,\dots ,m)$ the integration contours $\mathcal{C}_{\underline{%
\theta }}$ consists of several pieces (see figure~\ref{f5.1}):

\begin{itemize}
\item[a)]  A line from $-\infty $ to $\infty $ avoiding all poles such that $%
\mathop{\rm Im}\theta _{i}-\pi -\epsilon <\mathop{\rm Im}z_{j}<\mathop%
\mathrm{Im}\theta _{i}-\pi $.

\item[b)]  Clock wise oriented circles around the poles (of the $\phi
(\theta _{i}-z_{j})$) at $z_{j}=\theta _{i}$ $(i=1,\dots ,n)$.
\end{itemize}

Let the scalar function (c.f. \cite{BFKZ}) 
\begin{equation}
h(\underline{\theta },{\underline{z}})=\prod_{1\le i<j\le n}F(\theta
_{ij})\prod_{i=1}^{n}\prod_{j=1}^{m}\phi (\theta _{i}-z_{j})\prod_{1\le
i<j\le m}\tau (z_{i}-z_{j})\,,  \label{1.3}
\end{equation}
be given by 
\begin{equation}
\tau (z)=\frac{1}{\phi (z)\,\phi (-z)}~,~~~~\phi (z)=\frac{1}{F(z)\,F(z+i\pi
)}  \label{1.4}
\end{equation}
and 
\begin{equation}
F(\theta )=\sin \frac{1}{2i}\theta \,\,\exp \int_{0}^{\infty }\frac{dt}{t}%
\frac{\sinh \frac{1}{2}(1-\nu )t}{\sinh \frac{1}{2}\nu t\,\cosh \frac{1}{2}t}%
\frac{1-\cosh t(1-\theta /(i\pi ))}{2\sinh t}.  \label{1.7}
\end{equation}
The function $F(\theta )$ is the soliton-soliton form factor. It is a
solution of Watson's equations 
\begin{equation}
F(\theta )=F(-\theta )\,\dot{a}(\theta )=F(2\pi i-\theta )  \label{2.12}
\end{equation}
with $\,\dot{a}(\theta )=-a(\theta )$ where $a(\theta )$ is the
soliton-soliton scattering amplitude. It is the uniquely defined `minimal'
solution \cite{KW} which has no poles and no zeroes in the `physical strip' $%
0<\mathop{\rm Im}\theta \leq \pi $ and at most a simple zero at $\theta =0$.

\subparagraph{Remarks:}

\begin{itemize}
\item  Using Watson's equations (\ref{2.12}) for $F(z)$, crossing (\ref{1.9}%
) and unitarity (\ref{1.8}) for the sine-Gordon amplitudes one derives the
following identities for the scalar functions $\phi (z)$ and $\tau (z)$ from
the definitions (\ref{1.4}) 
\begin{equation}
\phi (z)=\phi (i\pi -z)=\frac{1}{\dot{b}(z)}\,\phi (z-i\pi )=\frac{\dot{a}%
(z-2\pi i)}{\dot{b}(z)}\,\phi (z-2\pi i)~,  \label{1.6}
\end{equation}
\[
\tau (z)=\tau (-z)=\frac{b(z)}{a(z)}\,\frac{a(2\pi i-z)}{b(2\pi i-z)}\,\tau
(z-2\pi i) 
\]
where $b(z)$ is the soliton-anti-soliton scattering amplitude related to $%
a(z)$ by crossing $b(z)=a(i\pi -z)$.

\item  The functions $\phi (z)$ and $\tau (z)$ are of the form 
\[
\phi (z)=const.~\frac{1}{\sinh z}\exp \int_{0}^{\infty }\frac{dt}{t}\frac{%
\sinh \frac{1}{2}(1-\nu )t\,\left( \cosh t(\frac{1}{2}-z/(i\pi ))-1\right) }{%
\sinh \frac{1}{2}\,\nu t\sinh t} 
\]
\[
\tau (z)=const.~\sinh z\sinh z/\nu 
\]

\item  The function $h(\underline{\theta },{\underline{z}})$ and the state $%
\Psi _{1\dots n}(\underline{\theta },{\underline{z}})$ are completely
determined by the S-matrix.
\end{itemize}

In contrast to the functions $F(z),\phi (z)$ and $\tau (z)$ the `p-function' 
$p^{\mathcal{O}}(\underline{\theta },{\underline{z}})$ in the integral
representation (\ref{1.2}) depends on the local operator $\mathcal{O}(x)$,
in particular on the spin, the charge and the statistics. The number of the
particles $n$ and the number of integrations $m$ are related by $q=n-2m$
where $q$ is the charge of the operator $\mathcal{O}(x)$. The p-functions is
an entire function in the $z_{j}~(j=1,\dots ,m)$ and in order that the form
factor satisfies the properties $(i)-(v)$ it has to satisfy the following

\paragraph{Conditions:\label{p}}

The p-function $p_{n}^{\mathcal{O}}(\underline{\theta },\underline{z})$
(where $n$ is the number of particles and the number of variables $\theta $)
satisfies

\begin{itemize}
\item[$(i^{\prime })$]  $p_{n}^{\mathcal{O}}(\underline{\theta },\underline{z%
})$ is symmetric with respect to the $\theta $'s and the $z$'s.

\item[$(ii^{\prime })$]  $p_{n}^{\mathcal{O}}(\underline{\theta },\underline{%
z})=\sigma _{\mathcal{O}i}p_{n}^{\mathcal{O}}(\dots ,\theta _{i}-2\pi
i,\dots ,\underline{z})$ and it is a polynomial in $e^{\pm z_{j}}~(j=1,\dots
,m)$.

The statistics factor $\sigma _{\mathcal{O}i}$ is $-1$ if the operator $%
\mathcal{O}(x)$ and the particle $i$ are both fermionic and $+1$ otherwise.

\item[$(iii^{\prime })$]  $\left\{ 
\begin{array}{l}
p_{n}^{\mathcal{O}}(\theta _{1}=\theta _{n}+i\pi ,\tilde{\underline{\theta }}%
,\theta _{n};\tilde{\underline{z}},z_{m}=\theta _{n}{)}=\dfrac{\varkappa }{m}%
\,p_{n-2}^{\mathcal{O}}(\tilde{\underline{\theta }},\tilde{\underline{z}})+%
\tilde{p}^{(1)}({\underline{\theta }}) \\ 
p_{n}^{\mathcal{O}}(\theta _{1}=\theta _{n}+i\pi ,\tilde{\underline{\theta }}%
,\theta _{n};\tilde{\underline{z}},z_{m}=\theta _{1}{)}=\sigma _{\mathcal{O}%
1}\dfrac{\varkappa }{m}\,p_{n-2}^{\mathcal{O}}(\tilde{\underline{\theta }},%
\tilde{\underline{z}})+\tilde{p}^{(2)}({\underline{\theta }})
\end{array}
\right. $

where $\tilde{\underline{\theta }}=(\theta _{2},\dots ,\theta _{n-1}),\;%
\tilde{\underline{z}}=(z_{1},\dots z_{m-1})$ and where $\tilde{p}^{(1,2)}({%
\underline{\theta }})$ are independent of the $z$'s and non-vanishing only
for charge-less operators $\mathcal{O}(x)$. The constant $\varkappa $
depends on the coupling and is given by (see formula (B.7) in \cite{BFKZ}) 
\begin{equation}
\varkappa =-\left( F^{\prime }(0)\right) ^{2}/\pi \,.  \label{3.1}
\end{equation}

\item[$(iv^{\prime })$]  the bound state p-functions are investigated in
section \ref{s6}

\item[$(v^{\prime })$]  $p_{n}^{\mathcal{O}}(\underline{\theta }+\mu ,%
\underline{z}+\mu )=e^{s\mu }p_{n}^{\mathcal{O}}(\underline{\theta },%
\underline{z})$ where $s$ is the `spin' of the operator $\mathcal{O}(x)$.
\end{itemize}

As an extension of theorem 4.1 in \cite{BFKZ} we prove the following theorem
which allows to construct generalized form factors.

\begin{theorem}
\label{t} The co-vector valued function $\mathcal{O}_{1\dots n}({\underline{%
\theta }})$ defined by (\ref{1.2}) fulfills the properties $(i),\,(ii)$ and $%
(iii)$ on page \pageref{pf} if the functions $F(\theta ),\phi (z)$ and $\tau
(z)$ are given by definition (\ref{1.3}) -- (\ref{1.7}) and if the
p-function $p_{n}^{\mathcal{O}}(\underline{\theta },\underline{z})$
satisfies the conditions $(i^{\prime })-(iii^{\prime })$.
\end{theorem}

\proof%
The properties $(i)$ and $(ii)$ follow as in \cite{BFKZ}. The proof of $%
(iii) $ is also the same as in \cite{BFKZ}, if the functions $\tilde{p}%
^{(1)}({\underline{\theta }})$ and $\tilde{p}^{(2)}({\underline{\theta }})$
in $(iii^{\prime })$ vanish. For the case of charge-less operators they are
in general non-vanishing. Then the proof of $(iii)$ in the form of 
\begin{multline*}
\mathop{\rm Res}_{\theta _{1n}=i\pi }\mathcal{O}_{1\dots n}(\theta
_{1},\dots ,\theta _{n})=-2i\,\mathbf{C}_{1n}\,\mathcal{O}_{2\dots
n-1}(\theta _{2},\dots ,\theta _{n-1}) \\
\times \left( \mathbf{1}_{2\dots n-1}-S_{2n}\cdots S_{n-1n}\right)
\end{multline*}
has to be modified as follows: Considering the $z_{m}$-integration as in 
\cite{BFKZ} the terms involving $p_{n-2}^{\mathcal{O}}(\tilde{\underline{%
\theta }},\tilde{\underline{z}})$ yield the desired result and those
proportional to $\tilde{p}^{(1,2)}({\underline{\theta }})$ yield terms with
a factor $\mathbf{1}_{2\dots n-1}+S_{2n}\cdots S_{n-1n}$ which, however,
vanish due to the following lemma.

\begin{lemma}
\label{l1} The integral given by (\ref{1.2}) and (\ref{1.3}) vanishes if the
p-function $p(\underline{\theta },\underline{z})$ is independent of the
integration variables $z_{i}$ and if the number of particles $n$ and the
number of C-operators $m$ are related by $n=2m$ which means that the charge $%
q=n-2m$ vanishes.
\end{lemma}

This lemma is proven in appendix \ref{a1}.

\subparagraph{Remarks:}

\begin{itemize}
\item  The number of C-operators $m$ depends on the charge $q=n-2m$ of the
operator $\mathcal{O}$, e.g. $m=(n-1)/2$ for the soliton field $\psi (x)$
with charge $q=1$ and $m=n/2$ for charge-less operators like $\overline{\psi 
}\psi $ or the energy momentum tensor $T^{\mu \nu }$.

\item  Note that other sine-Gordon form factors can be calculated from the
general formula (\ref{1.2}) using the bound state formula $(iv)$.

\item  The general representation of form factors by formula (\ref{1.2}) is
not specific to the sine-Gordon model. It may be applied to all integrable
quantum field theoretic model. So the main task is to solve the
corresponding Bethe Ansatz.
\end{itemize}

\section{Examples}

\label{s4}

In this section we propose the p-functions corresponding to some charge-less
local operators. For two-particle form factors the single integration is
performed explicitly and compared with known results of \cite{KW} and with
Feynman graph expansion in lowest order. For 4-particle form factors the
double integrals are calculated approximately by expansion in small
couplings to lowest order. The results are again checked against Feynman
graph expansions. In all cases agreement is obtained.

For the massive Thirring model with the Lagrangians and field equation 
\begin{gather*}
\mathcal{L}^{MT}=\overline{\psi }(i\gamma \partial -M)\psi -\tfrac{1}{2}%
g\,j^{\mu }j_{\mu }\,\,,\quad \left( j^{\mu }=\overline{\psi }\gamma ^{\mu
}\psi \right) \\
(i\gamma \partial -M)\psi (x)-g\,j^{\mu }(x)\gamma _{\mu }\psi (x)=0
\end{gather*}
we are looking for all matrix elements of the quantum operators
corresponding to the following classical fields

\begin{itemize}
\item[1)]  $\overline{\psi }(x)\psi (x)$

\item[2)]  $\overline{\psi }(x)\gamma ^{5}\psi (x)$

\item[3)]  $j^{\mu }(x)=\overline{\psi }(x)\gamma ^{\mu }\psi (x)$ the
topological (electro magnetic) current

\item[4)]  $T^{\mu \nu }(x)=\tfrac{i}{2}\overline{\psi }\gamma ^{\mu }%
\overleftrightarrow{\partial ^{\nu }\rule{0in}{0.14in}}\psi -g^{\mu \nu }%
\mathcal{L}^{MT}$ the energy momentum tensor.

The light cone components of this tensor are 
\begin{eqnarray*}
T^{\pm \pm }=T^{00}\pm 2T^{01}+T^{11}=\overline{\psi }\gamma ^{\pm }\frac{i}{%
2}\overleftrightarrow{\partial ^{\pm }}\psi \\
T^{+-}=T^{-+}=T^{00}-T^{11}=\overline{\psi }\gamma ^{+}\frac{i}{2}%
\overleftrightarrow{\partial ^{-}}\psi -g\,j^{\mu }j_{\mu }=M\overline{\psi }%
\psi
\end{eqnarray*}
where $\partial ^{\pm }=\partial ^{0}\pm \partial ^{1}$ and $\gamma ^{\pm
}=\gamma ^{0}\pm \gamma ^{1}$. For the last equality the field equation has
been used .

\item[5)]  The higher conserved currents \cite{BKT,KN} for $L=3,5,\dots $%
\[
J_{L}^{\pm }(x)=\left\{ 
\begin{array}{l}
i\psi _{1}^{\dagger }\left( \partial ^{+}\right) ^{L}\psi _{1}+h.c.+O(\psi
^{4}) \\ 
M\psi _{2}^{\dagger }\left( \partial ^{+}\right) ^{L-1}\psi _{1}+h.c.+O(\psi
^{4})
\end{array}
\right. \quad \psi =\left( 
\begin{array}{l}
\psi _{1} \\ 
\psi _{2}
\end{array}
\right) . 
\]
A second set of higher conserved currents is obtained by exchanging $%
\partial ^{+}\leftrightarrow \partial ^{-}$ and $\psi _{1}\leftrightarrow
\psi _{2}$ which we associate to $L=-3,-5,\dots $ .
\end{itemize}

\subsection{Examples of `p-functions'}

\label{s41}

In this subsection we propose the p-functions for various local operators.
Since the charge of the operators which we consider is zero, the number of
integrations $m$ and the number of particles $n$ are related by $m=n/2$ and
form factors are non-vanishing only for even number of particles $%
n=2,4,\dots $. We consider p-functions of the form 
\begin{equation}
p_{n}^{\mathcal{O}}(\underline{\theta },{\underline{z}})=N_{n}^{\mathcal{O}%
}\left( p_{+}^{\mathcal{O}}(P^{\mu })\sum_{j=1}^{m}e^{Lz_{j}}+p_{-}^{%
\mathcal{O}}(P^{\mu })\sum_{j=1}^{m}e^{-Lz_{j}}\right)  \label{4.1}
\end{equation}
where $P^{\mu }$ is the total energy momentum vector of all particles. The
integrals in (\ref{2.1}) converge for $L<(1/\nu +1)(n/2-m+1)+1/\nu $. For
large values of $L$ the form factors are defined in general as analytic
continuations of the integral representation from sufficiently small values
of $\nu $ to other values. Obviously the p-functions (\ref{4.1}) satisfy the
conditions $(i^{\prime })-(iii^{\prime })$ on page \pageref{p}. From the
property $(iii^{\prime })$ we obtain the recursion relation for the
normalization constants 
\begin{equation}
N_{n}^{\mathcal{O}}=N_{n-2}^{\mathcal{O}}\frac{\varkappa }{m}\quad
\Rightarrow \quad N_{n}^{\mathcal{O}}=N_{2}^{\mathcal{O}}\frac{1}{m!}%
\varkappa ^{m-1}.  \label{4.2}
\end{equation}
The absolute normalizations follow from the two-particle form factors (see
eq.~(\ref{4.6}) below).

We propose that the p-functions of equations (\ref{4.3}) on page \pageref
{4.3} are associated to the local operators $\mathcal{N}\left[ \,\overline{%
\psi }\psi \right] (x),\,\mathcal{N}\left[ \,\overline{\psi }\gamma ^{5}\psi
\right] (x),j^{\pm }(x),\,T^{\pm \pm }(x),\,T^{+-}(x)$ and $J_{L}^{\pm }(x)$
where $\pm $ denote the light cone components (e.g. $j^{\pm }=j^{0}\pm j^{1}$%
). The vector operator $j^{\mu }(x)=\mathcal{N}\left[ \,\overline{\psi }%
\gamma ^{\mu }\psi \right] (x)$ is the topological (electro-magnetic)
current, $T^{\mu \nu }(x)$ is the energy momentum tensor and the $J_{L}^{\pm
}(x)$ are the higher conserved currents. The normalization constants are
obtained in the next subsection, where we calculate the exact two-particle
form factors. The fundamental sine-Gordon bose field $\varphi (x)$ which
correspond to the lowest breather is related to the current by Coleman's
formula \cite{Co} 
\begin{equation}
\epsilon ^{\mu \nu }\partial _{\nu }\varphi =-\frac{2\pi }{\beta }j^{\mu
}\quad \text{or}\quad \partial ^{\pm }\varphi =\pm \frac{2\pi }{\beta }%
j^{\pm }.  \label{4.4}
\end{equation}
This implies the following representation for the p-function 
\[
p_{n}^{\varphi }(\underline{\theta },\underline{z})=N_{n}^{j}\frac{2\pi i}{%
\beta M}\left( \sum_{i=1}^{n}e^{\theta }\sum_{i=1}^{n}e^{-\theta }\right)
^{-1}\left( \sum_{i=1}^{n}e^{-\theta
}\sum_{i=1}^{m}e^{z}+\sum_{i=1}^{n}e^{\theta }\sum_{i=1}^{m}e^{-z}\right) . 
\]

Using the integral representation (\ref{1.2}) with these p-functions we
calculate in the following subsections the exact two-particle form factors
and the four-particle form factors in lowest order with respect to the
coupling $g$.

\paragraph{Conservation of higher charges}

The higher currents satisfy in terms of matrix elements $\mathcal{O}_{1\dots
n}(\underline{\theta })=\langle \,0\,|\,\mathcal{O}\,|\,p_{1},\dots
,p_{n}\,\rangle _{1\dots n}^{in}$ for all $L\in \Bbb{Z}$ the equation 
\[
\partial \,^{+}J_{L}^{-}(x)+\partial \,^{-}J_{L}^{+}(x)=0, 
\]
such that the higher charges are conserved 
\[
\frac{d}{dt}Q_{L}=\frac{d}{dt}\int dxJ_{L}^{0}(x)=0. 
\]

\proof%
>From the definition we get the correspondence of operators and p-functions 
\[
\partial ^{+}J_{L}^{-}+\partial ^{-}J_{L}^{+}\leftrightarrow
N_{n}^{J_{L}}iM\left( \sum_{i=1}^{n}e^{\theta _{i}}\sum_{i=1}^{n}e^{-\theta
_{i}}-\sum_{i=1}^{n}e^{-\theta _{i}}\sum_{i=1}^{n}e^{\theta _{i}}\right)
\sum_{i=1}^{m}e^{Lz_{i}}=0. 
\]
\endproof%

In the following subsection we calculate the charges on 1-particles states
and in section \ref{s5} on arbitrary n-particles states. It turns out that
for even $L$ the charges vanish as in the classical case. The energy
momentum tensor $T^{\mu \nu }$ is given by $J_{\pm 1}^{\pm }$ and the
momentum operator by $P^{0}\pm P^{1}=Q_{\pm 1}$.

\subsection{Examples of two particle form factors}

Two particle form factors may in general be obtained by diagonalization of
the two-particle S-matrix \cite{KW}. For several examples we shall show of
p-functions that the integral representation in this case reduces to known
results for two particle form factors for specific operators. This allows us
to confirm the association of p-functions and local operators as proposed
above. Also the normalization constants can be calculated. Moreover we check
our results in lowest order of perturbation theory.

We consider the form factor for charge-less operators 
\[
\mathcal{O}_{12}(\theta _{1},\theta _{2})=\langle \,0\,|\,\mathcal{O}%
\,|\,p_{1},p_{2}\,\rangle _{12}^{in}. 
\]
Non-vanishing matrix elements contain one soliton and one anti-soliton.
Choosing $n=2$ and $m=1$ in the general formula (\ref{1.2}) we obtain 
\begin{equation}
\mathcal{O}_{12}(\underline{\theta })=F(\theta _{12})\int_{\mathcal{C}_{%
\underline{\theta }}}dz\prod_{i=1}^{2}\phi (\theta _{i}-z)\,p^{\mathcal{O}}(%
\underline{\theta },z)\,\Omega _{12}\,C_{12}(\underline{\theta },z).
\label{1.5}
\end{equation}
The integration can be performed. As a special case we take a p-function of
the form (\ref{4.1}) for $L=1$ and prove the following lemma:

\begin{lemma}
\label{l3}For the simple p-functions $e^{\pm z}$ the integral in (\ref{1.5})
can be performed yielding the result 
\[
f_{12}^{\pm }(\underline{\theta })=\frac{2\sinh \tfrac{1}{2}\theta _{12}}{%
\nu \varkappa }\,e^{\pm \frac{1}{2}(\theta _{1}+\theta _{2})}\left( \pm 
\frac{f_{+}(\theta _{12})E_{12}^{+}}{\cosh \tfrac{1}{2}\theta _{12}}-\frac{%
f_{-}(\theta _{12})E_{12}^{-}}{\sinh \tfrac{1}{2}\theta _{12}}\right) 
\]
where $E^{\pm }=\left( s\otimes \bar{s}\pm \bar{s}\otimes s\right) $ is the
symmetric (anti-symmetric) soliton anti-soliton state and the constant $%
\varkappa $ is defined in eq.~(\ref{3.1}). The functions $f_{\pm }(\theta )$
are the positive and negative C-parity two-particle sine-Gordon form factors
calculated in \cite{KW} 
\[
\left( f_{+}(\theta )\,,\,f_{-}(\theta )\right) =\left( \frac{\tanh \frac{1}{%
2}(i\pi -\theta )}{\sinh \frac{1}{2\nu }(i\pi -\theta )}\,,\,\frac{1}{\cosh 
\frac{1}{2\nu }(i\pi -\theta )}\right) F(\theta ) 
\]
with $f_{ss}^{(0)}(\theta )$ given by eq.~(\ref{1.7}).
\end{lemma}

\proof%
We consider the expression 
\begin{eqnarray*}
I^{\pm } &=&\frac{a(\theta _{12})}{c(\theta _{12})}\int_{C_{\underline{%
\theta }}}dz\,\tilde{I}(z)\,e^{\pm z}\,\Omega \tilde{C}(\underline{\theta }%
,z) \\
&=&\frac{a(\theta _{12})}{c(\theta _{12})}\int_{C_{\underline{\theta }}}%
\tilde{I}(z)\,e^{\pm z}\left\{ \frac{b(\theta _{1}-z)}{a(\theta _{1}-z)}%
\frac{c(\theta _{2}-z)}{a(\theta _{2}-z)}\,s\otimes \bar{s}+\frac{c(\theta
_{1}-z)}{a(\theta _{1}-z)}\,\bar{s}\otimes s\right\}
\end{eqnarray*}
with $\tilde{I}(z)=\prod_{i=1,2}\tilde{\phi}(\theta _{i}-z)$ , $\tilde{\phi}%
(\theta )=\phi (\theta )a(\theta )$ and \newline
$\tilde{C}(\underline{\theta },z)=C(\underline{\theta },z)\prod_{i=1,2}1/a(%
\theta _{i}-z)$. Inserting the identities (which follow from Yang-Baxter
relations for the soliton S-matrix) 
\begin{eqnarray*}
\frac{a(\theta _{12})}{c(\theta _{12})} &=&\frac{a(\theta _{1}-z)}{c(\theta
_{1}-z)}\,\frac{a(\theta _{2}-z)}{c(\theta _{2}-z)}-\frac{b(\theta
_{1}-z-2\pi i)}{c(\theta _{1}-z-2\pi i)}\,\frac{b(\theta _{2}-z)}{c(\theta
_{2}-z)} \\
&=&-\frac{a(\theta _{1}-z)}{c(\theta _{1}-z)}\,\frac{a(\theta _{2}-z+2\pi i)%
}{c(\theta _{2}-z+2\pi i)}+\frac{b(\theta _{1}-z)}{c(\theta _{1}-z)}\,\frac{%
b(\theta _{2}-z)}{c(\theta _{2}-z)}
\end{eqnarray*}
into the two components of the integral, respectively and using the shift
property (\ref{1.6}) 
\[
\tilde{\phi}(\theta -z-2\pi i)=\frac{b(\theta -z)}{a(\theta -z)}\,\tilde{\phi%
}(\theta -z) 
\]
we obtain 
\begin{multline*}
I^{\pm }=\left[ \int_{\mathcal{C}_{\underline{\theta }}}-\int_{\mathcal{C}_{%
\underline{\theta }}+2\pi i}\right] dz\,\tilde{I}(z)\,e^{\pm z}\left\{ \frac{%
b(\theta _{1}-z)}{c(\theta _{1}-z)}\,s\otimes \bar{s}-\frac{a(\theta
_{2}-z+2\pi i)}{c(\theta _{2}-z+2\pi i)}\,\bar{s}\otimes s\right\} \\
=-2\pi i\left[ \tilde{I}(z)\,e^{\pm z}\left\{ \frac{b(\theta _{1}-z)}{%
c(\theta _{1}-z)}\,s\otimes \bar{s}-\frac{a(\theta _{2}-z+2\pi i)}{c(\theta
_{2}-z+2\pi i)}\,\bar{s}\otimes s\right\} \right] _{-\infty }^{\infty }.
\end{multline*}
There are no poles inside the integration contour. However, there are
contributions at $\pm \infty $. With the asymptotic formulae for $%
\mathop{\rm Re}z\to \pm \infty $ 
\begin{eqnarray*}
a(\theta -z)/b(\theta -z) &\approx &e^{\pm i\pi (1/\nu -1)} \\
b(\theta -z)/c(\theta -z) &\approx &\frac{\mp 1}{2i\sin (\pi /\nu )}\,e^{\mp
(\theta -z)/\nu } \\
\tilde{\phi}(\theta -z) &\approx &\frac{4}{\sqrt{4\pi \nu \varkappa }}e^{\pm 
\frac{i}{2}\pi (1/\nu -1)}\,e^{\pm \frac{1}{2}(1/\nu +1)(\theta -z-i\pi /2)}.
\end{eqnarray*}
which are derived in appendix \ref{a2} and with $c(\theta )/a(\theta )=i\sin 
\frac{\pi }{\nu }/\sinh \frac{\pi }{\nu }(i\pi -\theta )$ we obtain the
claim.%
\endproof%

\subsubsection{The exact form factors}

We use the following conventions for the $\gamma $-matrices and the spinors 
\begin{gather*}
\gamma ^{0}=\left( 
\begin{array}{cc}
0 & 1 \\ 
1 & 0
\end{array}
\right) ,~\gamma ^{1}=\left( 
\begin{array}{cc}
0 & 1 \\ 
-1 & 0
\end{array}
\right) ,~\gamma ^{5}=\gamma ^{0}\gamma ^{1} \\
u(p)=\sqrt{M}\left( 
\begin{array}{c}
e^{-\theta /2} \\ 
e^{\theta /2}
\end{array}
\right) ,~v(p)=\sqrt{M}\,i\left( 
\begin{array}{c}
e^{-\theta /2} \\ 
-e^{\theta /2}
\end{array}
\right) .
\end{gather*}
For the examples 1) -- 4) above we calculate the two-particle form factors 
\[
\mathcal{O}_{s\bar{s}}(\theta _{1},\theta _{2})=\langle \,0\,|\,\mathcal{O}%
(0)\,|\,p_{1},p_{2}\,\rangle _{s\bar{s}}^{in} 
\]
applying lemma \ref{l3} and using the p-functions (\ref{4.3}) explicitly%
\footnote{%
For the cases 3) and 4) these results agree with those of \cite{W} and \cite
{Sm}, respectively, which have been obtained by solving the scalar Watson's
equations due to diagonalization of the S-matrix.} 
\begin{equation}
\begin{array}{llrcl}
1) &  & \langle \,0\,|\,\mathcal{N}\left[ \overline{\psi }\psi \right]
(0)\,|\,p_{1},p_{2}\,\rangle _{s\bar{s}}^{in} & = & \bar{v}(\theta
_{2})u(\theta _{1})\,f_{+}(\theta _{12})/\nu \\[2mm] 
2) &  & \langle \,0\,|\,\mathcal{N}\left[ \overline{\psi }\gamma ^{5}\psi
\right] (0)\,|\,p_{1},p_{2}\,\rangle _{s\bar{s}}^{in} & = & \bar{v}(\theta
_{2})\gamma ^{5}u(\theta _{1})\,f_{-}(\theta _{12})/\nu \\[2mm] 
3) &  & \langle \,0\,|\,j^{\pm }(0)\,|\,p_{1},p_{2}\,\rangle _{s\bar{s}}^{in}
& = & \bar{v}(\theta _{2})\gamma ^{\pm }u(\theta _{1})\,f_{-}(\theta _{12})
\\[2mm] 
4) &  & \langle \,0\,|\,T^{\rho \sigma }(0)\,|\,p_{1},p_{2}\,\rangle _{s\bar{%
s}}^{in} & = & \bar{v}(\theta _{2})\gamma ^{\rho }u(\theta _{1})\,\tfrac{1}{2%
}(p_{1}^{\sigma }-p_{2}^{\sigma })\,f_{+}(\theta _{12})/\nu \,.
\end{array}
\label{4.5}
\end{equation}
For the higher conserved currents see the next paragraph. The normalization
constants $N_{2}^{\mathcal{O}}$ in the expressions for the two-particle form
factors of (\ref{4.5}) have been determined by the following normalization
conditions 
\[
\begin{array}{llrcl}
1) &  & _{s}\langle \,p\,|\,\mathcal{N\,}\left[ \,\overline{\psi }\psi
\right] (0)\,|\,p\,\rangle _{s} & = & \bar{u}(\theta )u(\theta )=2M \\[2mm] 
3) &  & _{s}\langle \,p\,|\,\mathcal{N\,}\left[ \,\overline{\psi }\gamma
^{\mu }\psi \right] (0)\,|\,p\,\rangle _{s} & = & \bar{u}(\theta )\gamma
^{\mu }u(\theta )=2p^{\mu } \\[2mm] 
4) &  & _{s}\langle \,p\,|\,T^{\mu \nu }(0)\,|\,p\rangle _{s}^{in} & = & 
\bar{u}(\theta )\gamma ^{\mu }u(\theta )p^{\nu }=2p^{\mu }p^{\nu }
\end{array}
\]
which are the free field values and which are natural due to the
corresponding `charges' of the operators. The crossing relations and $%
f_{+}(i\pi )=\nu $ and $f_{-}(i\pi )=1$ have been used. Since there is no
`charge' for $\,\mathcal{N}\left[ \,\overline{\psi }\gamma ^{5}\psi \right]
(x)$ we take $N_{2}^{\overline{\psi }\gamma ^{5}\psi }=-N_{2}^{\overline{%
\psi }\psi }$ which implies the natural relation $_{s}\langle \,p^{\prime
}\,|\,\mathcal{N\,}\left[ \overline{\psi }\gamma ^{5}\psi \right]
(0)\,|\,p\,\rangle _{s}\approx \bar{u}(\theta ^{\prime })\gamma ^{5}u(\theta
)$ for $\theta ^{\prime }\approx \theta $ for small couplings. This
normalization is also consistent with the desired identification (see \cite
{Co}) 
\[
\mathcal{N}\left[ \,\overline{\psi }\left( 1\pm \gamma ^{5}\right) \psi
\right] (x)=-\frac{2\alpha }{M\beta ^{2}}\left( 1-\frac{\beta ^{2}}{8\pi }%
\right) :e^{\pm \beta \varphi (x)}: 
\]
where $:\dots :$ means normal ordering with respect to the physical vacuum
(see also \cite{BK1,BK}). The constant on the right hand side is obtained
from the trace of the energy momentum tensor calculated in \cite{BK1,BK}.
For the cases 3) of the topological (electro-magnetic) current and 4) the
energy momentum tensor the given normalization are equivalent to the
eigenvalue relations 
\begin{eqnarray*}
\int dxj^{0}(x)|\,p\,\rangle _{s} &=&|\,p\,\rangle _{s} \\
\int dxT^{\mu 0}(x)|\,p\,\rangle _{s} &=&p^{\mu }|\,p\,\rangle _{s}.
\end{eqnarray*}
Finally we obtain from the recursion relation (\ref{4.2}) and lemma \ref{l3}
the normalization constants: 
\begin{equation}
N_{n}^{\overline{\psi }\psi }=-N_{n}^{\overline{\psi }\gamma ^{5}\psi }=%
\frac{1}{2\nu }N_{n}^{j}=-\frac{1}{M}N_{n}^{T}=-iM\frac{1}{4m!}\varkappa ^{m}
\label{4.6}
\end{equation}
where the constant $\varkappa $ is defined in eq.~(\ref{3.1}). Note that
these normalizations together with the proposed p-functions (\ref{4.3})
imply in particular the quantum version of the classical field operator
relation 
\[
\mathop{\rm tr}T(x)=T^{+-}(x)=M\mathcal{N}\left[ \overline{\psi }\psi
\right] (x) 
\]
for all matrix elements. For the sine-Gordon field Coleman's relation (\ref
{4.4}) yields \cite{KW} 
\[
\varphi _{s\bar{s}}(\theta _{1},\theta _{2})=\langle \,0\,|\,\varphi
(0)\,|\,p_{1},p_{2}\,\rangle _{s\bar{s}}^{in}=-\frac{2\pi }{\beta }\frac{%
f_{-}(\theta _{12})}{\cosh \tfrac{1}{2}\theta _{12}}. 
\]

\paragraph{The higher conserved currents:}

We will use a generalized version of lemma \ref{l3} where $e^{\pm z}$ is
replaced by $e^{Lz}$. The two particle form factors then turn out to be 
\begin{multline}
\left[ J_{L}^{\pm }\right] _{12}(\theta _{1},\theta _{2})=\langle
\,0\,|\,J_{L}^{\pm }\,|\,p_{1},p_{2}\,\rangle _{12}^{in}=\pm
N_{2}^{J_{L}}\,\sinh \tfrac{1}{2}\theta _{12}\left( \sum_{i=1}^{2}e^{\pm
\theta _{i}}\right) e^{\frac{1}{2}L(\theta _{1}+\theta _{2})}  \label{4.63}
\\
\times k_{L}(\theta _{12})\left( \frac{\mathop{\rm sgn}L}{\cosh \frac{1}{2}%
\theta _{12}}\,f_{+}(\theta _{12})E_{12}^{+}-\frac{1}{\sinh \frac{1}{2}%
\theta _{12}}\,f_{-}(\theta _{12})E_{12}^{-}\right) .
\end{multline}
The asymptotic behavior of $\phi (\theta )$ yields the functions (see
appendix \ref{a2}) 
\[
k_{L}(\theta )=const\sum_{i=0}\sum_{j=0}\delta _{i+j,|L|-1}\,A_{i}A_{j}e^{%
\frac{1}{2}(i-j)\theta } 
\]
where the constants $A_{i}$ are also given in appendix \ref{a2}. Only the
first term proportional to $f_{+}(\theta _{12})$ contributes to the higher
charges $Q_{L}=\int dxJ_{L}^{0}$. With a suitable normalization ($%
N_{2}^{J_{L}}\,k_{L}(i\pi )\,\nu =M^{L+1}$) and for odd $L=\pm 1,\pm 3,\dots
.$ one obtains 
\begin{align}
~_{s}\langle \,p_{2}\,|\,Q_{L}\,|\,p_{1}\,\rangle _{s}& =2\pi \delta
(p_{1}^{1}-p_{2}^{1})\tfrac{1}{2}\left( \left[ J_{L}^{+}\right] _{s\bar{s}%
}(\theta _{1},\theta _{1}-i\pi )+\left[ J_{L}^{-}\right] _{s\bar{s}}(\theta
_{1},\theta _{1}-i\pi )\right)  \nonumber \\
& =\langle \,p_{2}\,|\,p_{1}\,\rangle \,\left( p_{1}^{+}\right) ^{L}.
\label{4.65}
\end{align}
Note that $k_{L}(i\pi )=0$ for $L$ even.

\subsubsection{Feynman graph expansion}

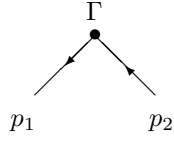
\begin{figure}[tbh]
\[
\unitlength4mm%
\begin{picture}(4,4)
\put(0,1){\line(1,1){2}} \put(1,2){\vector(-1,-1){.0}}
\put(4,1){\line(-1,1){2}} \put(3,2){\vector(-1,1){.0}}
\put(2,3){\makebox(0,0){$\bullet$}}
\put(1.7,3.5){$\Gamma$}
\put(-.8,0){$p_1$} \put(3.8,0){$p_2$}
\end{picture}
\]
\caption{The Feynman graph for two-particle form factors of charge-less
operators.}
\label{f1}
\end{figure}

From the Lagrangian (\ref{0.2}) we get the Feynman rules of figure~\ref{fa1}. 
\begin{figure}[tbh]
\[
\begin{array}{c}
\unitlength3mm \begin{picture}(4,4) \put(0,0){\line(1,1){4}}
\put(0,0){\vector(1,1){1}} \put(2,2){\vector(1,1){1.5}}
\put(4,0){\line(-1,1){4}} \put(4,0){\vector(-1,1){1}}
\put(2,2){\vector(-1,1){1.5}} \put(2,2){\makebox(0,0){$\bullet$}}
\end{picture}
\end{array}
=-ig\gamma^\mu\otimes\gamma_\mu~,~~~ 
\begin{array}{c}
\unitlength4mm \begin{picture}(6,1) \put(0,0){\line(1,0){6}}
\put(2.5,0){\vector(-1,0){0}} \put(3,.5){$k^\mu$} \end{picture}
\end{array}
=\frac i{\gamma k-M}. 
\]
\caption{\textit{The Feynman rules for the massive Thirring model. }}
\label{fa1}
\end{figure}
The two-particle form factors for charge-less operators of the form $%
\mathcal{O}(x)=\mathcal{N\,}\overline{\psi }\Gamma \psi $ are given in
lowest order perturbation theory by the Feynman graph depicted in figure \ref
{f1}. 
We consider several examples for $\Gamma $: 
\[
\begin{array}{llrcl}
1) &  & \langle \,0\,|\,\overline{\psi }\psi \,|\,p_{1},p_{2}\,\rangle _{%
\bar{s}s} & = & -\,\bar{v}(\theta _{1})u(\theta _{2})=-2Mi\sinh \tfrac{1}{2}%
\theta _{12} \\[2mm] 
2) &  & \langle \,0\,|\,\overline{\psi }\gamma ^{5}\psi
\,|\,p_{1},p_{2}\,\rangle _{\bar{s}s} & = & -\,\bar{v}(\theta _{1})\gamma
^{5}u(\theta _{2})=2Mi\cosh \tfrac{1}{2}\theta _{12} \\[2mm] 
3) &  & \langle \,0\,|\,\overline{\psi }\gamma ^{\pm }\psi
\,|\,p_{1},p_{2}\,\rangle _{\bar{s}s} & = & -\,\bar{v}(\theta _{1})\gamma
^{\pm }u(\theta _{2})=\mp 2Mie^{\pm \frac{1}{2}(\theta _{1}+\theta _{2})} \\%
[2mm] 
4) &  & \langle \,0\,|\,\overline{\psi }\gamma ^{+}\tfrac{i}{2}%
\overleftrightarrow{\partial ^{+}\rule{0in}{0.15in}}\psi
\,|\,p_{1},p_{2}\,\rangle _{\bar{s}s} & = & 
\begin{array}[t]{l}
-\bar{v}(\theta _{1})\gamma ^{+}u(\theta _{2})\tfrac{1}{2}(p_{1}-p_{2})^{+}
\\[2mm] 
=2M^{2}i\sinh \tfrac{1}{2}\theta _{12}e^{\pm (\theta _{1}+\theta _{2})}
\end{array}
\end{array}
\]
Note that the energy momentum tensor light cone components are $T^{\pm \pm }=%
\tfrac{i}{2}\mathcal{N\,}\left[ \overline{\psi }\gamma ^{\pm }%
\overleftrightarrow{\partial ^{\pm }\rule{0in}{0.15in}}\psi \right] $ and $%
T^{\pm \mp }=M\mathcal{N\,}\left[ \overline{\psi }\psi \right] .$ For all
these examples we have agreement with the exact expressions (\ref{4.5}) when 
$g\rightarrow 0$ or $\nu \rightarrow 1$.

\subsection{Examples of 4-particle form factors}

\subsubsection{Expansion of the exact formula}

We investigate the integral (\ref{1.2}) (for $n=$ 4 and $m=2$) 
\[
\mathcal{O}_{\bar{s}\bar{s}ss}(\underline{\theta })=\int_{\mathcal{C}_{%
\underline{\theta }}}dz_{1}\int_{\mathcal{C}_{\underline{\theta }}}dz_{2}\,h(%
\underline{\theta },{\underline{z}})\,p^{\mathcal{O}}(\underline{\theta },{%
\underline{z}})\,\Psi _{\bar{s}\bar{s}ss}(\underline{\theta },{\underline{z}}%
) 
\]
with the scalar function 
\[
h(\underline{\theta },{\underline{z}})=\prod_{1\le i<j\le 4}F(\theta
_{ij})\prod_{i=1}^{4}\prod_{j=1}^{2}\phi (\theta _{i}-z_{j})\,\tau
(z_{1}-z_{2})\,, 
\]
and the Bethe Ansatz state component 
\begin{multline*}
\Psi _{\bar{s}\bar{s}ss}(\underline{\theta },{\underline{z}})=\left( \Omega
C({\underline{\theta }},z_{1})C({\underline{\theta }},z_{2})\right) _{\bar{s}%
\bar{s}ss}=\prod_{i=1}^{4}\prod_{j=1}^{2}a(\theta _{i}-z_{j}) \\
\times \left( \tilde{c}(\theta _{1}-z_{1})\tilde{c}(\theta _{2}-z_{2})+%
\tilde{b}(\theta _{1}-z_{1})\tilde{c}(\theta _{2}-z_{1})\tilde{c}(\theta
_{1}-z_{2})\tilde{b}(\theta _{2}-z_{2})\right) .
\end{multline*}
with $\tilde{b}=b/a,\,\tilde{c}=c/a$ for small couplings. We consider first
the simple p-functions $\sum_{i=1}^{2}e^{\pm z_{i}}$ and using the $%
z_{1}\leftrightarrow z_{2}$ symmetry we calculate the following integral in
lowest order with respect to the coupling constant $g$ 
\begin{eqnarray}
I^{\pm } &=&\int_{\mathcal{C}_{\underline{\theta }}}dz_{1}\int_{\mathcal{C}_{%
\underline{\theta }}}dz_{2}\left( \prod_{i=1}^{4}\prod_{j=1}^{2}\tilde{\phi}%
(\theta _{i}-z_{j})\right) \tilde{c}(\theta _{1}-z_{1})\,\tilde{c}(\theta
_{2}-z_{2})  \nonumber \\
&&\times \left( 1+\tilde{b}(\theta _{1}-z_{2})\tilde{b}(\theta
_{2}-z_{1})\right) \tau (z_{1}-z_{2})\left( e^{\pm z_{1}}+e^{\pm
z_{2}}\right)  \nonumber \\
&=&\pm \frac{8ig\sinh \tfrac{1}{2}\theta _{12}\sinh \tfrac{1}{2}\theta _{34}%
}{\prod_{i<j}(\sinh \frac{1}{2}\theta _{ij}\cosh \frac{1}{2}\theta _{ij})}%
\frac{e^{\pm \frac{1}{2}(\theta _{1}+\theta _{2})}}{e^{\pm \frac{1}{2}%
(\theta _{3}+\theta _{4})}}\sum_{i=1}^{4}e^{\pm \theta _{i}}+O(g^{2})\,.
\label{4.7}
\end{eqnarray}
The derivation of this result is quite involved. A sketch of the calculation
is delegated to appendix \ref{a3}.

Finally we use the p-functions (\ref{4.3}) and the normalization constants
given by eq.~(\ref{4.6}) and $F(\theta )=-i\sinh \frac{1}{2}\theta +O(g)$
and obtain the four particle form factors 
\[
\mathcal{O}_{\bar{s}\bar{s}ss}(\underline{\theta })=\langle \,0\,|\,\mathcal{%
O}\,|\,p_{1},p_{2,}p_{3},p_{4}\,\rangle _{\bar{s}\bar{s}ss}^{in} 
\]
for the various operators in lowest order in the coupling $g$ as 
\begin{align*}
\left[ \overline{\psi }\psi \right] _{\bar{s}\bar{s}ss}(\underline{\theta }%
)& =-\frac{1}{2}gM\frac{\sinh \tfrac{1}{2}\theta _{12}\sinh \tfrac{1}{2}%
\theta _{34}}{\prod_{i<j}\cosh \tfrac{1}{2}\theta _{ij}}\cosh \tfrac{1}{2}%
\left( \theta _{13}+\theta _{24}\right) \sum_{i=1}^{4}e^{\theta
_{i}}\sum_{i=1}^{4}e^{-\theta _{i}} \\
\left[ \overline{\psi }\gamma ^{5}\psi \right] _{\bar{s}\bar{s}ss}(%
\underline{\theta })& =\frac{1}{2}gM\frac{\sinh \tfrac{1}{2}\theta
_{12}\sinh \tfrac{1}{2}\theta _{34}}{\prod_{i<j}\cosh \tfrac{1}{2}\theta
_{ij}}\sinh \tfrac{1}{2}\left( \theta _{13}+\theta _{24}\right)
\sum_{i=1}^{4}e^{\theta _{i}}\sum_{i=1}^{4}e^{-\theta _{i}} \\
\left[ j^{\pm }\right] _{\bar{s}\bar{s}ss}(\underline{\theta })& =\mp gM%
\frac{\sinh \tfrac{1}{2}\theta _{12}\sinh \tfrac{1}{2}\theta _{34}}{%
\prod_{i<j}\cosh \tfrac{1}{2}\theta _{ij}}\sinh \tfrac{1}{2}\left( \theta
_{13}+\theta _{24}\right) \sum_{i=1}^{4}e^{\pm \theta _{i}} \\
\left[ T^{\pm \pm }\right] _{\bar{s}\bar{s}ss}(\underline{\theta })& =\frac{1%
}{2}gM^{2}\frac{\sinh \tfrac{1}{2}\theta _{12}\sinh \tfrac{1}{2}\theta _{34}%
}{\prod_{i<j}\cosh \tfrac{1}{2}\theta _{ij}}\cosh \tfrac{1}{2}\left( \theta
_{13}+\theta _{24}\right) \left( \sum_{i=1}^{4}e^{\pm \theta _{i}}\right)
^{2} \\
\left[ T^{+-}\right] _{\bar{s}\bar{s}ss}(\underline{\theta })& =-\frac{1}{2}%
gM^{2}\frac{\sinh \tfrac{1}{2}\theta _{12}\sinh \tfrac{1}{2}\theta _{34}}{%
\prod_{i<j}\cosh \tfrac{1}{2}\theta _{ij}}\cosh \tfrac{1}{2}\left( \theta
_{13}+\theta _{24}\right) \sum_{i=1}^{4}e^{\theta
_{i}}\sum_{i=1}^{4}e^{-\theta _{i}}
\end{align*}
which agrees with the Feynman graph result calculated in the next subsection.

\subsubsection{Feynman graph expansion}

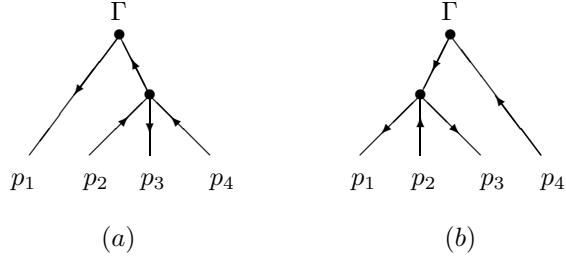
\begin{figure}[tbh]
\[
\unitlength4mm%
\begin{picture}(19,8.6)(0,-2)
\put(3,1){\line(1,1){2}} \put(4,2){\vector(1,1){.3}}
\put(5,1){\line(0,1){2}} \put(5,2){\vector(0,-1){.3}}
\put(5,3){\line(-1,2){1}} \put(5,3){\vector(-1,2){.6}}
\put(7,1){\line(-1,1){2}} \put(6,2){\vector(-1,1){.3}}
\put(5,3){\makebox(0,0){$\bullet$}}
\put(4,5){\makebox(0,0){$\bullet$}} \put(4,5){\line(-3,-4){3}}
\put(4,5){\vector(-3,-4){1.5}} \put(3.7,5.5){$\Gamma$}
\put(.4,0){$p_1$} \put(2.8,0){$p_2$} \put(3.4,-2){$(a)$}
\put(4.7,0){$p_3$} \put(7,0){$p_4$} \put(12,1){\line(1,1){2}}
\put(13,2){\vector(-1,-1){.3}} \put(14,1){\line(0,1){2}}
\put(14,2){\vector(0,1){.3}} \put(16,1){\line(-1,1){2}}
\put(15,2){\vector(1,-1){.3}} \put(14,3){\makebox(0,0){$\bullet$}}
\put(15,5){\makebox(0,0){$\bullet$}} \put(15,5){\line(-1,-2){1}}
\put(15,5){\vector(-1,-2){.6}} \put(18,1){\line(-3,4){3}}
\put(18,1){\vector(-3,4){1.5}} \put(14.7,5.5){$\Gamma$}
\put(11.7,0){$p_1$} \put(13.7,0){$p_2$} \put(14.8,-2){$(b)$}
\put(16,0){$p_3$} \put(18,0){$p_4$}
\end{picture}
\]
\caption{Feynman graphs for four-particle form factors of charge-less
operators.}
\label{f2}
\end{figure}

The graphs of figure \ref{f2} give the lowest order contributions to the
matrix element 
\begin{multline*}
\langle \,0\,|\,\overline{\psi }\Gamma \psi
\,|\,p_{1},p_{2},p_{3},p_{4}\,\rangle _{\bar{s}s\bar{s}s}^{in}=-i\tfrac{1}{2}%
g\,\langle \,0\,|\,\overline{\psi }\Gamma \psi \int d^{2}xj^{\mu }j_{\mu
}|\,p_{1},p_{2},p_{3},p_{4}\,\rangle _{\bar{s}s\bar{s}s} \\
=-ig\left[ \left( G_{a}+G_{b}\right) -\left( 1\leftrightarrow 3\right)
\right] -\left( 2\leftrightarrow 4\right)
\end{multline*}
up to terms of order $O(g^{2})$. We obtain 
\begin{eqnarray*}
G_{a} &=&\bar{v}_{1}\Gamma \frac{i}{\gamma (p_{2}+p_{3}+p_{4})-M}\mathbf{1}%
\gamma ^{\mu }u_{2}\;\bar{v}_{3}\gamma _{\mu }u_{4} \\
&=&i\frac{1}{2M}\bar{v}_{1}\Gamma \frac{\gamma (p_{2}+p_{3}+p_{4})+M}{%
(p_{2}+p_{3}+p_{4})^{2}-M^{2}}u_{3}\;\bar{u}_{3}\gamma ^{\mu }u_{2}\;\bar{v}%
_{3}\gamma _{\mu }u_{4}+symm \\
G_{b} &=&\bar{v}_{1}\gamma ^{\mu }u_{2}\;\bar{v}_{3}\gamma _{\mu }\mathbf{1}%
\frac{i}{\gamma (-p_{1}-p_{2}-p_{3})-M}\Gamma u_{4} \\
&=&-i\frac{1}{2M}\bar{v}_{1}\gamma ^{\mu }u_{2}\;\bar{v}_{3}\gamma _{\mu
}v_{2}\;\bar{v}_{2}\frac{\gamma (-p_{1}-p_{2}-p_{3})+M}{%
(p_{1}+p_{2}+p_{3})^{2}-M^{2}}\Gamma u_{4}+symm
\end{eqnarray*}
with $u_{i}=u(\theta _{i})$ etc. The relation $u\bar{u}-v\bar{v}=2M\mathbf{1}
$ has been used. The terms $symm$ vanish after anti-symmetrization. We
obtain for $\Gamma =\mathbf{1,}\gamma ^{5},\gamma ^{\pm },\gamma ^{\rho }%
\frac{i}{2}\overleftrightarrow{\partial ^{\sigma }}$ 
\begin{eqnarray*}
G_{a} &=&Mi\tanh \tfrac{1}{2}\theta _{24}\frac{\sinh \tfrac{1}{2}\theta _{12}%
}{\cosh \tfrac{1}{2}\theta _{34}}-(2\leftrightarrow 4) \\
G_{b} &=&Mi\tanh \tfrac{1}{2}\theta _{13}\frac{\sinh \tfrac{1}{2}\theta _{14}%
}{\cosh \tfrac{1}{2}\theta _{23}}-(1\leftrightarrow 3)
\end{eqnarray*}
\begin{eqnarray*}
G_{a}^{5} &=&-Mi\tanh \tfrac{1}{2}\theta _{24}\frac{\cosh \tfrac{1}{2}\theta
_{12}}{\cosh \tfrac{1}{2}\theta _{34}}-(2\leftrightarrow 4) \\
G_{b}^{5} &=&-Mi\tanh \tfrac{1}{2}\theta _{13}\frac{\cosh \tfrac{1}{2}\theta
_{14}}{\cosh \tfrac{1}{2}\theta _{23}}-(1\leftrightarrow 3),
\end{eqnarray*}
\begin{eqnarray*}
G_{a}^{\pm } &=&\pm Mi\tanh \tfrac{1}{2}\theta _{24}\frac{e^{\pm \frac{1}{2}%
(\theta _{1}+\theta _{2})}}{\cosh \tfrac{1}{2}\theta _{34}}%
-(2\leftrightarrow 4) \\
G_{b}^{\pm } &=&\pm Mi\tanh \tfrac{1}{2}\theta _{13}\frac{e^{\pm \frac{1}{2}%
(\theta _{1}+\theta _{4})}}{\cosh \tfrac{1}{2}\theta _{23}}%
-(1\leftrightarrow 3),
\end{eqnarray*}
\begin{gather*}
G_{a}^{\rho \sigma }=\rho M^{2}i\tanh \tfrac{1}{2}\theta _{24}e^{\rho \frac{1%
}{2}(\theta _{1}+\theta _{2})}\left( -\sigma \frac{\sinh \tfrac{1}{2}\theta
_{12}}{\cosh \tfrac{1}{2}\theta _{34}}e^{\sigma \frac{1}{2}(\theta
_{1}+\theta _{2})}+e^{\sigma \frac{1}{2}(\theta _{3}+\theta _{4})}\right) \\
-(2\leftrightarrow 4)~~~~~~~~~~~~~~ \\
G_{b}^{\rho \sigma }=\rho M^{2}i\tanh \tfrac{1}{2}\theta _{13}e^{\rho \frac{1%
}{2}(\theta _{1}+\theta _{4})}\left( -\sigma \frac{\sinh \tfrac{1}{2}\theta
_{14}}{\cosh \tfrac{1}{2}\theta _{23}}e^{\sigma \frac{1}{2}(\theta
_{1}+\theta _{4})}-e^{\sigma \frac{1}{2}(\theta _{2}+\theta _{3})}\right) \\
-(1\leftrightarrow 3)~~~~~~~~~~~~~~
\end{gather*}
which gives after anti-symmetrization the same expression as the one
calculated from the integral representation of the exact form factors
(exchanging $2\leftrightarrow 3$ and the sign due fermi statistics). In
particular for the energy momentum tensor we obtain with $T^{++}=\overline{%
\psi }\gamma ^{+}\frac{i}{2}\overleftrightarrow{\partial ^{+}\rule%
{0in}{0.15in}}\psi $ in lowest order in the coupling $g$%
\begin{multline*}
\langle \,0\,|\,T^{++}\,|\,p_{1},p_{2},p_{3},p_{4}\,\rangle _{\bar{s}s\bar{s}%
s}^{in} \\
=-\frac{1}{2}gM^{2}\frac{\sinh \frac{1}{2}\theta _{13}\sinh \frac{1}{2}%
\theta _{24}\cosh \frac{1}{2}(\theta _{12}+\theta _{34})}{\prod_{i<j}\cosh 
\frac{1}{2}\theta _{ij}}\left( \sum_{i=1}^{4}e^{\theta _{i}}\right) ^{2}.
\end{multline*}
Furthermore we have agreement with the classical relation for the trace of
the energy momentum tensor: $T^{+-}=\overline{\psi }\gamma ^{+}\frac{i}{2}%
\overleftrightarrow{\partial ^{-}\rule{0in}{0.15in}}\psi -g\,j^{\mu }j_{\mu
}=M\overline{\psi }\psi $ since the Feynman graph calculation implies 
\begin{eqnarray*}
\langle \,0\,|\,\overline{\psi }\gamma ^{+}\frac{i}{2}\overleftrightarrow{%
\partial ^{-}}\psi -M\overline{\psi }\psi
\,|\,p_{1},p_{2},p_{3},p_{4}\,\rangle _{\bar{s}s\bar{s}s}^{in}
&=&16gM^{2}\sinh \frac{1}{2}\theta _{13}\sinh \frac{1}{2}\theta _{24} \\
&=&g\langle \,0\,|\,j^{\mu }j_{\mu }\,|\,p_{1},p_{2},p_{3},p_{4}\,\rangle _{%
\bar{s}s\bar{s}s}^{in}\,
\end{eqnarray*}
in the coupling $g$ as it should be.

\subsubsection{Asymptotic behavior}

We are interested in the asymptotic behavior of form factors when one or
more rapidities tend to infinity. In perturbation theory for pure bosonic
models one may use Weinberg's power counting theorem for Feynman graphs \cite
{BK}\footnote{%
This type of arguments has also been used in \cite{KW,FMS,KM,MS}.}. For the
exponentials of the boson field $\mathcal{O=N}e^{i\gamma \varphi }$ this
yields in particular the asymptotic behavior 
\begin{equation}
\mathcal{O}_{n}(\theta _{1,}\theta _{2,}\dots )=\mathcal{O}_{1}(\theta
_{1})\,\mathcal{O}_{n-1}(\theta _{2,}\dots )+O(e^{-\mathop{\rm Re}\theta
_{1}})  \label{4.8}
\end{equation}
as $\mathop{\rm Re}\theta _{1}\rightarrow \infty $ in any order of
perturbation theory. This behavior is also assumed to hold for the exact
form factors and it was used e.g. in \cite{BK1} to obtain the normalization
of exponentials of fields. For fermionic models the asymptotic behavior is
more complicated. As an example we investigate a component of the
four-particle form factor of the operators $\mathcal{O}^{\pm }=\overline{%
\psi }\left( 1\pm \gamma ^{5}\right) \psi $ for the massive Thirring model 
\[
\mathcal{O}_{\bar{s}\bar{s}ss}^{\pm }(\underline{\theta })=\langle \,0\,|\,%
\mathcal{O}^{\pm }(0)\,|\underline{p}\,\,\rangle _{\bar{s}\bar{s}%
ss}^{in}=\int_{\mathcal{C}_{\underline{\theta }}}dz_{1}\int_{\mathcal{C}_{%
\underline{\theta }}}dz_{2}\,h(\underline{\theta },{\underline{z}})\,p^{\pm
}(\underline{\theta },{\underline{z}})\,\Psi _{\bar{s}\bar{s}ss}(\underline{%
\theta },{\underline{z}}) 
\]
with the p-functions of eqs. (\ref{4.3},\ref{4.6}) 
\[
p^{\pm }(\underline{\theta },\underline{z})=\mp N\sum\limits_{i=1}^{4}e^{\pm
\theta _{i}}\sum\limits_{i=1}^{2}e^{\mp z_{i}}\,. 
\]
After some calculation we finally obtain the asymptotic behavior for $%
\mathop{\rm Re}\theta _{1}\rightarrow \infty $%
\[
\mathcal{O}_{\bar{s}\bar{s}ss}^{\pm }(\underline{\theta })\approx const^{\pm
}\,e^{\frac{1}{4}(3-1/\nu )\theta _{1}}\times \int_{\mathcal{C}_{\underline{%
\theta }^{\prime }}}dz\,h(\underline{\theta }^{\prime },z)\,q^{\pm }(%
\underline{\theta }^{\prime },{z})\,\Psi _{\bar{s}ss}(\underline{\theta }%
^{\prime },{z}) 
\]
with $\underline{\theta }^{\prime }=(\theta _{2},\theta _{3},\theta _{4})$
and 
\begin{eqnarray*}
const^{\pm } &=&const\int_{\mathcal{C}_{0}}dz\,e^{-\frac{1}{2}(1/\nu \pm
1)z}\phi (-z)\,c(-z) \\
q^{\pm }(\underline{\theta }^{\prime },{z}) &=&e^{-\frac{1}{2}(1/\nu
+1)z}\prod_{i=2}^{4}e^{\frac{1}{4}(1/\nu +1)\theta _{i}}\left\{ 
\begin{array}{l}
e^{-z} \\ 
e^{-\theta _{2}}+e^{-\theta _{3}}+e^{-\theta _{4}}
\end{array}
\right. \\
\Psi _{\bar{s}ss}(\underline{\theta }^{\prime },{z}) &=&c(\theta
_{2}-z)\,a(\theta _{3}-z)\,a(\theta _{4}-z)
\end{eqnarray*}
Obviously this is not of the form given by relation (\ref{4.8}), in
particular the functions $q^{\pm }(\underline{\theta }^{\prime },{z})$ are
not valid p-functions since they do not satisfy the conditions on page 
\pageref{p}. This means that they do not correspond to local operators.

\section{``Crossing''}

\label{s5}

\subsection{General crossing relations}

In order to obtain the general matrix element $\,^{out}\langle \,\phi
^{\prime }\,|\,\mathcal{O}(x)\,|\,\phi \,\rangle ^{in}$ of a local operator
from that case where the state $\phi ^{\prime }$ is the vacuum one uses
`crossing'. This means one shifts in the matrix element particles from the
right hand side to the left hand side. It is well known from the theory of
Feynman graphs and more general also from LSZ-reduction formulas, that these
shifts are related to analytic continuation. We will now derive by means of
LSZ-assumptions and `maximal analyticity' a formula\footnote{%
In \cite{Sm} a similar formula was proposed which differs from the results
of this paper by sign factors. Our proof of the crossing formula follows
that of \cite{Q}.} which gives a general matrix element of a local operator $%
\mathcal{O}(x)$ in terms of an analytic continuation of the form factor
function $\mathcal{O}_{1\dots n}(\underline{\theta })$. As a generalization
of the co-vector valued function $\mathcal{O}_{1\dots n}(\underline{\theta }%
) $ we introduce the short notation $\mathcal{O}_{I}^{J}(\underline{\theta }%
_{J}^{\prime };\underline{\theta }_{I})$ given as follows.

Let the array of indices $I=(i_{1},\dots ,i_{|I|})$ denote the factors of
the tensor product of vector spaces $V_{I}=V_{i_{1}\dots
i_{|I|}}=V_{i_{1}}\otimes \dots \otimes V_{i_{|I|}}$ and correspondingly for 
$J$. For a local operator $\mathcal{O}(x)$ and for ordered sets of
rapidities $\theta _{i_{1}}>\dots >\theta _{i_{|I|}}$ and $\theta
_{j_{1}}^{\prime }<\dots <\theta _{j_{m}}^{\prime }$ we write 
\begin{equation}
\mathcal{O}_{I}^{J}(\underline{\theta }_{J}^{\prime };\underline{\theta }%
_{I}):=\,^{~out}\langle j_{|J|}(\,p_{j_{|J|}}^{\prime }),\dots
,j_{1}(p_{j_{1}}^{\prime })\,|\,\mathcal{O}(0)\,|\,i_{1}(p_{i_{1}}),\dots
,i_{|I|}(p_{i_{|I|}})\,\rangle ^{in}  \label{dcr}
\end{equation}
where $\underline{\theta }_{I}=(\theta _{i_{1}},\dots ,\theta _{i_{|I|}})$
and $\underline{\theta }_{J}^{\prime }=(\theta _{j_{1}}^{\prime },\dots
,\theta _{j_{|J|}}^{\prime })$. The function $\mathcal{O}_{I}^{J}(\underline{%
\theta }_{J}^{\prime };\underline{\theta }_{I})$ intertwines the spaces $%
V_{I}\rightarrow V_{J}$ and may be depicted as in figure \ref{fcr1}. 
\begin{figure}[tbh]
\[
\mathcal{O}_{I}^{J}(\underline{\theta }_{J}^{\prime };\underline{\theta }%
_{I})~=~~~ 
\begin{array}{c}
\unitlength .35mm\begin{picture}(60,90) \put(30,45){\oval(60,30)[]}
\put(30,45){\makebox(0,0)[cc]{$\cal O$}} \put(10,10){\line(0,1){20}}
\put(50,10){\line(0,1){20}} \put(10,60){\line(0,1){20}}
\put(50,60){\line(0,1){20}} \put(-4,16){$\theta_{i_1}$} \put(6,0){${i_1}$}
\put(24,18){$\dots$} \put(54,16){$\theta_{i_{|I|}}$} \put(48,0){${i_{|I|}}$}
\put(-4,71){$\theta'_{j_1}$} \put(6,86){${j_1}$} \put(24,70){$\dots$}
\put(54,70){$\theta'_{j_{|J|}}$} \put(48,86){${j_{|J|}}$} \end{picture}
\end{array}
\]
\caption{The general matrix element of a local operator.}
\label{fcr1}
\end{figure}
Similar to $\mathcal{O}_{1\dots n}(\underline{\theta })$ this function is
given for general order of the rapidities by the symmetry property $(i)$ for
both the $in$- and $out$-states which takes the general form: 
\[
\,\mathcal{O}_{I}^{J}(\underline{\theta }_{J}^{\prime };\underline{\theta }%
_{I})=\mathcal{O}_{K}^{J}(\underline{\theta }_{J}^{\prime };\underline{%
\theta }_{K})\dot{S}_{I}^{K}(\underline{\theta }_{I})=\dot{S}_{K}^{J}(%
\underline{\theta }_{K}^{\prime })\mathcal{O}_{I}^{K}(\underline{\theta }%
_{K}^{\prime };\underline{\theta }_{I}). 
\]
For the \emph{connected }contributions this follows again from analytic
continuation $\theta _{i_{1}}>\theta _{i_{2}}\rightarrow \theta
_{i_{1}}<\theta _{i_{2}}$ which can be proven similarly as for $\,\mathcal{O}%
_{1\dots n}(\underline{\theta })$ \cite{BFKZ}. For the disconnected
contributions this may be considered as a convenient definition. As a
generalization of the two-particle S-matrix (including the statistics
factor) $\dot{S}_{12}(\theta _{1},\theta _{2})$ we have introduced the more
general object $\dot{S}_{I}^{J}(\underline{\theta }_{I})$ given by the
following definition.

\begin{definition}
Let $J=\pi (I)$ be a permutation of $I$. Then $\dot{S}_{I}^{J}(\underline{%
\theta }_{I})$ is the matrix representation of the permutation group $%
\mathcal{S}_{|I|}$ generated by the simple transpositions $\sigma
_{ij}:i\leftrightarrow j$ for any pair of nearest neighbor indices $i,j\in I$
as\footnote{%
Note that this definition is quite analogous to that of representations of
the braid group by means of spectral parameter independent R-matrices.} 
\[
\sigma _{ij}\rightarrow \dot{S}_{I}^{\sigma _{ij}(I)}(\underline{\theta }%
_{I})=\dot{S}_{ij}(\theta _{ij})\, 
\]
Because of the Yang-Baxter relation and unitarity of the S-matrix the
representation $\dot{S}_{I}^{\pi (I)}(\underline{\theta }_{I})$ for all $\pi
\in \mathcal{S}_{|I|}$ is well defined. We will also use the notation 
\[
\dot{S}_{I}^{KM}(\underline{\theta }_{I})=\dot{S}_{I}^{\pi (I)}(\underline{%
\theta }_{I}) 
\]
if $\pi $ is that permutation which reorders the array $I$ such that it
coincides with the combined arrays of $K$ and $M$ which means that 
\[
\pi (I)=KM=(k_{1},\dots ,k_{|K|},m_{1},\dots ,m_{|M|})\,. 
\]
\end{definition}

As an example consider the case $I=(1,2,3,4),K=(2,3)$ and $M=(1,4)$ 
\begin{eqnarray*}
\dot{S}_{1234}^{2314}(\theta _{1},\theta _{2},\theta _{3},\theta _{4}) &=&%
\dot{S}_{13}(\theta _{13})\dot{S}_{12}(\theta _{12}) \\
\begin{array}{l}
\unitlength6mm\begin{picture}(4,4) \put(0,1.5){\framebox(4,1){$\dot S$}}
\put(3.5,1){\line(0,1){.5}} \put(3.5,2.5){\line(0,1){.5}}
\put(.5,1){\line(0,1){.5}} \put(.5,2.5){\line(0,1){.5}}
\put(1.5,1){\line(0,1){.5}} \put(1.5,2.5){\line(0,1){.5}}
\put(2.5,1){\line(0,1){.5}} \put(2.5,2.5){\line(0,1){.5}} \put(.3,.1){1}
\put(1.3,.1){2} \put(2.3,.1){3} \put(3.3,.1){4} \put(.3,3.3){2}
\put(1.3,3.3){3} \put(2.3,3.3){1} \put(3.3,3.3){4} \end{picture}
\end{array}
&=& 
\begin{array}{l}
\unitlength6mm\begin{picture}(4,3) \put(0,1){\line(2,1){2}}
\put(1,1){\line(-1,1){1}} \put(2,1){\line(-1,1){1}} \put(3,1){\line(0,1){1}}
\put(-.2,0){1} \put(0.8,0){2} \put(1.8,0){3} \put(2.8,0){4} \put(-.2,2.4){2}
\put(0.8,2.4){3} \put(1.8,2.4){1} \put(2.8,2.4){4} \end{picture}
\end{array}
\end{eqnarray*}
Similarly $\dot{S}_{LM}^{J}(\underline{\theta }_{L},\underline{\theta }_{M})$
is defined as an inverse by the formula \newline
$\dot{S}_{LM}^{J}(\underline{\theta }_{L},\underline{\theta }_{M})\dot{S}%
_{J}^{LM}(\underline{\theta }_{J})=\delta _{J}^{J}$ where $\delta _{J}^{J}$
is the unit matrix in the vector space $V_{J}$.

Analogously to eq.~(\ref{dcr}) for the general matrix $\mathcal{O}_{I}^{J}(%
\underline{\theta }_{J}^{\prime };\underline{\theta }_{I})$ for local
operators $\mathcal{O}(x)$ we write for the unit operator 
\begin{eqnarray*}
\mathbf{1}_{M}^{N}(\underline{\theta }_{N}^{\prime };\underline{\theta }%
_{M}) &=&\,^{out}\langle n_{|N|}(\,p_{n_{|N|}}^{\prime }),\dots
,n_{1}(p_{n_{1}}^{\prime })\,|\,\,m_{1}(p_{m_{1}}),\dots
,m_{|M|}(p_{m_{|M|}})\,\rangle ^{in} \\
&=&\dot{S}_{M}^{N}(\underline{\theta }_{M})\prod_{i=1}^{|M|}4\pi \delta
(\theta _{n_{|N|}-n_{i}+1}^{\prime }-\theta _{m_{i}})
\end{eqnarray*}
if the rapidities are ordered as $\theta _{m_{1}}>\dots >\theta _{m_{|M|}}$
and $\theta _{n_{1}}^{\prime }<\dots <\theta _{n_{|N|}}^{\prime }$ and if $N$
is the completely reordered array $M$. This object has obviously no analytic
properties, since it is completely disconnected. However, we define it for
other orders of the rapidities again by the form factor property $(i)$ 
\[
\mathbf{1}_{M}^{N}(\underline{\theta }_{N}^{\prime };\underline{\theta }%
_{M})=\mathbf{1}_{K}^{N}(\underline{\theta }_{N}^{\prime };\underline{\theta 
}_{K})\dot{S}_{M}^{K}(\underline{\theta }_{M})=\dot{S}_{K}^{N}(\underline{%
\theta }_{K}^{\prime })\mathbf{1}_{M}^{K}(\underline{\theta }_{K}^{\prime };%
\underline{\theta }_{M}). 
\]
This implies in particular that 
\[
\mathbf{1}_{M}^{N}(\underline{\theta }_{N}^{\prime };\underline{\theta }%
_{M})=\delta _{M}^{N}\prod_{i=1}^{|M|}4\pi \delta (\theta _{n_{i}}^{\prime
}-\theta _{m_{i}}) 
\]
for $\theta _{m_{1}}>\dots >\theta _{m_{|M|}}$ and $\theta _{n_{1}}^{\prime
}>\dots >\theta _{n_{|N|}}^{\prime }$. Here $\delta _{M}^{N}$ is the unit
matrix in the space $V_{M}=V_{N}$. For example if $\theta _{1}>\theta _{2}$
and $\theta _{1}^{\prime }>\theta _{2}^{\prime }$ one has the two cases 
\begin{align*}
\mathbf{1}_{\alpha \beta }^{\alpha ^{\prime }\beta ^{\prime }}(\theta
_{1}^{\prime },\theta _{2}^{\prime };\theta _{1},\theta _{2})&
=\,^{in}\langle \,\beta ^{\prime }(\theta _{2}^{\prime }),\alpha ^{\prime
}(\theta _{1}^{\prime })\,|\,\alpha (\theta _{1}),\beta (\theta
_{2})\,\rangle ^{in} \\
& =\delta _{\alpha \alpha ^{\prime }}\delta _{\beta \beta ^{\prime
}}\prod_{i=1}^{2}4\pi \delta (\theta _{i}^{\prime }-\theta _{i}) \\
\mathbf{1}_{\alpha \beta }^{\beta ^{\prime }\alpha ^{\prime }}(\theta
_{2}^{\prime },\theta _{1}^{\prime };\theta _{1},\theta _{2})&
=\,^{out}\langle \,\alpha ^{\prime }(\theta _{1}^{\prime }),\beta ^{\prime
}(\theta _{2}^{\prime })\,|\,\alpha (\theta _{1}),\beta (\theta
_{2})\,\rangle ^{in} \\
& =\dot{S}_{\alpha \beta }^{\beta ^{\prime }\alpha ^{\prime }}(\theta
_{1},\theta _{2})\prod_{i=1}^{2}4\pi \delta (\theta _{i}^{\prime }-\theta
_{i}).
\end{align*}

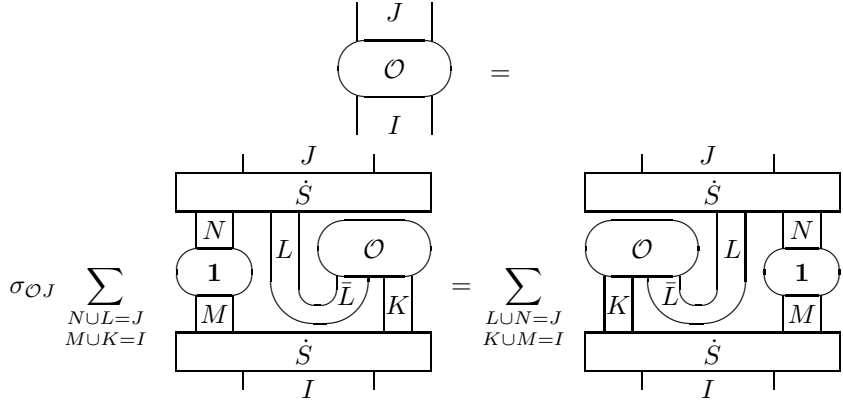
\begin{figure}[tbh]
\begin{gather*}
\begin{array}{l}
\unitlength .25mm\begin{picture}(60,70)(10,0) \put(30,35){\oval(60,30)[]}
\put(10,0){\line(0,1){20}} \put(50,0){\line(0,1){20}}
\put(10,50){\line(0,1){20}} \put(50,50){\line(0,1){20}}
\put(30,35){\makebox(0,0)[cc]{$\cal O$}} \put(30,5){\makebox(0,0)[cc]{$I$}}
\put(30,65){\makebox(0,0)[cc]{$J$}} \end{picture}
\end{array}
= \\
\sigma _{\mathcal{O}J}\sum_{\genfrac{}{}{0pt}1{N\cup L=J}{M\cup K=I}} 
\begin{array}{l}
\unitlength 0.25mm\begin{picture}(140,135) \put(110,85){\oval(60,30)[]}
\put(110,85){\makebox(0,0)[cc]{$\cal O$}} \put(81,70){\oval(52,50)[b]}
\put(80,70){\oval(20,30)[b]} \put(130,40){\line(0,1){30}}
\put(115,40){\line(0,1){30}} \put(70,70){\line(0,1){35}}
\put(55,70){\line(0,1){35}} \put(110,10){\line(0,1){10}}
\put(40,10){\line(0,1){10}} \put(110,125){\line(0,1){10}}
\put(40,125){\line(0,1){10}} \put(75,10){\makebox(0,0)[cc]{$I$}}
\put(75,135){\makebox(0,0)[cc]{$J$}} \put(122.50,55){\makebox(0,0)[cc]{$K$}}
\put(95,61){\makebox(0,0)[cc]{$\bar L$}} \put(62,85){\makebox(0,0)[cc]{$L$}}
\put(5,20){\framebox(135,20)[cc]{$\dot S$}}
\put(5,105){\framebox(135,20)[cc]{$\dot S$}} \put(25,72.50){\oval(40,25)[]}
\put(25,72.50){\makebox(0,0)[cc]{$\bf 1$}} \put(15,40){\line(0,1){20}}
\put(35,40){\line(0,1){20}} \put(35,85){\line(0,1){20}}
\put(15,85){\line(0,1){20}} \put(25,50){\makebox(0,0)[cc]{$M$}}
\put(25,95){\makebox(0,0)[cc]{$N$}} \end{picture}
\end{array}
=\sum_{\genfrac{}{}{0pt}1{L\cup N=J}{K\cup M=I}} 
\begin{array}{l}
\unitlength 0.25mm\begin{picture}(135,135) \put(30,85){\oval(60,30)[]}
\put(30,85){\makebox(0,0)[cc]{$\cal O$}} \put(59,70){\oval(52,50)[b]}
\put(60,70){\oval(20,30)[b]} \put(10,40){\line(0,1){30}}
\put(25,40){\line(0,1){30}} \put(70,70){\line(0,1){35}}
\put(85,70){\line(0,1){35}} \put(30,10){\line(0,1){10}}
\put(100,10){\line(0,1){10}} \put(30,125){\line(0,1){10}}
\put(100,125){\line(0,1){10}} \put(65,10){\makebox(0,0)[cc]{$I$}}
\put(65,135){\makebox(0,0)[cc]{$J$}} \put(17.50,55){\makebox(0,0)[cc]{$K$}}
\put(45,61){\makebox(0,0)[cc]{$\bar L$}} \put(78,85){\makebox(0,0)[cc]{$L$}}
\put(0,20){\framebox(135,20)[cc]{$\dot S$}}
\put(0,105){\framebox(135,20)[cc]{$\dot S$}} \put(115,72.50){\oval(40,25)[]}
\put(115,72.50){\makebox(0,0)[cc]{$\bf 1$}} \put(125,40){\line(0,1){20}}
\put(105,40){\line(0,1){20}} \put(105,85){\line(0,1){20}}
\put(125,85){\line(0,1){20}} \put(115,50){\makebox(0,0)[cc]{$M$}}
\put(115,95){\makebox(0,0)[cc]{$N$}} \end{picture}
\end{array}
\end{gather*}
\caption{The general crossing relations}
\label{fcr}
\end{figure}

\begin{theorem}
\label{tcros}For any local operator $\mathcal{O}(x)$ the 'intertwiner
valued' function \newline
$\mathcal{O}_{I}^{J}(\underline{\theta }_{J}^{\prime };\underline{\theta }%
_{I})$ defined by eq.~(\ref{dcr}) satisfies the general crossing relations%
\footnote{%
These crossing formulae are generalizations of those in \cite{Q}.} 
\begin{align}
\mathcal{O}_{I}^{J}(\underline{\theta }_{J}^{\prime };\underline{\theta }%
_{I})  \label{cross} \\
=\sigma _{\mathcal{O}J}\sum_{\genfrac{}{}{0pt}1{L\cup N=J}{K\cup M=I}}\dot{S}%
_{NL}^{J}(\underline{\theta }_{N}^{\prime },\underline{\theta }_{L}^{\prime
})\,\mathbf{1}_{M}^{N}(\underline{\theta }_{N}^{\prime },\underline{\theta }%
_{M})\,\mathbf{C}^{L\bar{L}}\mathcal{O}_{\bar{L}K}(\underline{\theta }_{\bar{%
L}}^{\prime }+i\pi _{-},\underline{\theta }_{K})\,\dot{S}_{I}^{MK}(%
\underline{\theta }_{I})  \nonumber \\
=\sum_{\genfrac{}{}{0pt}1{L\cup N=J}{K\cup M=I}}\dot{S}_{LN}^{J}(\underline{%
\theta }_{L}^{\prime },\underline{\theta }_{N}^{\prime })\,\mathcal{O}_{K%
\bar{L}}(\underline{\theta }_{K},\underline{\theta }_{\bar{L}}^{\prime
}-i\pi _{-})\mathbf{C}^{\bar{L}L}\,\mathbf{1}_{M}^{N}(\underline{\theta }%
_{N}^{\prime },\underline{\theta }_{M})\,\dot{S}_{I}^{KM}(\underline{\theta }%
_{I})  \nonumber
\end{align}
where $K,L,M,N$ and $\underline{\theta }_{K},\underline{\theta }_{L},%
\underline{\theta }_{M},\underline{\theta }_{N}$ are defined analogously to $%
I$ and $\underline{\theta }_{I}$. However, $\bar{L}=(\bar{l}_{|L|},\dots ,%
\bar{l}_{1})$ and $\underline{\theta }_{\bar{L}}=(\theta _{\bar{l}%
_{|L|}},\dots ,\theta _{\bar{l}_{1}})$ where the bar denotes the
anti-particles and $\mathbf{C}^{\bar{L}L}$ is a multi-particle charge
conjugation matrix. The general crossing relations are depicted in figure 
\ref{fcr}.
\end{theorem}

\noindent{\bfseries Proof. } In \cite{BFKZ} the crossing formula was proven
for the case of only one out-going particle $|J|=1$ using LSZ-reduction
formulae and the assumption of maximal analyticity. General LSZ-reduction
formulae for bosons take the form 
\begin{multline*}
^{out}\langle \,\phi ^{\prime }\,|\,\mathcal{O}\,|\,p,\phi \,\rangle
_{\alpha }^{in}=\,^{out}\langle \,\phi ^{\prime }\,|\,a_{\alpha
}^{out\,\dagger }(p)\,\mathcal{O}\,|\,\phi \,\rangle ^{in} \\
+\,i\int d^{2}x~^{out}\langle \,\phi ^{\prime }\,|\,T\left[ \mathcal{O}%
j_{\alpha }^{\dagger }(x)\right] \,|\,\phi \,\rangle ^{in}\,e^{-ipx}
\end{multline*}
\begin{multline*}
_{~~\bar{\alpha}}^{out}\langle \,\phi ^{\prime },p^{\prime }\,|\,\mathcal{O}%
\,|\,\phi \,\rangle ^{in}=\,^{out}\langle \,\phi ^{\prime }\,|\,\mathcal{O}%
\,a_{\alpha }^{in}(p^{\prime })\,|\,\phi \,\rangle ^{in} \\
+\,i\int d^{2}x~^{out}\langle \,\phi ^{\prime }\,|\,T\left[ \mathcal{O}%
j_{\alpha }^{\dagger }(x)\right] \,|\,\phi \,\rangle ^{in}\,e^{ip^{\prime }x}
\end{multline*}
and for fermions the form 
\begin{multline*}
^{out}\langle \phi ^{\prime }\,|\,\mathcal{O}\,|\,p,\phi \,\rangle _{\alpha
}^{in}=\sigma _{\mathcal{O}\alpha }~^{out}\langle \phi ^{\prime
}\,|\,a_{\alpha }^{out\,\dagger }(p)\,\mathcal{O}\,|\,\phi \,\rangle ^{in} \\
+i\int d^{2}x~^{out}\langle \,\phi ^{\prime }\,|\,T\left[ \mathcal{O}\bar{j}%
_{\alpha }(x)\right] \,|\,\phi \,\rangle ^{in}\,u(\theta )\,e^{-ipx}
\end{multline*}
\begin{multline*}
_{~~\bar{\alpha}}^{out}\langle \phi ^{\prime },p^{\prime }\,|\,\mathcal{O}%
\,|\,\phi \rangle ^{in}=\sigma _{\mathcal{O}\bar{\alpha}}~^{out}\langle
\,\phi ^{\prime }\,|\,\mathcal{O}\,a_{\alpha }^{in}(p^{\prime })\,|\,\phi
\,\rangle ^{in} \\
-i\,\sigma _{\mathcal{O}\bar{\alpha}}\int d^{2}x~^{out}\langle \,\phi
^{\prime }\,|\,T\left[ \mathcal{O}\bar{j}_{\alpha }(x)\right] \,|\,\phi
\,\rangle ^{in}\,v(\theta ^{\prime })\,e^{ip^{\prime }x}\,.
\end{multline*}
Using these relations one can derive similar crossing formulae for the
general case that $\phi ^{\prime }\neq \emptyset $. In these more general
formulae one particle (from several particles) in the $out$-state is crossed
to the $in$-state. Iterating these relations one obtains the general
crossing relations.%
\endproof%

\paragraph{Remarks:}

\begin{enumerate}
\item  Note that the equivalence of both formulations of the crossing
relation follows from the properties $(i)-(iii)$.

\item  As was shown in \cite{BFKZ} the properties $(ii)$ and $(iii)$ on page 
\pageref{p} follow from the crossing formula for $|J|=1$.
\end{enumerate}

\subsection{Eigenvalues of higher charges}

In this subsection and as a simple application of the general crossing
formula we investigate the higher conservation laws given by the currents $%
J_{L}^{\mu }(x)$.

\begin{proposition}
In terms of the matrix elements \newline
$\mathcal{O}_{1\dots n}(\underline{\theta })=\langle \,0\,|\,\mathcal{O}%
\,|\,p_{1},\dots ,p_{n}\,\rangle _{1\dots n}^{in}$ the higher charges 
\[
Q_{L}=\int dxJ_{L}^{0}(x)=\int dx\tfrac{1}{2}\left(
J_{L}^{+}(x)+J_{L}^{-}(x)\right) 
\]
satisfy for odd $L=\pm 1,\pm 3,\dots $ the eigenvalue equation 
\[
\left( Q_{L}-\sum_{i=1}^{n}\left( p_{i}^{+}\right) ^{L}\right)
|\,p_{1},\dots ,p_{n}\rangle _{1\dots n}^{in}=0 
\]
for suitable normalizations constants $N_{n}^{J_{L}}$. For even $L$ the
charges vanish as in the classical case.
\end{proposition}

\proof%
As a generalization of equation (\ref{4.65}) for $n=1$ we prove for
arbitrary $n=|I|$ and $n^{\prime }=|J|$ that 
\begin{equation}
\left[ Q_{L}\right] _{I}^{J}(\underline{\theta }^{\prime };\underline{\theta 
})=\sum_{i=1}^{n}\left( p_{i}^{+}\right) ^{L}\mathbf{1}_{I}^{J}(\underline{%
\theta }^{\prime };\underline{\theta }).  \label{4.50}
\end{equation}
First we show that for $n+n^{\prime }>2$ the connected part of the matrix
element $\left[ Q_{L}\right] _{1\dots n}^{1^{\prime }\dots n^{\prime }}(%
\underline{\theta ^{\prime }},\underline{\theta })$ vanishes. This connected
part is obtained by the analytic continuation $\left[ J_{L}^{0}\right]
_{n^{\prime }\dots 1^{\prime }1\dots n}(\underline{\theta ^{\prime }}+i\pi ,%
\underline{\theta })$. From the correspondence of operator and p-function 
\[
Q_{L}=\int dxJ_{L}^{0}(x)\leftrightarrow 2\pi \delta (P^{\prime
}-P)N_{n^{\prime }+n}^{J_{L}}\sum_{\pm }\tfrac{\mp 1}{2}\left(
\sum_{i=1}^{n^{\prime }}e^{\pm \theta _{i}^{\prime }}-\sum_{i=1}^{n}e^{\pm
\theta _{i}}\right) \sum_{i=1}^{r}e^{Lz_{i}} 
\]
the claim follows since for $n^{\prime }+n>2$ there are no poles which may
cancel the zero at $P^{\prime }=P$ where $P^{(\prime )}=\sum p_{i}^{(\prime
)}$. Note that only for $n=n^{\prime }=1$ the factor $1/\cosh \frac{1}{2}%
\theta _{12}$ in (\ref{4.63}) cancels the zero. Therefore contributions to (%
\ref{4.50}) come from disconnected parts which contain (analytically
continuated) two-particle form factors (c.f. eq.~(\ref{4.65})) 
\[
\left[ Q_{L}\right] _{i}^{j}(\theta _{j},\theta _{i})=\left(
p_{i}^{+}\right) ^{L}\mathbf{1}_{i}^{j}(\theta _{j},\theta _{i})\,. 
\]
It follows that in the general crossing formula only those terms with $%
K=\left\{ i\right\} ,M=I\setminus \left\{ i\right\} ,L=\left\{ j\right\}
,N=J\setminus \left\{ j\right\} $ contribute 
\begin{eqnarray*}
\left[ Q_{L}\right] _{I}^{J}(\underline{\theta }^{\prime };\underline{\theta 
}) &=&\sum_{i=1}^{n}\sum_{j=1}^{n}\dot{S}_{LN}^{J}(\underline{\theta }%
_{L}^{\prime },\underline{\theta }_{N}^{\prime })\,\left( p_{i}^{+}\right)
^{L}\,\mathbf{1}_{i}^{j}(\theta _{j},\theta _{i})\,\mathbf{1}_{M}^{N}(%
\underline{\theta }_{N}^{\prime },\underline{\theta }_{M})\,\dot{S}_{I}^{KM}(%
\underline{\theta }) \\
&=&\sum_{i=1}^{n}\left( p_{i}^{+}\right) ^{L}\mathbf{1}_{I}^{J}(\underline{%
\theta }^{\prime };\underline{\theta })
\end{eqnarray*}
where the obvious relation $\sum_{j=1}^{n}\dot{S}_{LN}^{J}(\underline{\theta 
}_{L}^{\prime },\underline{\theta }_{N}^{\prime })\,\mathbf{1}%
_{i}^{j}(\theta _{j},\theta _{i})\,\mathbf{1}_{M}^{N}(\underline{\theta }%
_{N}^{\prime },\underline{\theta }_{M})\,\dot{S}_{I}^{KM}(\underline{\theta }%
)=\mathbf{1}_{I}^{J}(\underline{\theta }^{\prime };\underline{\theta })$ has
been used.%
\endproof%

\section{Bound states}

\label{s6}

Before we define and investigate the properties of bound state form factors
we recall some facts on bound state S-matrices and define the ``bound state
intertwiners''. The two-particle S-matrix satisfies real analyticity (\ref
{1.65}) unitarity (\ref{1.8}), crossing (\ref{1.9}), the Yang-Baxter
relations (\ref{1.68}) and the permutation property at vanishing argument (%
\ref{1.10}).

Let the two particles labeled by $\alpha $ and $\beta $ of mass $m_{\alpha }$
and $m_{\beta }$, respectively form a bound state labeled by $\gamma $ of
mass $m_{\gamma }$. The mass of the bound state $\gamma $ is given by 
\[
m_{\gamma }=\sqrt{m_{\alpha }^{2}+m_{\beta }^{2}+2m_{\alpha }m_{\beta }\cos
u_{\alpha \beta }^{\gamma }}~~,~~~(0<u_{\alpha \beta }^{\gamma }<\pi ). 
\]
where $u_{\alpha \beta }^{\gamma }$ is the so called fusion angle and $%
iu_{\alpha \beta }^{\gamma }$ is equal to the pure imaginary relative
rapidity of the constituents $\alpha $ and $\beta $. The two-particle
S-matrix may be diagonalized 
\begin{equation}
\dot{S}_{\alpha \beta }^{\delta \gamma }(\theta )=\sum_{\epsilon }\varphi
_{\epsilon }^{\delta \gamma }(\theta )\,\dot{S}(\alpha ,\beta ,\epsilon
,\theta )\,\varphi _{\alpha \beta }^{\epsilon }(\theta )  \label{b1.30}
\end{equation}
where the projections onto the eigenspaces (labeled by $\epsilon $) are
given by the intertwiners $\varphi _{\epsilon }^{\delta \gamma }(\theta )$
and $\varphi _{\alpha \beta }^{\epsilon }(\theta )$ with 
\[
\sum_{\epsilon }\varphi _{\epsilon }^{\delta \gamma }(\theta )\varphi
_{\alpha \beta }^{\epsilon }(\theta )=\delta _{\alpha \gamma }\delta _{\beta
\delta }\,,\quad \sum_{\alpha \beta }\varphi _{\alpha \beta }^{\epsilon
^{\prime }}(\theta )\varphi _{\epsilon }^{\alpha \beta }(-\theta )=\delta
_{\epsilon ^{\prime }\epsilon }\,. 
\]
The eigenvalue of the S-matrix $\dot{S}(\alpha ,\beta ,\gamma ,\theta )$
which correspond to a bound state $(\alpha \beta )=\gamma $ has a pole at $%
\theta =iu_{\alpha \beta }^{\gamma }$ , a fact which will be used to define
the `bound state intertwiners'.

\subsection{Bound state intertwiners}

Following the investigations of \cite{K1} (see also \cite{Q}) we use in
addition to the intertwines $\varphi _{\alpha \beta }^{\epsilon }$ which are
defined for all eigenstates $\epsilon $ of the S-matrix also similar ones $%
\Gamma _{\alpha \beta }^{\gamma }(u_{\alpha \beta }^{\gamma })$ which are
defined for all fusion angles. They are therefore only defined for an
eigenstate $\gamma $ which correspond to bound states i.e. an to eigenvalue
of the two-particle S-matrix $S(\theta )$ which has a pole at $\theta
=iu_{\alpha \beta}^{\gamma }$.

\begin{definition}
The matrix elements $\Gamma _{\alpha \beta }^{\gamma }(\theta _{\alpha \beta
}^{\gamma })$ of the \textbf{bound state intertwiner} are defined by the
residue of the S-matrix 
\begin{equation}
i\mathop{\rm Res}_{\theta =iu_{\alpha \beta }^{\gamma }}\dot{S}_{\alpha
\beta }^{\beta ^{\prime }\alpha ^{\prime }}(\theta )=\Gamma _{\gamma
}^{\beta ^{\prime }\alpha ^{\prime }}(u_{\alpha \beta }^{\gamma })\Gamma
_{\alpha \beta }^{\gamma }(u_{\alpha \beta }^{\gamma }):\quad 
\begin{array}{l}
\unitlength3mm%
\begin{picture}(12,8) \put(0,3.5){$i\limfunc{Res}$} \put(4,2){\line(1,2){2}} \put(6,2){\line(-1,2){2}} \put(3.4,.5){$\alpha$} \put(5.6,.5){$\beta$} \put(5.6,6.5){$\alpha'$} \put(3.4,6.5){$\beta'$} \put(7.7,3.7){$=$} \put(9.4,-.5){$\alpha$} \put(11.6,-.5){$\beta$} \put(11.6,7.5){$\alpha'$} \put(9.4,7.5){$\beta'$} \put(11,1){\oval(2,4)[t]} \put(11,3){\line(0,1){2}} \put(11,3){\makebox(0,0){$\bullet$}}
\put(11,5){\makebox(0,0){$\bullet$}} \put(11,7){\oval(2,4)[b]} \end{picture}
\end{array}
\label{b1.40}
\end{equation}
where the dual intertwiner is defined by the crossing relation 
\begin{equation}
\Gamma _{\gamma }^{\beta \alpha }(u_{\alpha \beta }^{\gamma })=\mathbf{C}%
_{\gamma \gamma ^{\prime }}\Gamma _{\alpha ^{\prime }\beta ^{\prime
}}^{\gamma ^{\prime }}(u_{\alpha \beta }^{\gamma })\mathbf{C}^{\beta
^{\prime }\beta }\mathbf{C}^{\alpha ^{\prime }\alpha }:\quad 
\begin{array}{l}
\unitlength2.5mm%
\begin{picture}(14,8) \put(1,7){\oval(2,6)[b]} \put(1,1){\line(0,1){3}} \put(1,4){\makebox(0,0){$\bullet$}} \put(.6,0){$\gamma$}
\put(-.4,7.5){$\beta$} \put(1.6,7.5){$\alpha$} \put(3,4){$=$}
\put(6,1){\line(0,1){4}} \put(7.5,5){\oval(3,4)[t]}
\put(9,5){\makebox(0,0){$\bullet$}} \put(9,4){\oval(2,2)[t]} \put(11,4){\oval(2,2)[b]} \put(11,4){\oval(6,6)[b]} \put(12,4){\line(0,1){3}} \put(14,4){\line(0,1){3}} \put(5.6,0){$\gamma$} \put(11.6,7.5){$\beta$} \put(13.6,7.5){$\alpha$} \end{picture}
\end{array}
\label{b1.43}
\end{equation}
with the charge conjugation matrix $\mathbf{C}$ (e.g. for the sine-Gordon
model $\mathbf{C}^{\alpha ^{\prime }\alpha }=\mathbf{C}_{\alpha ^{\prime
}\alpha }=\delta _{\alpha ^{\prime }\bar{\alpha}}$).
\end{definition}

\paragraph{Remarks:}

\begin{enumerate}
\item  For the bound state intertwiner given by the matrix elements $\Gamma
_{\alpha \beta }^{\gamma }(u_{\alpha \beta }^{\gamma })$ and defined by eq.~(%
\ref{b1.40}) we will also use the notation 
\[
\Gamma _{12}^{(12)}(u_{12}^{(12)})=~ 
\begin{array}{l}
\unitlength3mm\begin{picture}(3,7) \put(2,1.5){\oval(2,5)[t]}
\put(2,4){\line(0,1){2}} \put(2,4){\makebox(0,0){$\bullet$}} \put(.5,0){1}
\put(2.7,0){2} \put(1,6.5){(12)} \end{picture}
\end{array}
\]
where $u_{12}^{(12)}=u_{\alpha \beta }^{\gamma }$. It intertwines the spaces 
$V_{1}\otimes V_{2}$ and $V_{(12)}$ 
\begin{equation}
\Gamma _{12}^{(12)}:V_{1}\otimes V_{2}\rightarrow V_{(12)}\,.  \label{b1.37}
\end{equation}

\item  Note that the bound state intertwiners $\Gamma
_{12}^{(12)}(u_{12}^{(12)})$ and the dual ones $\Gamma
_{(12)}^{21}(u_{12}^{(12)})$ are defined for $u_{12}^{(12)}>0$. In addition
one may define the `inverse' ones $\Gamma _{21}^{(12)}(u_{21}^{(12)})$ with $%
u_{21}^{(12)}=-u_{12}^{(12)}<0$ such that 
\begin{eqnarray}
\Gamma _{(12)}^{12}(u_{21}^{(12)})\Gamma
_{12}^{(12)}(u_{12}^{(12)})=P_{12}(12)  \nonumber \\
\Gamma _{12}^{(12)}(u_{12}^{(12)})\Gamma _{(12)}^{12}(u_{21}^{(12)})=\delta
_{(12)}  \label{b1.52}
\end{eqnarray}
where $P_{12}(12)$ projects onto that subspace of $V_{1}\otimes V_{2}$
defined by (\ref{b1.37}) on which $\Gamma _{12}^{(12)}(u_{12}^{(12)})$ is
nonzero and $\delta _{(12)}$ is the unit matrix in $V_{(12)}$. The dual
`inverse' intertwiner is again defined by a crossing relation analogously to
(\ref{b1.43}).

\item  Obviously the bound state intertwiners are defined only up to some
phase factors. From eqs.~(\ref{b1.30},\ref{b1.40}) follows that the
components are given as 
\[
\Gamma _{\alpha \beta }^{\gamma }=\varepsilon (\alpha ,\beta ,\gamma
)i\left| \mathop{\rm Res}_{\theta =iu_{12}^{(12)}}\dot{S}(\alpha ,\beta
,\gamma ,\theta )\right| ^{1/2}\varphi _{\alpha \beta }^{\gamma }\,. 
\]
Here $\dot{S}(\alpha ,\beta ,\gamma ,\theta )$ is the eigenvalue of the
S-matrix $\dot{S}_{\alpha \beta }^{\beta ^{\prime }\alpha ^{\prime }}$ which
correspond to the bound state $\gamma $ and the $\varepsilon $'s are phase
factors. In \cite{K1} it was shown that $i\mathop{\rm Res}\dot{S}_{\gamma
}(\theta )$ is real where the sign is related to the parity of the particles
involved. The phase factors $\varepsilon $ fulfill $\varepsilon (\alpha
,\beta ,\gamma )\varepsilon (\bar{\alpha},\bar{\beta},\bar{\gamma}%
)\allowbreak =-\mathop{\rm sgn}\left( i\mathop{\rm Res}\dot{S}(\alpha ,\beta
,\gamma ,\theta )\right) .$
\end{enumerate}

\begin{example}
For the sine-Gordon model we fix the phase factors by $\varepsilon (s,\bar{s}%
,k)=(-1)^{k}$ and $\varphi _{s\bar{s}}^{k}=(-1)^{k}\varphi _{\bar{s}s}^{k}=1/%
\sqrt{2}$ \footnote{%
This choice is motivated by the fact that with this convention the breather
matrix elements $\langle 0|:\varphi ^{k}:(0)|b_{k}\rangle $ are positive
(see \cite{BK,BK1}).}. The two-particle S-matrix has poles at $\theta
=iu^{(k)}=i\pi (1-k\nu )$ which correspond to the soliton anti-soliton bound
states alias breathers $b_{k}$. Using the short notation $\Gamma _{s\bar{s}%
}^{k}=\Gamma _{s\bar{s}}^{b_{k}}(u^{(k)})$ we obtain 
\begin{equation}
\Gamma _{s\bar{s}}^{k}=(-1)^{k}\Gamma _{\bar{s}s}^{k}=(-1)^{k}i\left| \tfrac{%
1}{2}\mathop{\rm Res}_{\theta =i\pi (1-k\nu )}\dot{S}_{\pm }(\theta )\right|
^{1/2}  \label{b1.41}
\end{equation}
where $+$ and $-$ correspond to even or odd $k$, respectively. The residues
have been calculated in \cite{KT} 
\begin{eqnarray}
\tfrac{1}{2}\mathop{\rm Res}_{\theta =i\pi (1-k\nu )}\dot{S}_{\pm }(\theta )=%
\mathop{\rm Res}_{\theta =i\pi (1-k\nu )}\dot{b}(\theta )=(-1)^{k}%
\mathop{\rm Res}_{\theta =i\pi (1-k\nu )}\dot{c}(\theta )(-1)^{k}  \nonumber
\\
=2i(-1)^{k}\cot \tfrac{\pi }{2}k\nu \prod_{l=1}^{k-1}\cot ^{2}\tfrac{\pi }{2}%
l\nu .  \label{b1.42}
\end{eqnarray}
All bound state intertwiners $\Gamma $ involving solitons are uniquely given
by the crossing relations 
\begin{align*}
\Gamma _{s}^{ks}=\Gamma _{ks}^{s}=\Gamma _{k}^{s\bar{s}}=\Gamma _{\bar{s}k}^{%
\bar{s}}=\Gamma _{\bar{s}}^{\bar{s}k}=\Gamma _{s\bar{s}}^{k} \\
\Gamma _{\bar{s}}^{k\bar{s}}=\Gamma _{k\bar{s}}^{\bar{s}}=\Gamma _{k}^{\bar{s%
}s}=\Gamma _{sk}^{s}=\Gamma _{\bar{s}}^{sk}=\Gamma _{\bar{s}%
s}^{k}=(-1)^{k}\Gamma _{s\bar{s}}^{k}.
\end{align*}
Proposition \ref{pb1} on page \pageref{pb1} and the general relation (\ref
{b1.40}) imply up to a sign 
\[
\mathop{\rm Res}_{\theta =i\pi (1-k\nu )}\dot{b}(\theta )=\mathop{\rm Res}%
_{\theta =i\pi \frac{1}{2}(1+k\nu )}S_{ks}(\theta ) 
\]
which may easily checked directly.
\end{example}

\begin{example}
For the breather-breather bound states we fix the phase factors $\varepsilon
(k,l,k+l)=1$ and $\varphi _{kl}^{k+l}=1$ (for $k+l<1/\nu $) and get 
\[
\Gamma _{kl}^{k+l}=i\left| \mathop{\rm Res}_{\theta =i\pi \frac{1}{2}%
(k+l)\nu }S_{kl}(\theta )\right| ^{1/2} 
\]
where for $k\leq l$ 
\[
\mathop{\rm Res}_{\theta =i\pi \frac{1}{2}(k+l)\nu }S_{kl}(\theta )=2i\tan 
\tfrac{\pi }{2}(k+l)\nu \frac{\tan \tfrac{\pi }{2}l\nu }{\tan \tfrac{\pi }{2}%
k\nu }\prod_{j=1}^{k-1}\frac{\tan ^{2}\tfrac{\pi }{2}(k+l-j)\nu }{\tan ^{2}%
\tfrac{\pi }{2}j\nu }. 
\]
All further breather bound state intertwiners $\Gamma $ are again uniquely
given by the crossing relations using that the breather are self-conjugate 
\[
\Gamma _{k+lk}^{l}=\Gamma _{k+l}^{lk}=\Gamma _{lk+l}^{k}=\Gamma
_{l}^{kk+l}=\Gamma _{kl}^{k+l}=\Gamma _{lk}^{k+l}. 
\]
Again proposition \ref{pb1} and the general relation (\ref{b1.40}) imply
that up to a sign 
\[
\mathop{\rm Res}_{\theta =i\pi \frac{1}{2}(k+l)\nu }S_{kl}(\theta )=%
\mathop{\rm Res}_{\theta =i\pi \frac{1}{2}(1-l\nu )}S_{kk+l}(\theta )\,, 
\]
which also may easily checked directly.
\end{example}

\subsubsection{The \textbf{bootstrap principle}}

If there exist bound states in a quantum field theoretic model, the
bootstrap principle means that all particles are to be considered on the
same footing, in particular:

\begin{enumerate}
\item  The space of all kinds of particles $V$ is closed under bound state
fusion which means that there exist for each fusion angle a bound state
intertwiner $\Gamma _{12}^{(12)}(u_{12}^{(12)})$ such that 
\[
\Gamma _{12}^{(12)}:V_{1}\otimes V_{2}\rightarrow V_{(12)}\,,\quad \mathrm{%
with}\quad V_{1}=V_{2}=V,\ V_{(12)}\subseteq V. 
\]

\item  If the fusion process $\alpha +\beta \rightarrow \gamma $ exists then
also the fusions $\beta +\bar{\gamma}\rightarrow \bar{\alpha}$ and $\bar{%
\gamma}+\alpha \rightarrow \bar{\beta}$ must exist. The corresponding fusion
angles may be read off figure~\ref{fb3} where the euclidean momenta $(p^{0},%
\mathop{\rm Im}p^{1})$ are depicted and where $\hat{u}_{\alpha \beta
}^{\gamma }=\pi -u_{\alpha \beta }^{\gamma }$ etc. In particular the
rapidities of the constituents are given in terms of the rapidity of the
bound state by 
\begin{eqnarray}
\theta _{\alpha } &=&\theta _{\gamma }+i\hat{u}_{\beta \bar{\gamma}}^{\bar{%
\alpha}}  \nonumber \\
\theta _{\beta } &=&\theta _{\gamma }-i\hat{u}_{\bar{\gamma}\alpha }^{\bar{%
\beta}}  \label{b1.38}
\end{eqnarray}
and similar for the other fusion processes. The fusion angles satisfy the
relation $u_{\alpha \beta }^{\gamma }+u_{\beta \bar{\gamma}}^{\bar{\alpha}%
}+u_{\bar{\gamma}\alpha }^{\bar{\beta}}=2\pi $ or $\hat{u}_{\alpha \beta
}^{\gamma }+\hat{u}_{\beta \bar{\gamma}}^{\bar{\alpha}}+\hat{u}_{\bar{\gamma}%
\alpha }^{\bar{\beta}}=\pi $. 
\begin{figure}[tbh]
\[
\unitlength5mm%
\begin{picture}(9,6)
\put(8,0){\vector(0,1){6}}
\put(8,0){\vector(-4,1){8}}
\put(0,2){\vector(2,1){8}}
\put(3.3,0){$\vec p_\alpha$}
\put(3,4.5){$\vec p_\beta$}
\put(8.3,3){$\vec p_\gamma$}
\bezier{80}(2.7,1.3)(3.5,2.3)(2.6,3.3)
\put(1.2,2){$\hat u_{\alpha\beta}^{\gamma}$}
\bezier{80}(5.4,.7)(6,2)(8,2)
\put(6.2,.8){$\hat u_{\beta\bar\gamma}^{\bar\alpha}$}
\bezier{80}(5.3,4.6)(6.5,3.5)(8,3.8)
\put(6.3,4.3){$\hat u_{\bar\gamma\alpha}^{\bar\beta}$}
\end{picture}
\]
\caption{\textit{The euclidian momenta }$(p^{0},\mathop{\rm Im}p^{1})$ 
\textit{of the fusion process }$\alpha +\beta \rightarrow \gamma $ \textit{%
in the center of mass system of }$\gamma $.}
\label{fb3}
\end{figure}
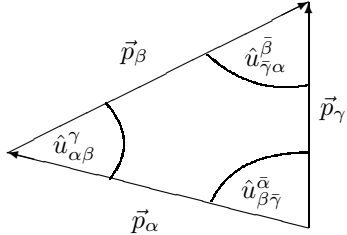

\item  The various \textbf{bound state intertwiners} are related by crossing
as depicted in figure~\ref{fb1} on page \pageref{fb1}.
\end{enumerate}

\subsection{Bound state S-matrix}

We show that the `bootstrap principle' provides a consistent scheme. In
particular we prove that the definition of the bound state intertwiners by
the relation (\ref{b1.40}) is consistent with crossing symmetry. First we
define the bound state S-matrix (see \cite{K1}). Using the Yang-Baxter
equation and the definition of the bound state intertwiners we have 
\begin{gather*}
\mathop{\rm Res}S_{12}S_{13}S_{23}=\mathop{\rm Res}S_{23}S_{13}S_{12} \\
\Gamma _{(12)}^{21}\Gamma _{12}^{(12)}S_{13}S_{23}=S_{23}S_{13}\Gamma
_{(12)}^{21}\Gamma _{12}^{(12)} \\
\unitlength=7.4mm
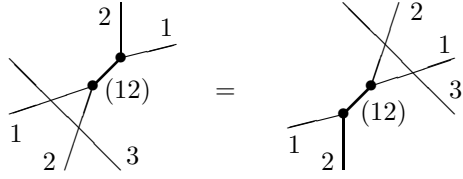
\begin{picture}(8,3.4) \put(1,0){\line(1,3){.5}}
\put(1.5,1.5){\line(-3,-1){1.5}} \put(2,0){\line(-1,1){2}}
\put(2,2){\line(4,1){1}} \put(2,3){\line(0,-1){1}} \put(0,.5){1}
\put(.6,0){2} \put(2.1,.1){3} \put(1.7,1.3){(12)} \put(2.7,2.4){1}
\put(1.6,2.6){2} \put(3.7,1.3){=} \put(6,0){\line(0,1){1}}
\put(6,1){\line(-4,-1){1}} \put(8,1){\line(-1,1){2}}
\put(6.5,1.5){\line(3,1){1.5}} \put(7,3){\line(-1,-3){.5}} \put(5,.3){1}
\put(5.6,0){2} \put(7.9,1.3){3} \put(6.3,.9){(12)} \put(7.7,2.1){1}
\put(7.2,2.7){2} \put(1.5,1.5){\makebox(0,0){$\bullet$}}
\put(2,2){\makebox(0,0){$\bullet$}} \put(6,1){\makebox(0,0){$\bullet$}}
\put(6.5,1.5){\makebox(0,0){$\bullet$}} \thicklines
\put(1.5,1.5){\line(1,1){.5}} \put(6,1){\line(1,1){.5}} \end{picture}
\end{gather*}
Therefore the following definition of the bound state S-matrix is natural.

\begin{definition}
\cite{K1}The bound state S-matrix which describes the scattering of a bound
state $(12)$ with another particle $3$ is given by 
\begin{eqnarray}
\dot{S}_{(12)3}(\theta _{(12)3})\Gamma _{12}^{(12)} &=&\Gamma _{12}^{(12)}%
\dot{S}_{13}(\theta _{13})\dot{S}_{23}(\theta _{23})\,\Big|_{\theta
_{12}=iu_{12}^{(12)}}  \label{b1.60} \\
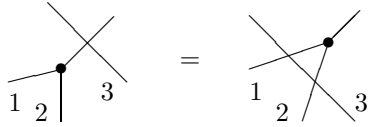
\begin{array}{c}
\unitlength3.5mm\begin{picture}(5,5) \put(2,2){\line(1,1){2}}
\put(4.5,1.5){\line(-1,1){3}} \put(2,2){\line(-4,-1){2}}
\put(2,0){\line(0,1){2}} \put(0,.4){$1$} \put(1,0){$2$} \put(3.5,.8){$3$}
\put(2,2){\makebox(0,0){$\bullet$}} \end{picture}
\end{array}
&=&~ 
\begin{array}{c}
\unitlength3.5mm\begin{picture}(5,5) \put(3,3){\line(1,1){1.2}}
\put(4,0){\line(-1,1){4}} \put(0,2){\line(3,1){3}} \put(2,0){\line(1,3){1}}
\put(0,.8){$1$} \put(1,0){$2$} \put(4,.4){$3$}
\put(3,3){\makebox(0,0){$\bullet$}} \end{picture}
\end{array}
\nonumber
\end{eqnarray}
where the rapidity of the bound state $\theta _{(12)}$ is defined by the
relation of the 2-momenta $p_{1}+p_{2}=p_{(12)}$ (see also eqs.~(\ref{b1.38}%
)).
\end{definition}

It was shown in \cite{K1} that the bound state S-matrix defined by (\ref
{b1.60}) fulfills unitarity, crossing and the Yang-Baxter equation. The
equation (\ref{b1.60}) is also called a 'bootstrap equation' \cite{K2} (also
referred to as a 'pentagon equation'). It relates different two-particle
S-matrices and therefore implies strong restrictions on the complete
two-particle S-matrix which may be used to calculate the S-matrix for
integrable models \cite{K2}.

We show that the definition of the bound state intertwiners by means of the
residue of the two-particle S-matrix (\ref{b1.40}) is consistent with the
`bootstrap principle' and crossing symmetry. 
\begin{figure}[tbh]
\[
\unitlength2.5mm%
\begin{picture}(25,7)
\put(2,1){\oval(2,6)[t]}
\put(2,4){\line(0,1){2}}
\put(2,4){\makebox(0,0){$\bullet$}}
\put(.5,-.5){$\alpha$}
\put(2.6,-.5){$\beta$}
\put(1.6,6.7){$\gamma$}
\put(5,3){$=$}
\put(8,2){\line(0,1){4}}
\put(9,2){\oval(2,2)[b]}
\put(11,2){\oval(2,4)[t]}
\put(12.5,4){\oval(3,4)[t]}
\put(12,1){\line(0,1){1}}
\put(14,1){\line(0,1){3}}
\put(11,4){\makebox(0,0){$\bullet$}}
\put(7.5,6.7){$\gamma$}
\put(13.6,-.5){$\beta$}
\put(11.6,-.5){$\alpha$}
\put(16,3){$=$}
\put(25,2){\line(0,1){4}}
\put(24,2){\oval(2,2)[b]}
\put(22,2){\oval(2,4)[t]}
\put(20.5,4){\oval(3,4)[t]}
\put(21,1){\line(0,1){1}}
\put(19,1){\line(0,1){3}}
\put(22,4){\makebox(0,0){$\bullet$}}
\put(24.5,6.7){$\gamma$}
\put(20.6,-.5){$\beta$}
\put(18.6,-.5){$\alpha$}
\end{picture}
\]
\caption{\textit{Crossing relations of the fusion intertwiners}}
\label{fb1}
\end{figure}
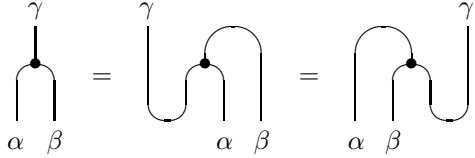

\begin{proposition}
\label{pb1}If the bound state intertwiners satisfy the crossing relations 
\begin{equation}
\Gamma _{\alpha \beta }^{\gamma }=\mathbf{C}^{\gamma \bar{\gamma}^{\prime
}}\Gamma _{\bar{\gamma}^{\prime }\alpha }^{\bar{\beta}^{\prime }}\mathbf{C}_{%
\bar{\beta}^{\prime }\beta }=\mathbf{C}_{\alpha \bar{\alpha}^{\prime
}}\Gamma _{\beta \bar{\gamma}^{\prime }}^{\bar{\alpha}^{\prime }}\mathbf{C}^{%
\bar{\gamma}^{\prime }\gamma }  \label{b1.70}
\end{equation}
(see figure~\ref{fb1}) and if the relation (\ref{b1.40}) holds for the
fusion process $\alpha +\beta \rightarrow \gamma $ then it also holds for
the fusion processes $\beta +\bar{\gamma}\rightarrow \bar{\alpha}$ i.e. 
\[
i\mathop{\rm Res}_{\theta _{\beta \bar{\gamma}}=iu_{\beta \bar{\gamma}}^{%
\bar{\alpha}}}\dot{S}_{\beta \bar{\gamma}}^{\bar{\gamma}^{\prime }\beta
^{\prime }}(\theta _{\beta \bar{\gamma}})=\Gamma _{\bar{\alpha}}^{\bar{\gamma%
}^{\prime }\beta ^{\prime }}\Gamma _{\beta \bar{\gamma}}^{\bar{\alpha}}\,. 
\]
and correspondingly for $\bar{\gamma}+\alpha \rightarrow \bar{\beta}$.
\end{proposition}

The proof of this proposition is delegated to appendix \ref{a30}.

\subsection{Bound state form factors}

If there are no bound states in the model there exist only the `annihilation
poles' according to property $(iii)$ on page \pageref{p} which follow from
the crossing formula. If there are also bound states there are additional
poles \cite{KW} and we have also property $(iv)$. Such additional poles have
been discussed for simple cases in \cite{KW,BFKZ}. Here we give the
arguments for more general cases. Let us consider a model with a bound state
(12) of two particles of 1 and 2 such that the attractive region is
connected analytically (by a coupling constant) to a repulsive region, where
the bound state decays.

We start in the repulsive region and consider the two-point Wightman
function (\ref{0.4}) $\langle 0|\mathcal{O}^{\prime }\mathcal{O}|\,0\rangle
=\langle 0|\mathcal{O}^{\prime }(x)\mathcal{O}(0)|\,0\rangle $. We use the
symmetry given by the statistics of the particles to express the $in$-state
matrix elements in terms of the form factor functions $\mathcal{O}_{1\dots
n}(\underline{\theta })$ for ordered rapidities 
\begin{eqnarray*}
\langle 0|\mathcal{O}^{\prime }\mathcal{O}|\,0\rangle &=&\sum_{n=0}^{\infty }%
\frac{1}{n!}\int \dots \int \frac{dp_{1}\ldots dp_{n}}{\,(2\pi )^{n}2\omega
_{1}\dots 2\omega _{n}} \\
&&\quad \times \langle 0|\mathcal{O}^{\prime }|p_{1},\ldots ,p_{n}\rangle
^{in}\,^{in}\langle \,p_{n},\ldots ,p_{1}|\mathcal{O}|0\rangle \\
&=&\sum_{n=0}^{\infty }\int_{-\infty }^{\infty }\frac{d\theta _{1}}{4\pi }%
\int_{-\infty }^{\theta _{1}}\frac{d\theta _{2}}{4\pi }\dots \int_{-\infty
}^{\theta _{n-1}}\frac{d\theta _{n}}{4\pi }\,\mathcal{O}_{1\dots n}^{\prime
}(\underline{\theta })\,\mathcal{O}^{1\dots n}(\underline{\theta })\,e^{-iPx}
\\
&=&\sum_{n=0}^{\infty }\frac{1}{n!}\int_{-\infty }^{\infty }\frac{d\theta
_{1}}{4\pi }\int_{-\infty }^{\infty }\frac{d\theta _{2}}{4\pi }\dots
\int_{-\infty }^{\infty }\frac{d\theta _{n}}{4\pi }\,\mathcal{O}_{1\dots
n}^{\prime }(\underline{\theta })\,\mathcal{O}^{1\dots n}(\underline{\theta }%
)\,e^{-iPx}
\end{eqnarray*}
where $P=p_{1}+\dots +p_{n}$. For the last equality the symmetry $(i)$ of
the form factor functions has been used, in particular we have 
\begin{equation}
\mathcal{O}_{12\dots n}^{\prime }\mathcal{O}^{12\dots n}=\mathcal{O}%
_{21\dots n}^{\prime }\dot{S}_{12}\mathcal{O}^{12\dots n}=\mathcal{O}%
_{12\dots n}^{\prime }\dot{S}_{21}\mathcal{O}^{21\dots n}\,.  \label{b1.80}
\end{equation}
In the repulsive case the S-matrices $\dot{S}_{12}$ and $\dot{S}_{21}$ have
poles at $\theta _{12}=\theta _{1}-\theta _{2}=iu_{12}^{(12)}$ and $\theta
_{21}=\theta _{2}-\theta _{1}=iu_{12}^{(12)}$, respectively, in the
`unphysical region' $u_{12}^{(12)}<0$. Therefore also the left hand side has
these poles. There are $\binom{n}{2}$ such pairs of poles. By analytic
continuation with respect to the coupling constant to the attractive region
where $u_{12}^{(12)}>0$ these poles cross the integration contours and we
obtain additional contributions from residues 
\begin{multline*}
\langle 0\,|\,\mathcal{O}^{\prime }\,\mathcal{O}\,|\,0\rangle
=\sum_{n=0}^{\infty }\frac{1}{n!}\left\{ \int_{-\infty }^{\infty }\frac{%
d\theta _{1}}{4\pi }\int_{-\infty }^{\infty }\frac{d\theta _{2}}{4\pi }\dots
\int_{-\infty }^{\infty }\frac{d\theta _{n}}{4\pi }\right. \\
\left. +\,\binom{n}{2}\left[
-\oint_{iu_{12}^{(12)}}+\oint_{-iu_{12}^{(12)}}\right] \frac{d\theta _{12}}{%
4\pi }\int_{-\infty }^{\infty }\frac{d\theta _{(12)}}{4\pi }\dots
\int_{-\infty }^{\infty }\frac{d\theta _{n}}{4\pi }\right\} \\
\times \mathcal{O}_{1\dots n}^{\prime }(\underline{\theta })\mathcal{O}%
^{1\dots n}(\underline{\theta })\,e^{-iPx}.
\end{multline*}
The substitution $(\theta _{1},\theta _{2})\rightarrow (\theta _{12}=\theta
_{1}-\theta _{2},\theta _{(12)})$ has been applied for the additional
residue terms (if the masses of 1 and 2 are equal the bound state rapidity
is $\theta _{(12)}=\frac{1}{2}(\theta _{1}+\theta _{2})$, for the general
case see eqs.~(\ref{b1.38})). Using (\ref{b1.80}) and the residue formula (%
\ref{b1.40}) we obtain 
\begin{eqnarray*}
-\oint_{iu_{12}^{(12)}}\frac{d\theta _{12}}{4\pi }\mathcal{O}_{12\dots
n}^{\prime }(\underline{\theta }) &=&-\frac{1}{2}\left. \mathcal{O}_{21\dots
n}^{\prime }(\underline{\theta })\Gamma _{(12)}^{21}\Gamma
_{12}^{(12)}\right| _{\theta _{12}=iu_{12}^{(12)}} \\
\oint_{-iu_{12}^{(12)}}\frac{d\theta _{12}}{4\pi }\mathcal{O}^{12\dots n}(%
\underline{\theta }) &=&-\frac{1}{2}\left. \Gamma _{(12)}^{21}\Gamma
_{12}^{(12)}\mathcal{O}^{21\dots n}(\underline{\theta })\right| _{\theta
_{12}=iu_{12}^{(12)}}
\end{eqnarray*}
which means that both residues give the same contribution. Therefore the
additional term may be written as 
\begin{multline*}
-\sum_{n=2}^{\infty }\frac{1}{n!}\binom{n}{2}\int_{-\infty }^{\infty }\frac{%
d\theta _{(12)}}{4\pi }\dots \int_{-\infty }^{\infty }\frac{d\theta _{n}}{%
4\pi }\left. \mathcal{O}_{21\dots n}^{\prime }\Gamma _{(12)}^{21}\Gamma
_{12}^{(12)}\mathcal{O}^{12\dots n}\right| _{\theta _{12}=iu_{12}^{(12)}} \\
=\int_{-\infty }^{\infty }\frac{d\theta _{(12)}}{4\pi }\sum_{n=2}^{\infty }%
\frac{1}{(n-2)!}\int_{-\infty }^{\infty }\frac{d\theta _{3}}{4\pi }\dots
\int_{-\infty }^{\infty }\frac{d\theta _{n}}{4\pi }\mathcal{O}_{(12)\dots
n}^{\prime }\mathcal{O}^{(12)\dots n}\,e^{-iPx}
\end{multline*}
where we have introduced the bound state form factor 
\begin{equation}
\mathcal{O}_{(12)3\dots n}(\theta _{(12)},\underline{\theta }^{\prime
})=\left. \frac{1}{i\sqrt{2}}\mathcal{O}_{21\dots n}(\underline{\theta }%
)\Gamma _{(12)}^{21}\right| _{\theta _{12}=iu_{12}^{(12)}}  \label{b2.10}
\end{equation}
with $\underline{\theta }^{\prime }=(\theta _{3},\dots ,\theta _{n})$ and
the rapidity $\theta _{(12)}$ of the bound state is given by $%
p_{1}+p_{2}=p_{(12)}$. This identification is obviously unique up to a sign
which may be absorbed into the bound state intertwiner. Using again the
property $(i)$ and the residue formula for the S-matrix (\ref{b1.40}) we
obtain the property $(iv)$ of form factors on page \pageref{pf} 
\begin{eqnarray*}
\mathop{\rm Res}_{\theta _{12}=iu_{12}^{(12)}}\mathcal{O}_{123\dots n}(%
\underline{\theta }) &=&\mathop{\rm Res}_{\theta _{12}=iu_{12}^{(12)}}%
\mathcal{O}_{213\dots n}(\underline{\theta })\dot{S}_{12} \\
&=&\left. -i\mathcal{O}_{213\dots n}(\underline{\theta })\Gamma
_{(12)}^{21}\Gamma _{12}^{(12)}\right| _{\theta _{12}=\theta _{12}^{(12)}} \\
&=&\mathcal{O}_{(12)3\dots n}(\theta _{(12)},\underline{\theta }^{\prime })\,%
\sqrt{2}\Gamma _{12}^{(12)}
\end{eqnarray*}

In order to show that the formula (\ref{b2.10}) for the bound state form
factor is consistent with the bootstrap principle we prove the following
proposition.

\begin{proposition}
\label{p1}The bound state form factor defined by the property $(iv)$ or by (%
\ref{b2.10}) satisfies properties $(i)$ $(ii)$ and $(iii)$ 
\[
\begin{array}{lllll}
(i) &  & \mathcal{O}_{(12)3\dots n}(\theta _{(12)},\theta _{3},\dots ) & = & 
\mathcal{O}_{3(12)\dots n}(\theta _{3},\theta _{(12)},\dots )S_{(12)3} \\%
[5pt] 
(ii) &  &  & = & \mathcal{O}_{3\dots n(12)}(\theta _{3},\dots ,\theta
_{(12)}-2\pi i)\sigma _{\mathcal{O}(12)} \\[5pt] 
(iii) &  &  & \approx & 
\begin{array}{r}
\frac{2i}{\theta _{(12)3}-i\pi }\mathbf{C}_{(12)3}\mathcal{O}_{4\dots
n}(\theta _{4},\dots ) \\ 
\times \left( \mathbf{1}-S_{3n}\dots S_{34}\right)
\end{array}
\end{array}
\]
\end{proposition}

\proof%
These relations follow directly from the corresponding relations for the
form factor before taking the residue, by using in addition the Yang-Baxter
equation, the fusion equation (\ref{b1.60}) and the crossing relations of
the bound state intertwiners. For $(i)$ and $(ii)$ the proofs are obvious.
For $(iii)$ the proof is quite involved. It is delegated to appendix \ref{a4}%
.

\section{Soliton Breather form factors}

\label{s7}

In this section we apply the results of the previous section to the
sine-Gordon model. We calculate breather form factors starting with the
general formula (\ref{1.2}) for the soliton form factors.

The $b_{k}$-breather-$(n-2)$-soliton form factor is obtained from $\mathcal{O%
}_{123\dots n}(\underline{\theta })$ by means of the fusion procedure $(iv)$
with the fusion angle given by $u_{12}^{(12)}=u^{(k)}=\pi (1-k\nu )$ the
bound state rapidity $\xi =\theta _{(12)}=\frac{1}{2}(\theta _{1}+\theta
_{2})$ and $\underline{\theta }^{\prime }=\theta _{3},\dots ,\theta _{n}$ 
\[
\mathop{\rm Res}_{\theta _{12}=iu^{(k)}}\mathcal{O}_{123\dots n}(\underline{%
\theta })=\,\mathcal{O}_{(12)3\dots n}(\xi ,\underline{\theta }^{\prime })%
\sqrt{2}\Gamma _{12}^{(12)}(iu^{(k)}) 
\]
where the bound state intertwiner is given by eqs.~(\ref{b1.41}) and (\ref
{b1.42}).

For $\theta _{12}\rightarrow iu^{(k)}$ there will be pinchings of the
integration contours in formula (\ref{1.2}) at the poles $%
z_{i}=z^{(l)}=\theta _{2}-i\pi l\nu =\xi -\frac{1}{2}i\pi (1-k\nu +2l\nu )$
for $l=0,\dots ,k$ and $i=1,\dots ,m$. Using the pinching rule of contour
integrals and the symmetry with respect to the $m~z$-integrations we obtain 
\begin{eqnarray*}
\mathop{\rm Res}_{\theta _{12}=iu^{(k)}}\mathcal{O}_{123\dots n}(\underline{%
\theta })\, &=&\mathop{\rm Res}_{\theta _{12}=iu^{(k)}}(-2\pi
i)\,m\sum_{l=0}^{k}\mathop{\rm Res}_{z_{1}=z^{(l)}}\int_{\mathcal{C}_{%
\underline{\theta }}}dz_{2}\cdots \int_{\mathcal{C}_{\underline{\theta }%
}}dz_{m} \\
&&\times h(\underline{\theta }{,\underline{z}})p^{\mathcal{O}}(\underline{%
\theta }{,\underline{z}})\,\Psi _{1\dots n}(\underline{\theta },{\underline{z%
}}).
\end{eqnarray*}
After a lengthy calculation (see appendix \ref{a5}) we obtain for the case
of the lowest breather ($k=1,\,u^{(1)}=\pi (1-\nu )$) the one-breather-$%
(n-2) $-soliton form factor 
\begin{eqnarray*}
\mathcal{O}_{3\dots n}(\xi ,\underline{\theta }^{\prime })
&=&\prod_{2<i}F_{sb}(\xi -\theta _{i})\prod_{2<i<j}F(\theta _{ij})\, \\
&&\times \sum_{l=0}^{1}(-1)^{l}\prod_{2<i}\rho (\xi -\theta _{i},l) \\
&&\times \int_{\mathcal{C}_{\underline{\theta }}}dz_{2}\cdots \int_{\mathcal{%
C}_{\underline{\theta }}}dz_{m}\,\prod_{1<j}\chi (\xi -z_{j},l) \\
&&\times \prod_{2<i}\prod_{1<j}\phi (\theta _{i}-z_{j})\prod_{1<i<j}\tau
(z_{ij})\,\tilde{p}^{\mathcal{O}}(\xi ,\underline{\theta }^{\prime },z^{(l)},%
{\underline{z}}^{\prime })\,\Psi _{3\dots n}(\underline{\theta }^{\prime },{%
\underline{z}}^{\prime })
\end{eqnarray*}
with ${\underline{z}}^{\prime }=\left( z_{2},\dots ,z_{m}\right) $. The
soliton-breather form factor has been introduced as 
\begin{eqnarray*}
F_{sb}(\theta ) &=&K_{sb}(\theta )\sin \tfrac{1}{2i}\theta \,\,\exp
\int_{0}^{\infty }\frac{dt}{t}\,2\frac{\cosh \frac{1}{2}\nu t}{\cosh \frac{1%
}{2}t}\,\frac{1-\cosh t(1-\theta /(i\pi ))}{2\sinh t}\, \\
K_{sb}(\theta ) &=&\frac{-\cos \tfrac{\pi }{4}(1-\nu )/E(\frac{1}{2}(1-\nu ))%
}{\sinh \frac{1}{2}(\theta -\frac{i\pi }{2}(1+\nu ))\sinh \frac{1}{2}(\theta
+\frac{i\pi }{2}(1+\nu ))}\,.
\end{eqnarray*}
The normalization has been chosen such that $F_{sb}(\infty )=1$. The
function $E(\nu )$ was used in \cite{KW,BFKZ} and is defined in appendix \ref
{a5}. Also we have introduced the short notations 
\begin{eqnarray*}
\rho (\xi ,l) &=&\frac{\sinh \tfrac{1}{2}\left( \xi -(-1)^{l}\tfrac{i\pi }{2}%
(1+\nu )\right) }{\sinh \tfrac{1}{2}\xi }=(-1)^{l}\frac{\sinh \tfrac{1}{2}%
\left( \xi -\tfrac{i\pi }{2}(1+(-1)^{l}\nu )\right) }{\sinh \tfrac{1}{2}\xi }
\\
\chi (\xi ,l) &=&(-1)^{l}\frac{\sinh \frac{1}{2}(\xi +\frac{i\pi }{2}%
(1+(-1)^{l}\nu ))}{\sinh \frac{1}{2}(\xi -\frac{i\pi }{2}(1+(-1)^{l}\nu ))}
\end{eqnarray*}
The following identities have been used 
\begin{eqnarray*}
F(\theta _{1}-\theta _{i})F(\theta _{2}-\theta _{i})\tilde{\phi}(\theta
_{i}-z^{(l)}) &=&\,F_{sb}(\xi -\theta _{i})\rho (\xi -\theta _{i},l) \\
\phi (\theta _{1}-z_{j})\phi (\theta _{2}-z_{j})S_{sb}(\xi -z_{j})\tau
(z^{(l)}-z_{j}) &=&\chi (\xi -z_{j},l)
\end{eqnarray*}
for $\theta _{1/2}=\xi \pm \frac{i\pi }{2}(1-\nu ),\,z^{(l)}=\xi -\frac{i\pi 
}{2}(1-(-1)^{l}\nu ).$ The new p-function is obtained from the old one by 
\[
\tilde{p}^{\mathcal{O}}(\xi ,\underline{\theta }^{\prime },z^{(l)},{%
\underline{z}}^{\prime })\,=m\,d(\nu )\,p^{\mathcal{O}}(\xi +\tfrac{1}{2}%
iu^{(1)},\xi -\tfrac{1}{2}iu^{(1)},\underline{\theta }^{\prime },z^{(l)},{%
\underline{z}}^{\prime }\,) 
\]
where the constant $d(\nu )$ is given by 
\[
d(\nu )=\frac{\sqrt{E(\nu )}}{\varkappa \sqrt{\sin \frac{1}{2}\pi \nu }} 
\]
(see appendix \ref{a5}).

Iterating the above procedure we obtain the $r$-breather-$s$-soliton form
factor with $2r+s=n,$ the breather rapidities $\underline{\xi }=(\xi
_{1},\dots ,\xi _{r})$ and the soliton rapidities$\,\underline{\theta }%
=(\theta _{1},\dots ,\theta _{s})$ 
\begin{eqnarray*}
\mathcal{O}_{1\dots s}(\underline{\xi },\underline{\theta }) &=&\prod_{1\leq
i<j\leq r}F_{bb}(\xi _{ij})\prod_{i=1}^{r}\prod_{j=1}^{s}F_{sb}(\xi
_{i}-\theta _{j})\prod_{1\leq i<j\leq s}F(\theta _{ij})\, \\
&&\times \sum_{l_{1}=0}^{1}\dots \sum_{l_{r}=0}^{1}(-1)^{l_{1}+\dots
+l_{r}}\prod_{1\leq i<j\leq r}\left( 1+(l_{i}-l_{j})\frac{i\sin \pi \nu }{%
\sinh \xi _{ij}}\right) \\
&&\times \prod_{i=1}^{r}\prod_{j=1}^{s}\rho (\xi _{i}-\theta _{j},l) \\
&&\times \int dz_{r+1}\cdots \int
dz_{m}\,\prod_{i=1}^{r}\prod_{j=r+1}^{m}\chi (\xi _{i}-z_{j},l) \\
&&\times \prod_{i=1}^{s}\prod_{j=r+1}^{m}\phi (\theta
_{i}-z_{j})\prod_{r<i<j\leq m}\tau (z_{ij})\,\tilde{p}(\underline{\xi },%
\underline{\theta },\underline{z^{(l)}},{\underline{z}})\,\Psi _{1\dots s}(%
\underline{\theta },{\underline{z}})
\end{eqnarray*}
again with $\,z_{i}^{(l_{i})}=\xi _{i}-\frac{i\pi }{2}(1-(-1)^{l_{i}}\nu
),(i=1,\dots ,r)$. The two-breather form factor has been introduced as 
\begin{eqnarray*}
F_{bb}(\theta ) &=&K_{bb}(\theta )\,\sin \tfrac{1}{2i}\theta \,\exp
\int_{0}^{\infty }\frac{dt}{t}\,2\frac{\cosh (\frac{1}{2}-\nu )t}{\cosh 
\frac{1}{2}t}\,\frac{1-\cosh t(1-\theta /(i\pi ))}{2\sinh t}\, \\
K_{bb}(\theta ) &=&\frac{-\cos \frac{1}{2}\pi \nu /E(\nu )}{\sinh \frac{1}{2}%
(\theta -i\pi \nu )\sinh \frac{1}{2}(\theta +i\pi \nu )}
\end{eqnarray*}
The normalization has been chosen such that $F_{bb}(\infty )=1$. The
relation 
\begin{multline*}
F_{sb}(\xi _{1}-\theta _{3})F_{sb}(\xi _{1}-\theta _{4})\,\rho (\xi
_{1}-\theta _{3},l_{1})\rho (\xi _{1}-\theta _{4},l_{1})\chi (\xi
_{1}-z_{2}^{(l_{2})},l_{1}) \\
=F_{bb}(\xi _{12})\left( 1+(l_{1}-l_{2})\frac{i\sin \pi \nu }{\sinh \xi _{12}%
}\right)
\end{multline*}
has been used for $\theta _{3/4}=\xi _{2}\pm \frac{i\pi }{2}(1-\nu )$. The
new p-function is obtained from the old one by 
\[
\,\tilde{p}(\underline{\xi },\underline{\theta }^{\prime \prime },\underline{%
z^{(l)}},{\underline{z}}^{\prime \prime })\,=\binom{m}{r}r!\,d^{r}(\nu
)\,p\left( \xi _{1}+\tfrac{1}{2}\theta ^{(1)},\xi _{1}-\tfrac{1}{2}\theta
^{(1)},\dots ,\underline{\theta }^{\prime \prime },z_{1}^{(l_{1})},\dots ,{%
\underline{z}}^{\prime \prime }\right) \,. 
\]
In particular for $n=2r=2m$ we get the pure lowest breather form factor 
\begin{multline*}
\mathcal{O}(\underline{\xi })=\prod_{i<j}F_{bb}(\xi
_{ij})\,\sum_{l_{1}=0}^{1}\dots \sum_{l_{r}=0}^{1}(-1)^{l_{1}+\dots +l_{r}}
\\
\times \prod_{1=i<j}^{r}\left( 1+(l_{i}-l_{j})\frac{i\sin \pi \nu }{\sinh
\xi _{ij}}\right) \,\tilde{p}(\underline{\xi },\underline{z^{(l)}})
\end{multline*}
and the pure breather p-function 
\[
\tilde{p}(\underline{\xi },\underline{z^{(l)}})\,=r!\,d^{r}(\nu )\,p\left(
\xi _{1}+\tfrac{1}{2}iu^{(1)},\xi _{1}-\tfrac{1}{2}iu^{(1)},\dots
,\,z_{1}^{(l_{1})},\dots \right) . 
\]

\section{Conclusion}

\label{s8}

In a forthcoming paper \cite{BK} we investigate extensively the pure
breather form factors of the sine-Gordon model. Some results have been
published previously \cite{BK1}. Furthermore other integrable models of
quantum field theory will be analyzed. In particular the $SU(N)$%
-chiral-Gross-Neveu \cite{BFKZ1} and the $O(N)$-Gross-Neveu \cite{BK3} model
is under investigation. In these models the particles possess anyonic
statistics. The form factor program as considered in the present article may
easily extended to anyonic fields and particles. Mainly the statistics
factors in the form factor equations $(i)-(v)$ have to be replaced by more
general phase factors. This has been discussed in \cite{Sm,YuZ,Lu,Lu2}.

\paragraph{Acknowledgments:}

We thank A.A. Belavin, J. Balog, V.A.Fateev, R. Flume, A. Fring,
S. Pakuliak, R.H.Poghossian, 
F.A. Smirnov, R. Schrader, B. Schroer and Al.B. Zamolodchikov
for discussions. One of authors (M.K.) thanks E. Seiler and P. Weisz for
discussions and hospitality at the Max-Planck Insitut f\"{u}r Physik
(M\"{u}nchen), where parts of this work have been performed. H.B. was
supported by DFG, Sonderforschungsbereich 288 `Differentialgeometrie und
Quantenphysik' and partially by grants INTAS 99-01459 and INTAS 00-561.

\appendix

\section*{Appendix}

\renewcommand{\theequation}{\mbox{\Alph{section}.\arabic{equation}}} %
\setcounter{equation}{0}

\section{Proof of the lemma \ref{l1}}

\label{a1}%
\proof%
We start the proof of lemma \ref{l1} by establishing some algebraic
identities. The formula 
\[
\prod_{1\le i<j\le m}(x_{j}-x_{i})=\left| 
\begin{array}{ccccc}
1 & x_{1} & x_{1}^{2} & \dots & x_{1}^{m-1} \\ 
1 & x_{2} & x_{2}^{2} & \dots & x_{2}^{m-1} \\ 
\vdots & \vdots & \vdots & \ddots & \vdots \\ 
1 & x_{m} & x_{m}^{2} & \dots & x_{m}^{m-1}
\end{array}
\right| 
\]
follows from the fact that both sides are polynomials in $x_{m}$ of degree $%
m-1$ with the same $m-1$ zeros and the same asymptotic behavior. Further
taking the anti-symmetric sum $\left[ \cdots \right] _{a}$ with respect to $%
x_{1},\dots ,x_{m}$ one gets the relations 
\begin{multline*}
\left[ x_{m}^{k}\prod_{1\le i<j\le m}(x_{j}-x_{i})\right] _{a}=\left| 
\begin{array}{ccccc}
1 & x_{1} & \dots & x_{1}^{m-2} & x_{1}^{k} \\ 
1 & x_{2} & \dots & x_{2}^{m-2} & x_{2}^{k} \\ 
\vdots & \vdots & \vdots & \ddots & \vdots \\ 
1 & x_{m} & \dots & x_{m}^{m-2} & x_{m}^{k}
\end{array}
\right| \\
=\delta _{k,m-1}\prod_{1\le i<j\leq m}(x_{j}-x_{i})
\end{multline*}
for $k=0,1,\dots ,m-1$. Therefore for any set of constants $A_{k}$ one has 
\[
\left[ \left( \sum_{k=0}^{m-1}A_{k}x_{m}^{k}\right) \prod_{1\le
i<j<m}(x_{j}-x_{i})\right] _{a}=A_{m-1}\prod_{1\le i<j\le m}(x_{j}-x_{i}) 
\]
In particular with $x_{i}=e^{2z_{i}}$ and $a_{j}=e^{2\theta _{j}}$ we may
finally write 
\begin{multline*}
\prod_{1\leq i<j\leq m}\sinh z_{ij}\varpropto \\
\left[ \left( \prod_{i=1}^{2m}\cosh \tfrac{1}{2}(\theta
_{i}-z_{m})-\prod_{i=1}^{2m}\sinh \tfrac{1}{2}(\theta _{i}-z_{m})\right)
\prod_{i=1}^{m-1}e^{-z_{i}}\prod_{1\leq i<j<m}\sinh z_{ij}\right] _{a}
\end{multline*}
because the difference on the right hand side is of the form $%
\sum_{k=0}^{m-1}A_{k}e^{2kz_{m}}$. Using $\tau (z)\varpropto \sinh z\sinh
(z/\nu )$ the right hand side of formula (\ref{2.1}) for $\,p^{\mathcal{O}}(%
\underline{\theta },{\underline{z}})=$ independent of ${\underline{z}}$ and $%
n=2m$ is proportional to 
\begin{multline*}
\int_{\mathcal{C}_{\underline{\theta }}}dz_{1}\cdots \int_{\mathcal{C}_{%
\underline{\theta }}}dz_{m}\,\prod_{i=1}^{n}\prod_{j=1}^{m}\phi (\theta
_{i}-z_{j})\prod_{1\le i<j\le m}\sinh \frac{1}{\nu }\left(
z_{i}-z_{j}\right) \,\Psi _{1\dots n}(\underline{\theta },{\underline{z}}) \\
\times \left[ \left( \prod_{i=1}^{n}\cosh \tfrac{1}{2}(\theta
_{i}-z_{m})-\prod_{i=1}^{n}\sinh \tfrac{1}{2}(\theta _{i}-z_{m})\right)
\prod_{i=1}^{m-1}e^{-z_{i}}\prod_{1\leq i<j<m}\sinh z_{ij}\right] _{a}
\end{multline*}
The $z_{m}$-integral may be written as 
\begin{multline*}
\left( \int_{\mathcal{C}_{\underline{\theta }}}-\int_{\mathcal{C}_{%
\underline{\theta }}+i\pi \nu }\right) dz_{m}\prod_{i=1}^{n}\tilde{\phi}%
(\theta _{i}-z_{m})\,\prod_{i=1}^{n}\cosh \tfrac{1}{2}(\theta _{i}-z_{m}) \\
\times \prod_{i=1}^{m-1}\sinh \frac{z_{i}-z_{m}}{\nu }\,\tilde{\Psi}(%
\underline{\theta },{\underline{z}})=0
\end{multline*}
where $\tilde{\phi}(\theta )=a(\theta )\phi (\theta )$ and $\tilde{\Psi}(%
\underline{\theta },{\underline{z}})=\Psi (\underline{\theta },{\underline{z}%
})/\left( \prod_{i=1}^{n}\prod_{j=1}^{m}a(\theta _{i}-z_{j})\right) $. The
shift relations 
\[
\tilde{\phi}(\theta -i\pi \nu )\,\cosh \frac{1}{2}(\theta -i\pi \nu )=\tilde{%
\phi}(\theta )\,i\sinh \frac{1}{2}(\theta ) 
\]
\[
\tilde{\Psi}(\underline{\theta },z_{1},{\dots ,z}_{m}+i\pi \nu )=-\tilde{\Psi%
}(\underline{\theta },z_{1},{\dots ,z}_{m}) 
\]
have been used. They can easily be derived from (\ref{1.6}), the definition
of the Bethe Ansatz state (\ref{2.4}) and the S-matrix (\ref{s}). The
contour $\mathcal{C}_{\underline{\theta }}-\mathcal{C}_{\underline{\theta }%
}+i\pi \nu $ may be closed at $\pm $ infinity since the integrand behaves
like $e^{-2|z_{m}|/\nu }$ for $z_{2}\to \pm \infty $ (see appendix \ref{a2}%
). There are no poles inside the closed contour, therefore the integral
vanishes by Cauchy's theorem.%
\endproof%

\section{Asymptotic formulae}

\label{a2}The asymptotic behavior of the soliton-soliton scattering
amplitude is easily obtained from its integral representation 
\[
a(\theta )=\exp \int_{0}^{\infty }\frac{dt}{t}\frac{\sinh \frac{1}{2}(1-\nu
)t}{\sinh \frac{1}{2}\nu t\,\cosh \frac{1}{2}t}\sinh t(\theta /(i\pi
))=e^{\mp i\frac{1}{2}(1/\nu -1)}+o(1) 
\]
for $\mathop{\rm Re}\theta \to \pm \infty $, ($|\mathop{\rm Im}\theta |<%
\frac{\pi }{2}(1+\nu -|1-\nu |)$). This implies for the other amplitudes 
\[
b(\theta )=e^{\pm i\frac{1}{2}(1/\nu -1)}+o(1)~,~~~c(\theta )=\pm 2i\sin
(\pi /\nu )\,e^{\pm i\frac{1}{2}(1/\nu -1)}e^{\mp \theta /\nu }\left(
1+o(1)\right) . 
\]
The asymptotic behavior of the `minimal' two-soliton form factor function is
given by 
\begin{align*}
F(\theta )& =\cosh \tfrac{1}{2}\left( i\pi -\theta \right) \exp
\int_{0}^{\infty }\frac{dt}{t}\frac{\sinh \frac{1}{2}(1-\nu )t}{\sinh \frac{1%
}{2}\nu t\,\cosh \frac{1}{2}t}\frac{1-\cosh t(1-\theta /(i\pi ))}{2\sinh t}
\\
& =const\,e^{\pm \left( 1/\nu +1\right) (\theta -i\pi )}(1+o(1))
\end{align*}
for $\mathop{\rm Re}\theta \to \pm \infty $, ($|\mathop{\rm Im}\theta -\pi |<%
\frac{\pi }{2}(3+\nu -|1-\nu |)$) with the constant 
\[
const=\tfrac{1}{2}\exp \tfrac{1}{2}\int_{0}^{\infty }\frac{dt}{t}\left( 
\frac{\sinh \frac{1}{2}(1-\nu )t}{\sinh \frac{1}{2}\nu t\,\cosh \frac{1}{2}%
t\,\sinh t}-\frac{1-\nu }{\nu t}\right) 
\]
which satisfies 
\[
const^{4}=-\tfrac{1}{4}\nu \left( F^{\prime }(0)\right) ^{2}=\tfrac{1}{4}\nu
\pi \varkappa 
\]
(for $\varkappa $ see eq.~(\ref{3.1})). Similarly one has for $\mathop%
\mathrm{Re}\theta \to \pm \infty $, ($|\mathop{\rm Im}\theta -\frac{\pi }{2}%
|<\frac{\pi }{2}(2+\nu -|1-\nu |)$) the asymptotic expansion 
\[
\phi (\theta )=\frac{4}{\sqrt{4\pi \nu \varkappa }}\,e^{\mp \frac{1}{2}%
(1/\nu +1)(\theta -i\pi /2)}\left( \sum_{n=0}^{\tilde{n}}A_{n}e^{\mp n\theta
}+o\left( e^{\mp \tilde{n}\theta }\right) \right) 
\]
for any integer $\tilde{n}$ and $\nu <2/\tilde{n}$. The constants $A_{n}$
are determined by the expansion 
\begin{align*}
& \exp \tfrac{1}{2}\left[ \int_{-\infty }^{\infty }\frac{dt}{t}\left( \frac{%
\sinh \frac{1}{2}(1-\nu )t}{\sinh \frac{1}{2}\nu t\,\sinh t}-\frac{1-\nu }{%
\nu t}\right) e^{it\left( \theta -i\pi /2\right) /\pi }\right] \\
& =\exp \left( \sum_{n=1}^{\tilde{n}}a_{n}e^{\mp n\theta }+o\left( e^{\mp 
\tilde{n}\theta }\right) +\sum_{m=1}^{\tilde{m}}b_{m}e^{\mp 2m\theta /\nu
}+o\left( e^{\mp 2\tilde{m}\theta /\nu }\right) \right) \\
& =\sum_{n=0}^{\tilde{n}}A_{n}e^{\mp n\theta }+o\left( e^{\mp \tilde{n}%
\theta }\right)
\end{align*}
where $a_{n}=(-i)^{n-1}\sin \frac{1}{2}\pi (1-\nu )n/\left( \pi n\sin \frac{1%
}{2}\pi \nu n\right) $. Note that $A_{0}=1$.

\section{Expansion of the integral representation}

\label{a3}In order to compare the exact result for 4-particle form factors
with the Feynman graph result the integral representation has to be expanded
for small couplings. We calculate the integral $I^{\pm }=\int_{\mathcal{C}_{%
\underline{\theta }}}dz_{1}\int_{\mathcal{C}_{\underline{\theta }%
}}dz_{2}I^{\pm }(z_{1},z_{2})$ to prove formula (\ref{4.7}). The integrands
are 
\begin{multline*}
I^{\pm }(z_{1},z_{2})=\left( \prod_{i=1}^{4}\prod_{j=1}^{2}\tilde{\phi}%
(\theta _{i}-z_{j})\right) \tilde{c}(\theta _{1}-z_{1})\tilde{c}(\theta
_{2}-z_{2}) \\
\times \left( 1+\tilde{b}(\theta _{1}-z_{2})\tilde{b}(\theta
_{2}-z_{1})\right) \tau (z_{1}-z_{2})\left( e^{\pm z_{1}}+e^{\pm
z_{2}}\right)
\end{multline*}
up to order $O(g^{2})$. The functions $I^{\pm }(z_{1},z_{2})$ have poles at $%
z_{i}=\theta _{1},\theta _{2},\theta _{3},\theta _{4}$ and $\theta _{i}-i\pi
,\theta _{i}+i\pi (\nu -1)$ for $i=1,2$. By means of Cauchy's theorem we
have 
\[
\int_{\mathcal{C}_{\underline{\theta }}}dz_{i}I(z_{1},z_{2})=\left[
\oint_{\theta _{i}-i\pi }-\sum_{j=1}^{4}\oint_{\theta _{j}}+\int_{\mathcal{C}%
_{0}}\right] dz_{i}I(z_{1},z_{2}) 
\]
where $\mathcal{C}_{0}$ goes from $-\infty $ to $\infty $ such that $%
\mathop{\rm Im}\theta _{i}-\pi <\mathop{\rm Im}z<\mathop{\rm Im}\theta _{i}$%
. Then shifting by the integration contour $\mathcal{C}_{0}$by $i\pi $ one
obtains 
\[
\int_{\mathcal{C}_{0}}dz_{i}I(z_{1},z_{2})=\left[
\sum_{j=1}^{4}\oint_{\theta _{j}}+\oint_{\theta _{i}+i\pi (\nu -1)}+\int_{%
\mathcal{C}_{0}+i\pi }\right] dz_{i}I(z_{1},z_{2}) 
\]
which implies finally 
\begin{multline*}
\int_{\mathcal{C}_{\underline{\theta }}}dz_{i}I(z_{1},z_{2})=\left[
-\oint_{\theta _{i}}+\oint_{\theta _{i}-i\pi }+\tfrac{1}{2}\left[
\oint_{\theta _{i}}+\oint_{\theta _{i}+i\pi (\nu -1)}\right] \right. \\
\left. -\,\tfrac{1}{2}\sum_{j\neq i}\oint_{\theta _{j}}\right]
dz_{i}I(z_{1},z_{2})+\,\tfrac{1}{2}\int_{\mathcal{C}_{0}}dz_{i}\left(
I(z_{i})+I(z_{i}+i\pi )\right) .
\end{multline*}
The last term $\int_{\mathcal{C}_{0}}dz_{i}\left( I(z_{i})+I(z_{i}+i\pi
)\right) $ is of order $O(g^{2})$. Therefore we may consider up to $O(g^{2})$%
\begin{multline*}
I^{\pm }=\left[ -\oint_{\theta _{1}}+\oint_{\theta _{1}-i\pi }+\tfrac{1}{2}%
\left[ \oint_{\theta _{1}}+\oint_{\theta _{1}+i\pi (\nu -1)}\right] -\tfrac{1%
}{2}\sum_{j\neq 1}\oint_{\theta _{j}}\right] dz_{1} \\
\times \left[ -\oint_{\theta _{2}}+\oint_{\theta _{2}-i\pi }+\tfrac{1}{2}%
\left[ \oint_{\theta _{2}}+\oint_{\theta _{2}+i\pi (\nu -1)}\right] -\tfrac{1%
}{2}\sum_{j\neq 2}\oint_{\theta _{j}}\right] dz_{2}I^{\pm }(z_{1},z_{2})\,.
\end{multline*}

The $\left[ -\mathop{\rm Res}_{\theta _{1}}+\mathop{\rm Res}_{\theta
_{1}-i\pi }\right] _{z_{i}}$ terms have $O(1)$ and $O(g)$ contributions the
other ones are of order $O(g)$ only. The $O(1)$-terms cancel and after a
long but straight forward calculation all $O(g)$-terms give formula (\ref
{4.7}).

\section{Proof of proposition \ref{pb1}}

\label{a30}In this appendix we prove that the `bootstrap principle' provides
a consistent scheme. In particular we prove proposition \ref{pb1} which
states that the definition of the bound state intertwiner is consistent with
crossing symmetry.

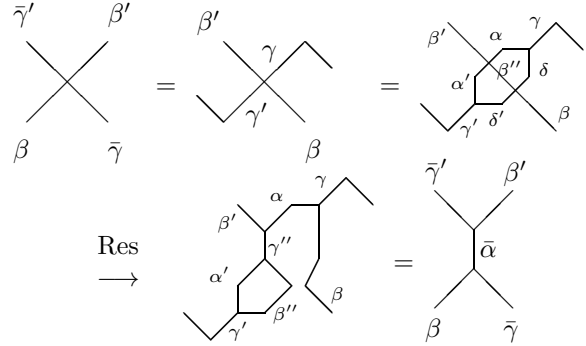
\begin{figure}[tbh]
\begin{gather*}
\begin{array}{l}
\unitlength 3.6mm\begin{picture}(4,5) \put(0,1){\line(1,1){3}}
\put(3,1){\line(-1,1){3}} \put(-.5,-.3){$\beta$} \put(3,-.3){$\bar\gamma$}
\put(-.5,4.7){$\bar\gamma'$} \put(3,4.7){$\beta'$} \end{picture}
\end{array}
= 
\begin{array}{l}
\unitlength 3.6mm\begin{picture}(6,5) \put(0,2){\line(1,-1){1}}
\put(1,1){\line(1,1){3}} \put(4,1){\line(-1,1){3}} \put(4,4){\line(1,-1){1}}
\put(0,4.4){$\beta'$} \put(4,-.3){$\beta$} \put(1.8,1){$\gamma'$}
\put(2.4,3.4){$\gamma$} \end{picture}
\end{array}
= 
\begin{array}{l}
\unitlength 3.60mm\linethickness{0.4pt}\begin{picture}(6,5.60)
\put(0,2){\line(1,-1){1}} \put(1,1){\line(1,1){1}} \put(2,2){\line(0,1){1}}
\put(2,3){\line(1,1){1}} \put(3,4){\line(1,0){1}} \put(4,4){\line(1,1){1}}
\put(5,5){\line(1,-1){1}} \put(4,4){\line(0,-1){1}}
\put(4,3){\line(-1,-1){1}} \put(3,2){\line(-1,0){1}}
\put(1,5){\line(1,-1){4}}
\put(1.8,1.10){\makebox(0,0)[cc]{$\scriptstyle\gamma'$}}
\put(1.50,2.83){\makebox(0,0)[cc]{$\scriptstyle\alpha'$}}
\put(2.77,4.5){\makebox(0,0)[cc]{$\scriptstyle\alpha$}}
\put(3.27,3.25){\makebox(0,0)[cc]{$\scriptstyle\beta''$}}
\put(2.82,1.45){\makebox(0,0)[cc]{$\scriptstyle\delta'$}}
\put(5.3,1.6){\makebox(0,0)[cc]{$\scriptstyle\beta$}}
\put(0.6,4.4){\makebox(0,0)[cc]{$\scriptstyle\beta'$}}
\put(4.55,3.27){\makebox(0,0)[cc]{$\scriptstyle\delta$}}
\put(4.23,5){\makebox(0,0)[cc]{$\scriptstyle\gamma$}} \end{picture}
\end{array}
\\
\begin{array}{c}
\mathop{\rm Res} \\ 
\longrightarrow
\end{array}
~~ 
\begin{array}{l}
\unitlength 3.60mm\linethickness{0.4pt}\begin{picture}(7,6)
\put(0,1){\line(1,-1){1}} \put(1,0){\line(1,1){1}} \put(2,1){\line(0,1){1}}
\put(2,2){\line(1,1){1}} \put(3,3){\line(0,1){1}} \put(3,4){\line(-1,1){1}}
\put(3,4){\line(1,1){1}} \put(4,5){\line(1,0){1}} \put(5,5){\line(1,1){1}}
\put(6,6){\line(1,-1){1}} \put(5,5){\line(0,-1){2}}
\put(5,3){\line(-1,-2){0.50}} \put(4.50,2){\line(1,-1){1}}
\put(3,3){\line(1,-1){1}} \put(4,2){\line(-1,-1){1}}
\put(3,1){\line(-1,0){1}}
\put(2.02,0.33){\makebox(0,0)[cc]{$\scriptstyle\gamma'$}}
\put(1.33,2.45){\makebox(0,0)[cc]{$\scriptstyle\alpha'$}}
\put(3.77,0.97){\makebox(0,0)[cc]{$\scriptstyle\beta''$}}
\put(3.58,3.48){\makebox(0,0)[cc]{$\scriptstyle\gamma''$}}
\put(1.62,4.33){\makebox(0,0)[cc]{$\scriptstyle\beta'$}}
\put(3.43,5.30){\makebox(0,0)[cc]{$\scriptstyle\alpha$}}
\put(5.08,5.87){\makebox(0,0)[cc]{$\scriptstyle\gamma$}}
\put(5.68,1.65){\makebox(0,0)[cc]{$\scriptstyle\beta$}} \end{picture}
\end{array}
= 
\begin{array}{l}
\unitlength 2.6mm\begin{picture}(4,7) \put(0,1){\line(1,1){2}}
\put(4,1){\line(-1,1){2}} \put(2,3){\line(0,1){2}} \put(0,7){\line(1,-1){2}}
\put(4,7){\line(-1,-1){2}} \put(-.5,-.5){$\beta$}
\put(3.5,-.5){$\bar\gamma$} \put(2.3,3.6){$\bar\alpha$}
\put(-.5,7.6){$\bar\gamma'$} \put(3.5,7.5){$\beta'$} \end{picture}
\end{array}
\end{gather*}
\caption{\textit{Proof of proposition \ref{pb1}.}}
\label{fb20}
\end{figure}

\proof%
The idea of the proof may read off figure \ref{fb20}: We use the crossing
property of the S-matrix (\ref{1.9}), the S-matrix bound state formula (\ref
{b1.60}), the residue formula (\ref{b1.40}) and the crossing relations of
the bound state intertwiners (\ref{b1.43}). 
We make the convention that rapidities as $\theta _{\alpha },\theta _{\alpha
^{\prime }}$ etc. belong to particles with the same mass and other invariant
quantum numbers. If $\delta $ denotes a particle with the same mass as that
of $\beta $ then we have (see fig. \ref{fb20}) 
\begin{eqnarray*}
i\mathop{\rm Res}_{\theta _{\beta \bar{\gamma}}=\theta _{\beta \bar{\gamma}%
}^{\bar{\alpha}}}\dot{S}_{\beta \bar{\gamma}}^{\bar{\gamma}^{\prime }\beta
^{\prime }}(\theta _{\beta \bar{\gamma}}) &=&i\mathop{\rm Res}_{\theta
_{\beta \bar{\gamma}}=\theta _{\beta \bar{\gamma}}^{\bar{\alpha}}}\mathbf{C}%
^{\bar{\gamma}^{\prime }\gamma ^{\prime }}\mathbf{C}_{\gamma \bar{\gamma}}%
\dot{S}_{\gamma ^{\prime }\beta }^{\beta ^{\prime }\gamma }(i\pi -\theta
_{\beta \bar{\gamma}}) \\
&=&i\mathop{\rm Res}_{\theta _{\beta \bar{\gamma}}=\theta _{\beta \bar{\gamma%
}}^{\bar{\alpha}}}\mathbf{C}^{\bar{\gamma}^{\prime }\gamma ^{\prime }}\Gamma
_{\gamma ^{\prime }}^{\alpha \delta }\Gamma _{\alpha \delta }^{\gamma
^{\prime \prime }}\mathbf{C}_{\gamma \bar{\gamma}}\dot{S}_{\gamma ^{\prime
\prime }\beta }^{\beta ^{\prime }\gamma }(i\pi -\theta _{\beta \bar{\gamma}})
\\
&=&-i\mathop{\rm Res}_{\theta _{\alpha \beta }=\theta _{\alpha \beta
}^{\gamma }}\mathbf{C}^{\bar{\gamma}^{\prime }\gamma ^{\prime }}\Gamma
_{\gamma ^{\prime }}^{\alpha \delta }\dot{S}_{\alpha \beta ^{\prime \prime
}}^{\beta ^{\prime }\alpha ^{\prime }}(\theta _{\alpha \beta })\dot{S}%
_{\delta \beta }^{\beta ^{\prime \prime }\delta ^{\prime }}(0)\Gamma
_{\alpha ^{\prime }\delta ^{\prime }}^{\gamma }\mathbf{C}_{\gamma \bar{\gamma%
}} \\
&=&\mathbf{C}^{\bar{\gamma}^{\prime }\gamma ^{\prime }}\Gamma _{\gamma
^{\prime }}^{\alpha \beta ^{\prime \prime }}\Gamma _{\gamma ^{\prime \prime
}}^{\beta ^{\prime }\alpha ^{\prime }}\Gamma _{\alpha \beta ^{\prime \prime
}}^{\gamma ^{\prime \prime }}\Gamma _{\alpha ^{\prime }\beta }^{\gamma }%
\mathbf{C}_{\gamma \bar{\gamma}} \\
&=&\mathbf{C}^{\bar{\gamma}^{\prime }\gamma ^{\prime }}\Gamma _{\gamma
^{\prime }}^{\beta ^{\prime }\alpha ^{\prime }}\Gamma _{\alpha ^{\prime
}\beta }^{\gamma }\mathbf{C}_{\gamma \bar{\gamma}}=\Gamma _{\bar{\alpha}}^{%
\bar{\gamma}^{\prime }\beta ^{\prime }}\Gamma _{\beta \bar{\gamma}}^{\bar{%
\alpha}}\,
\end{eqnarray*}
where the crossing relation of the S-matrix, the orthogonality relation of
the bound state intertwiners (\ref{b1.52}), the residue formula (\ref{b1.40}%
), the permutation property (\ref{1.10}) of the S-matrix at $\theta =0$ and
the crossing relations of the intertwiners (\ref{b1.70}) have been used.
Obviously the rapidities satisfy the equivalence 
\[
\left. 
\begin{array}{c}
\theta _{\beta \bar{\gamma}}=iu_{\beta \bar{\gamma}}^{\bar{\alpha}} \\ 
\theta _{\alpha \delta }=iu_{\alpha \delta }^{\gamma }
\end{array}
\right\} \Leftrightarrow \left\{ 
\begin{array}{c}
\theta _{\alpha \beta }=iu_{\alpha \beta }^{\gamma } \\ 
\theta _{\beta }=\theta _{\delta }
\end{array}
\right. 
\]
(note that $\theta _{\alpha \beta }=\theta _{\alpha }-\theta _{\beta }$
etc.). An analogous relation can be shown for the process $\bar{\gamma}%
+\alpha \rightarrow \bar{\beta}$. The relation of the three fusion angles
may be read off figure \ref{fb3}.%
\endproof%

\section{Proof of proposition \ref{p1}}

\label{a4}In this appendix we prove that the bound state form factors given
by property $(iv)$ on page \pageref{pf} satisfy property $(iii)$ which is
the recursion relation of $n$-particle form factors to $(n-2)$-particle form
factors. We consider $(iii)$ in the form 
\begin{multline*}
\mathop{\rm Res}_{\theta _{(12)(34)}=i\pi }\mathcal{O}_{(12)(34)5\dots
n}(\theta _{(12)},\theta _{(34)},\underline{\theta }^{\prime })=2i\mathbf{C}%
_{(12)(34)}\mathcal{O}_{5\dots n}(\underline{\theta }^{\prime }) \\
\times \left( 1-S_{(34)n}\dots S_{(34)5}\right)
\end{multline*}
with $\underline{\theta }^{\prime }=\theta _{5},\dots ,\theta _{n}$. First
we note that the bound state formula $(iv)$ on page \pageref{iv} for form
factors may also be written in a form without taking a residue. This follows
if we make the bound state pole at $\theta _{12}=iu$ explicit by using $(i)$
such that 
\begin{eqnarray}
\mathcal{O}_{(12)3\dots n}(\theta _{(12)},\underline{\theta }^{\prime }) &=&%
\frac{1}{\sqrt{2}}\mathop{\rm Res}_{\theta _{12}=iu}\mathcal{O}_{123\dots
n}(\theta _{1},\theta _{2},\dots ,\theta _{n})\Gamma _{(12)}^{12}(-u) 
\nonumber \\
&=&\frac{1}{\sqrt{2}}\mathop{\rm Res}_{\theta _{12}=iu}\mathcal{O}_{213\dots
n}(\underline{\theta })\dot{S}_{12}(\theta _{12})\Gamma _{(12)}^{12}(-u) 
\nonumber \\
&=&-i\frac{1}{\sqrt{2}}\mathcal{O}_{213\dots n}(\underline{\theta })\Big|%
_{\theta _{12}=iu}\Gamma _{(12)}^{21}(u).  \label{c1.10}
\end{eqnarray}
where also the definition (\ref{b1.40}) and the orthogonality relation (\ref
{b1.52}) of the bound state intertwiners have been used.

The form factor $\mathcal{O}_{1234\dots n}(\underline{\theta })$ has poles
at $\theta _{12}=iu,\,\theta _{34}=iu,\,\theta _{13}=i\pi ,\,\theta
_{24}=i\pi .$ We extract the two first poles by applying formula (\ref{c1.10}%
) twice and introduce the function 
\[
G(\theta _{12},\theta _{34},\theta )=\mathcal{O}_{21435\dots n}(\underline{%
\theta }^{\prime })\,\Gamma _{(12)}^{21}\Gamma _{(34)}^{43}\,,\quad \left(
\theta =\theta _{(12)(34)}=\theta _{(12)}-\theta _{(34)}\right) 
\]
where we suppress the co-vector structure and the variables $\theta
_{4},\dots ,\theta _{n}$. If for simplicity we assume that all masses are
equal then $\theta _{(12)}=\frac{1}{2}(\theta _{1}+\theta _{2})$ and $\theta
_{(34)}=\frac{1}{2}(\theta _{3}+\theta _{4})$ are the center-of-mass
rapidities. Extracting the poles at $\theta _{13}=i\pi $ and $\theta
_{24}=i\pi $, the remainder is analytic and can be expanded as 
\begin{equation}
G(\theta _{12},\theta _{34},\theta )=\frac{A+(\theta _{12}-iu)B+(\theta
_{34}-iu)C+(\theta -i\pi )D}{(\theta _{13}-i\pi )(\theta _{24}-i\pi )}+\dots
\label{c1.20}
\end{equation}
with certain constants $A,B,C,D$. We consider three limiting procedures:

\paragraph{1.}

Let first $\theta _{13}\rightarrow i\pi $, then $\theta _{34}\rightarrow iu$
and finally $\theta \rightarrow i\pi $:\newline
This means that also $\theta _{12}\rightarrow iu$ and $\theta
_{24}\rightarrow i\pi $. Due to $(i)$ we have $G(\theta _{12},\theta
_{34},\theta )=\mathcal{O}_{1324\dots n}\dot{S}_{23}\dot{S}_{21}\dot{S}%
_{43}\Gamma _{(12)}^{21}\Gamma _{(34)}^{43}$ and therefore with $(iii)$ for
the particle pair 13 
\begin{align*}
G_{1}(\theta )& =\left[ \mathop{\rm Res}_{\theta _{13}=i\pi }G(\theta
_{12},\theta _{34},\theta )\right] _{\theta _{34}=iu} \\
& =2i\left[ \mathbf{C}_{13}\mathcal{O}_{24\dots n}\left( \mathbf{1}%
-S_{3n}\dots S_{34}S_{32}\right) \dot{S}_{23}\dot{S}_{21}\dot{S}_{43}\Gamma
_{(12)}^{21}\Gamma _{(34)}^{43}\right] _{\theta _{34}=iu} \\
& =2i\mathbf{C}_{13}\mathcal{O}_{{24\dots n}}\sigma _{43}\left( -S_{3n}\dots
S_{35}S_{21}\right) \Gamma _{(12)}^{21}\Gamma _{(34)}^{43}
\end{align*}
\begin{align*}
& \approx \frac{(2i)^{2}}{\theta _{24}-i\pi }\mathbf{C}_{13}\mathbf{C}%
_{24}\sigma _{43}\mathcal{O}_{5\dots n}(1-S_{4n}\dots S_{45})\left(
-S_{3n}\dots S_{35}S_{21}\right) \Gamma _{(12)}^{21}\Gamma _{(34)}^{43} \\
& \approx (2i)^{2}i\frac{\theta _{12}-iu}{\theta _{24}-i\pi }\mathbf{C}_{13}%
\mathbf{C}_{24}\sigma _{43}\sigma _{12}\mathcal{O}_{5\dots n}(1-S_{4n}\dots
S_{45})\left( -S_{3n}\dots S_{35}\right) \Gamma _{(12)}^{12}\Gamma
_{(34)}^{43} \\
& \approx -(2i)^{2}i\mathbf{C}_{13}\mathbf{C}_{24}\mathcal{O}_{5\dots
n}\left( -S_{3n}\dots S_{35}\Gamma _{(34)}^{43}+\Gamma
_{(34)}^{43}S_{(34)n}\dots S_{(34)5}\right) \Gamma _{(12)}^{12}
\end{align*}
In deriving the third line unitarity (\ref{1.8}) and crossing (\ref{1.9}) of
the S-matrix has bee used. Also for $(i,j)=(3,4)$ and $(1,2)$ we have used
that 
\begin{equation}
S_{ji}\Gamma _{(ij)}^{ji}(u)\approx i(\theta _{ij}-iu)\sigma _{ij}\Gamma
_{(ji)}^{ij}(-u)  \label{c1.30}
\end{equation}
which follows from the unitarity (\ref{1.8}) of the S-matrix, the definition
(\ref{b1.40}) and the orthogonality (\ref{b1.52}) of the bound state
intertwiners. Further we have used again $(iii)$ for the particle pair 24
and the fact that $(\theta _{12}-iu)/(\theta _{24}-i\pi )=-1$ holds for $%
\theta _{13}=i\pi $ and $\theta _{34}=iu$. The statistics factors $\sigma
_{43}\sigma _{12}$ cancel since $1=\bar{2}$ and $3=\bar{4}$. Also the fusion
rule $\dot{S}_{4i}\dot{S}_{3i}\Gamma _{(34)}^{43}=\Gamma _{(34)}^{43}\dot{S}%
_{(34)i}$ has been applied. On the other hand we obtain from eq.~(\ref{c1.20}%
) using $\theta _{24}-i\pi =2(\theta -i\pi )$ for $\theta _{13}=i\pi $ and $%
\theta _{12}-iu=-2(\theta -i\pi )$ for $\theta _{13}=i\pi $ and $\theta
_{34}=iu$ 
\begin{eqnarray*}
G_{1}(\theta ) &=&\frac{A-2(\theta -i\pi )B+(\theta -i\pi )D}{2(\theta -i\pi
)}+\dots \\
&=&\frac{A}{2(\theta -i\pi )}-B+\tfrac{1}{2}D+\dots
\end{eqnarray*}
Since $G_{1}(\theta )$ is non-singular for $\theta \rightarrow i\pi $ we
conclude that $A=0$ and 
\begin{align*}
G_{1}(i\pi )& =-B+\tfrac{1}{2}D \\
& =-(2i)^{2}i\mathbf{C}_{13}\mathbf{C}_{24}\mathcal{O}_{5\dots n}\left(
-S_{3n}\dots S_{35}\Gamma _{(34)}^{43}+\Gamma _{(34)}^{43}S_{(34)n}\dots
S_{(34)5}\right) \Gamma _{(12)}^{12}
\end{align*}

\paragraph{2.}

Let now $\theta _{24}\rightarrow i\pi $, then $\theta _{34}\rightarrow iu$
and finally $\theta \rightarrow i\pi $:\newline
Due to $(i)$ we have $G(\theta _{12},\theta _{34},\theta )=\mathcal{O}%
_{2413\dots n}\dot{S}_{14}\Gamma _{(12)}^{21}\Gamma _{(34)}^{43}$ and
therefore similarly as above with $(iii)$ for the particle pair 24 
\begin{eqnarray*}
G_{2}(\theta ) &=&\left[ \mathop{\rm Res}_{\theta _{24}=i\pi }G(\theta
_{12},\theta _{34},\theta )\right] _{\theta _{34}=iu} \\
&=&2i\left[ \mathbf{C}_{24}\mathcal{O}_{13\dots n}\left( \mathbf{1}%
-S_{4n}\dots S_{45}S_{43}S_{41}\right) \dot{S}_{14}\Gamma _{(12)}^{21}\Gamma
_{(34)}^{43}\right] _{\theta _{34}=iu} \\
&\approx &2i\mathbf{C}_{24}\mathcal{O}_{13\dots n}\sigma _{21}S_{21}\Gamma
_{(12)}^{21}\Gamma _{(34)}^{43} \\
&\approx &(2i)^{2}i\frac{\theta _{12}-iu}{\theta _{13}-i\pi }\mathbf{C}_{13}%
\mathbf{C}_{24}\mathcal{O}_{5\dots n}(1-S_{3n}\dots S_{35})\Gamma
_{(12)}^{12}\Gamma _{(34)}^{43} \\
&\approx &(2i)^{2}i\mathbf{C}_{13}\mathbf{C}_{24}\mathcal{O}_{5\dots
n}(1-S_{3n}\dots S_{35})\Gamma _{(12)}^{12}\Gamma _{(34)}^{43}
\end{eqnarray*}
again we have used (\ref{c1.30}) and the fact that $(\theta
_{12}-iu)/(\theta _{13}-i\pi )=1$ holds for $\theta _{24}=i\pi $ and $\theta
_{34}=iu$. On the other hand we obtain from eq.~(\ref{c1.20}) using $\theta
_{13}-i\pi =2(\theta -i\pi )$ for $\theta _{24}=i\pi $ and $\theta
_{12}-iu=2(\theta -i\pi )$ for $\theta _{13}=i\pi $ and $\theta _{34}=iu$ 
\begin{eqnarray*}
G_{2}(\theta ) &=&\frac{A+2(\theta -i\pi )B+(\theta -i\pi )D}{2(\theta -i\pi
)}+\dots \\
&=&\frac{A}{2(\theta -i\pi )}+B+\tfrac{1}{2}D+\dots
\end{eqnarray*}
such that with $A=0$ 
\begin{align*}
G_{2}(i\pi )& =B+\tfrac{1}{2}D \\
& =(2i)^{2}i\mathbf{C}_{13}\mathbf{C}_{24}\mathcal{O}_{5\dots
n}(1-S_{3n}\dots S_{35})\Gamma _{(12)}^{12}\Gamma _{(34)}^{43}
\end{align*}

\paragraph{3.}

Let finally $\theta _{12}=\theta _{34}=iu$ and then $\theta \rightarrow i\pi 
$:\newline
Applying formula (\ref{c1.10}) twice the form factor of two bound states and
arbitrary other particles is obtained as $\mathcal{O}_{(12)(34)5\dots
n}(\theta _{(12)},\theta _{(34)},\underline{\theta }^{\prime })=-\tfrac{1}{2}%
G(iu,iu,\theta )$. Therefore using $\theta _{13}=\theta _{24}=\theta =\theta
_{(12)(34)}$ for $\theta _{12}=\theta _{34}$ we obtain 
\begin{multline*}
\mathop{\rm Res}_{\theta _{(12)(34)}=i\pi }\mathcal{O}_{(12)(34)5\dots
n}(\theta _{(12)},\theta _{(34)},\underline{\theta }^{\prime })=-\tfrac{1}{2}%
\mathop{\rm Res}_{\theta =i\pi }G(iu,iu,\theta ) \\
=-\tfrac{1}{2}D=-\tfrac{1}{2}(G_{1}(i\pi )+G_{2}(i\pi )) \\
=2i\mathbf{C}_{13}\mathbf{C}_{24}\Gamma _{(12)}^{12}\Gamma _{(34)}^{43}%
\mathcal{O}_{5\dots n}(\underline{\theta }^{\prime })(1-S_{(34)n}\dots
S_{(34)5})
\end{multline*}
which proves claim with the charge conjugation matrix for the bound states 
\[
\mathbf{C}_{(12)(34)}=\mathbf{C}_{13}\mathbf{C}_{24}\Gamma
_{(12)}^{12}\Gamma _{(34)}^{43}\qquad 
\begin{array}{l}
\unitlength=3.5mm\begin{picture}(15,5) \put(3,2){\oval(4,6)[t]}
\put(.2,1){(12)} \put(4.2,1){(34)} \put(7,3){=} \put(10,3){\oval(2,2)[b]}
\put(14,3){\oval(2,2)[b]} \put(12,3){\oval(2,2)[t]}
\put(12,3){\oval(6,5)[t]} \put(10,1){\line(0,1){1}}
\put(14,1){\line(0,1){1}} \put(9.2,0){(12)} \put(13.2,0){(34)}
\put(9.3,3.4){1} \put(10.3,3.4){2} \put(13.2,3.4){4} \put(14.2,3.4){3}
\end{picture}
\end{array}
\]
\endproof%

\section{Residue formula}

\label{a5}In order to obtain the breather form factors from the soliton form
factors we have to calculate the residue 
\begin{multline*}
\mathop{\rm Res}_{\theta _{12}=iu^{(k)}}\mathcal{O}_{123\dots n}(\underline{%
\theta })\,=\mathop{\rm Res}_{\theta _{12}=iu^{(k)}}(-2\pi
i)\,m\sum_{l=0}^{k}\mathop{\rm Res}_{z_{1}=z^{(l)}}\int_{\mathcal{C}_{%
\underline{\theta }}}dz_{2}\cdots \int_{\mathcal{C}_{\underline{\theta }%
}}dz_{m} \\
\times h(\underline{\theta }{,\underline{z}})p^{\mathcal{O}}(\underline{%
\theta }{,\underline{z}})\,\Psi _{1\dots n}(\underline{\theta },{\underline{z%
}})
\end{multline*}
where the pinching rule of contour integrals has been applied.

\begin{lemma}
For any analytic function $P(z)$ the following identity holds 
\begin{multline*}
\mathop{\rm Res}_{\theta _{12}=iu^{(k)}}\mathop{\rm Res}_{z_{1}=z^{(l)}}P(z)%
\,\phi (\theta _{1}-z_{1})\phi (\theta _{2}-z_{1})\,\Psi _{1\dots n}(%
\underline{\theta },{\underline{z}}) \\
=c(k,l)(-1)^{l}P(z^{(l)})\prod_{i=3}^{n}\dot{a}(\theta
_{i}-z^{(l)})\prod_{j=2}^{m}S_{b_{k}s}(\xi -z_{j})\Gamma _{12}^{b_{k}}\Psi
_{3\dots n}(\underline{\theta }^{\prime },{\underline{z}}^{\prime })
\end{multline*}
with $\underline{\theta }^{\prime }=\theta _{3},\dots ,\theta _{n},{%
\underline{z}}^{\prime }=z_{2},\dots ,z_{m}$ and 
\[
\Psi _{3\dots n}(\underline{\theta }^{\prime },{\underline{z}}^{\prime
})=\Omega _{3\dots n}C_{3\dots n}({\underline{\theta }}^{\prime
},z_{2})\cdots C_{3\dots n}({\underline{\theta }}^{\prime },z_{m}). 
\]
The constant is 
\[
c(k,l)=R_{k-l}R_{l}\,i\left| \tfrac{1}{2}\mathop{\rm Res}_{\theta =i\pi
(1-k\nu )}\dot{S}_{\pm }(\theta )\right| ^{-1/2} 
\]
with 
\[
R_{l}=\mathop{\rm Res}_{z=i\pi l\nu }\phi (z)\dot{a}(z). 
\]
\end{lemma}

\proof%
Using the decomposition 
\[
C_{1\dots n}=C_{12}A_{3\dots n}+D_{12}C_{3\dots n} 
\]
and the action on the pseudo ground state $\Omega $ 
\begin{eqnarray*}
\Omega _{12}C_{12} &=&\dot{b}(\theta _{1}-z)\dot{c}(\theta _{2}-z)\left(
s\otimes \bar{s}\right) _{12}+\dot{c}(\theta _{1}-z)\dot{a}(\theta
_{2}-z)\left( \bar{s}\otimes s\right) _{12} \\
\Omega _{12}D_{12} &=&\Omega _{12}\dot{b}(\theta _{1}-z)\dot{b}(\theta
_{2}-z).
\end{eqnarray*}
we obtain 
\begin{multline*}
\mathop{\rm Res}_{\theta _{12}=iu^{(k)}}\mathop{\rm Res}_{z_{1}=z^{(l)}}P(z)%
\,\phi (\theta _{1}-z_{1})\phi (\theta _{2}-z_{1})\,\Omega _{1\dots
n}C_{1\dots n}(\underline{\theta },{\underline{z}}) \\
=P(z^{(l)})R_{k-l}R_{l}(-1)^{l}\left( \left( s\otimes \bar{s}\right)
_{12}+(-1)^{k}\left( \bar{s}\otimes s\right) _{12}\right) \Omega _{3\dots
n}A_{3\dots n}
\end{multline*}
The relations 
\begin{eqnarray*}
\mathop{\rm Res}_{\theta _{12}=i\pi (1-k\nu )}\phi (\theta _{1}-z^{(l)})\dot{%
b}(\theta _{1}-z^{(l)}) &=&\mathop{\rm Res}_{\theta =i\pi (1-(k-l)\nu )}\phi
(\theta )\dot{b}(\theta ) \\
&=&-\mathop{\rm Res}_{\theta =i\pi (k-l)\nu }\phi (\theta )\dot{a}(\theta
)=-R_{k-l} \\
\mathop{\rm Res}_{z=z^{(l)}}\phi (\theta _{2}-z)\dot{a}(\theta _{2}-z) &=&%
\mathop{\rm Res}_{z=-i\pi l\nu }\phi (-z)\dot{a}(-z)=-R_{l}
\end{eqnarray*}
and 
\[
\frac{c(i\pi l\nu )}{a(i\pi l\nu )}=(-1)^{l},\quad \frac{c(i\pi (1-(k-l)\nu )%
}{b(i\pi (1-(k-l)\nu )}=(-1)^{k-l},\quad \frac{b(i\pi l\nu )}{a(i\pi l\nu )}%
=0 
\]
have been used. Further we use that the bound state intertwiner may be
written as 
\[
\Gamma _{12}^{b_{k}}=(-1)^{k}i\left| \tfrac{1}{2}\mathop{\rm Res}_{\theta
=i\pi (1-k\nu )}\dot{S}_{\pm }(\theta )\right| ^{1/2}\left( s\otimes \bar{s}%
+(-1)^{k}\,\bar{s}\otimes s\right) _{12}. 
\]
Applying further C-operators we use the fusion relation 
\begin{eqnarray*}
\Gamma _{12}^{(12)}S_{10}(\theta +\tfrac{1}{2}\theta
_{12}^{(12)})S_{20}(\theta -\tfrac{1}{2}\theta _{12}^{(12)})
&=&S_{(12)0}(\theta )\Gamma _{12}^{(12)} \\
\begin{array}{c}
\unitlength3.5mm\begin{picture}(5,5) \put(3,3){\line(1,1){1.2}}
\put(4,0){\line(-1,1){4}} \put(0,2){\line(3,1){3}} \put(2,0){\line(1,3){1}}
\put(0,.8){1} \put(1,0){2} \put(4,.4){0} \end{picture}
\end{array}
&=& 
\begin{array}{c}
\unitlength3.5mm\begin{picture}(5,5) \put(2,2){\line(1,1){2}}
\put(4.5,1.5){\line(-1,1){3}} \put(2,2){\line(-4,-1){2}}
\put(2,0){\line(0,1){2}} \put(0,.4){1} \put(1,0){2} \put(3.5,.8){0}
\end{picture}
\end{array}
\end{eqnarray*}
which implies 
\[
\Gamma _{12}^{b_{k}}C_{1\dots n}(\underline{\theta },z)|_{\theta
_{12}=iu^{(k)}}=S_{b_{k}s}(\theta _{(12)}-z)\Gamma _{12}^{b_{k}}C_{3\dots n}(%
\underline{\theta }^{\prime },z)\,. 
\]
Together with the eigenvalue equation 
\[
\Omega _{3\dots n}A_{3\dots n}(\underline{\theta }^{\prime
},z)=\prod_{i=3}^{n}\dot{a}(\theta _{i}-z)\Omega _{3\dots n} 
\]
the claim follows.%
\endproof%

Using the shift relation 
\[
\phi (z+i\pi \nu )\dot{a}(z+i\pi \nu )=-\frac{\cos \frac{1}{2i}z}{\sin \frac{%
1}{2i}(z+i\pi \nu )}\phi (z)\dot{a}(z) 
\]
the residues $R_{l}$ can be calculated 
\[
R_{l}=R_{0}\frac{(-1)^{l}}{\sin \frac{1}{2}\pi l\nu }\prod_{j=1}^{l-1}\cot 
\tfrac{1}{2}j\pi \nu 
\]
with 
\[
R_{0}=\mathop{\rm Res}_{z=0}\phi (z)\dot{a}(z)=\mathop{\rm Res}_{z=0}\frac{%
\dot{a}(z)}{F(z)F(i\pi -z)}=\frac{-1}{F^{\prime }(0)}. 
\]
In particular for the lowest breather $k=1$ we obtain the constants
independent of $l$%
\[
c(1,l)=R_{0}R_{1}\,i\left| \tfrac{1}{2}\mathop{\rm Res}_{\theta =i\pi
(1-k\nu )}\dot{S}_{\pm }(\theta )\right| ^{-1/2}=\frac{i}{\pi \varkappa 
\sqrt{\sin \pi \nu }} 
\]
where $\varkappa $ is defined by eq.~(\ref{3.1}).Using these results the
form factors for arbitrary numbers of lowest breathers and solitons are
calculated explicitly in section \ref{s7}. There we use the constant 
\begin{eqnarray*}
d(\nu ) &=&\frac{-2\pi i}{\sqrt{2}}F(i\pi (1-\nu ))\,c(1,0)=\frac{-2\pi i}{%
\sqrt{2}}\sqrt{\cos \tfrac{\pi }{2}\nu E(\nu )}\frac{i}{\pi \varkappa \sqrt{%
\sin \pi \nu }} \\
&=&\frac{\sqrt{E(\nu )}}{\varkappa \sqrt{\sin \frac{1}{2}\pi \nu }}.
\end{eqnarray*}
where the function $E(\nu )$ is defined as 
\[
E(\nu )=\exp \int_{0}^{\infty }\frac{dt}{t}\frac{\sinh \nu t}{2\cosh ^{2}%
\frac{1}{2}t}=\exp \frac{1}{\pi }\int_{0}^{\pi \nu }\frac{tdt}{\sin t}\,. 
\]


\begin{thebibliography}{99}
\bibitem{VG}  S. Vergeles and V. Gryanik, \emph{Sov. Journ. Nucl. Phys.} 
\textbf{23} (1976) 704.

\bibitem{W}  P. Weisz, \emph{Nucl. Phys.} \textbf{B122} (1977) 1.

\bibitem{KW}  M. Karowski and P. Weisz, \emph{Nucl. Phys.} \textbf{B139}
(1978) 445.

\bibitem{Sm}  F.A. Smirnov \emph{'Form Factors in Completely Integrable
Models of Quantum Field Theory', Adv. Series in Math. Phys.} \textbf{14},
World Scientific 1992.

\bibitem{BFKZ}  H. Babujian, A. Fring, M. Karowski and A. Zapletal, \emph{%
Nucl. Phys.} \textbf{B538} [FS] (1999) 535-586.

\bibitem{Lu}  S.Lukyanov, \emph{Mod.Phys.Lett.} \textbf{A 12} (1997)
2543-2550.

\bibitem{LZ}  S. Lukyanov and A.B. Zamolodchikov, hep-th /0102079.

\bibitem{KLP}  S. Khoroshkin, D. Lebedev and S. Pakuliak, \emph{Lett. Math.
Phys. }\textbf{41} (1997) 31-47.

\bibitem{NPT}  A. Nakayashiki, S. Pakuliak and V. Tarasov, \emph{Annales de
l'Institut Henri Poincar\'{e} }\textbf{71 }N4 (1999) 459-496.
\bibitem{NT}  A. Nakayashiki and Y. Takeyama, math-ph/0105040.

\bibitem{GNT}  A.O. Gogolin, A.A Nersesyan and A.M. Tsvelik, \emph{%
'Bosonization in Strongly Correlated Systems', }Cambridge University Press
(1999).

\bibitem{CET}  D. Controzzi, F.H.L. Essler and A.M. Tsvelik, \emph{%
'Dynamical Properties of one dimensional Mott Insulators', }%
cond-math/0011439.

\bibitem{BKZ}  H. Babujian, M. Karowski and A. Zapletal, \emph{J. Phys.} 
\textbf{A30} (1997) 6425.

\bibitem{BFKZ1}  H. Babujian, A. Fring, M. Karowski and A. Zapletal, \emph{%
Exact Form Factors in Integrable Quantum Field Theories: the }$SU(N)$\emph{%
-Chiral-Gross-Neveu Model,} in preparation.

\bibitem{BK3}  H. Babujian and M. Karowski, \emph{Exact Form Factors in }$%
O(N)$ \emph{Symmetric Integrable Quantum Field Theories,} in preparation.

\bibitem{KTTW}  M. Karowski, H.J. Thun, T.T. Troung, and P. Weisz, \emph{%
Phys. Lett.} \textbf{67B} (1977) 321.

\bibitem{KT}  M. Karowski and H.J. Thun, \emph{Nucl. Phys.} \textbf{B130}
(1977) 295.

\bibitem{K}  M. Karowski, \emph{Phys. Rep.} \textbf{49} (1979) 229.

\bibitem{ZZ}  A.B. Zamolodchikov and Al. B. Zamolodchikov, \emph{Ann. Phys.} 
\textbf{120} (1979) 253.

\bibitem{BKW}  B. Berg, M. Karowski and P. Weisz, \emph{Phys. Rev.} \textbf{%
D19} (1979) 2477.

\bibitem{Korepin}  V.E. Korepin and N.A. Slavnov, \emph{J. Phys.} \textbf{A31%
} (1998) 9283;\newline
V.E. Korepin and T.Oota, \emph{J. Phys.} \textbf{A31} (1998) L371;\newline
T. Oota, \emph{J. Phys.} \textbf{A31} (1998) 7611.

\bibitem{CM}  J.L. Cardy and G. Mussardo, \emph{Phys. Lett.} \textbf{B225}
(1989) 275. \emph{Nucl. Phys.} \textbf{B340} (1990) 387.

\bibitem{Zi}  W. Zimmermann, \emph{Ann. Phys.} \textbf{77} (1973) 536.

\bibitem{Q}  T. Quella, \emph{Formfaktoren und Lokalit\"{a}t in integrablen
Modellen der Quan\-ten\-feld\-theo\-rie in 1+1 Dimensionen,} Diploma thesis
FU-Berlin (1999) unpublished.

\bibitem{Co}  S. Coleman, \emph{Phys. Rev.} {D11} (1975) 2088.

\bibitem{Za}  A.B. Zamolodchikov, \emph{JEPT Lett.} \textbf{25} (1977) 468.

\bibitem{K1}  M. Karowski, \emph{Nucl. Phys.} \textbf{B153} (1979) 244.

\bibitem{La}  M.Yu. Lashkevich, \emph{Sectors of Mutually Local Fields in
Integrable Models of Quantum Field Theory, }LANDAU-94-TMP-4, hep-th/9406118.

\bibitem{B}  H.M. Babujian, \emph{Correlation functions in WZNW model as a
Bethe wave function for the Gaudin magnets}, in: 'Proc. XXIV Int. Symp.
Ahrenshoop', Gosen 1990.

\bibitem{B1}  H.M. Babujian, \emph{J. Phys.} \textbf{A26} (1993) 6981, \emph{%
J. Phys.} \textbf{A27} (1994) 7753.

\bibitem{BF}  H.M. Babujian and R. Flume, \emph{Mod. Phys. Lett.} \textbf{A9}
(1994) 2029.

\bibitem{BKT}  B. Berg, M. Karowski and H.J. Thun, Phys. Lett. \textbf{62B}
(1976) 187, \textbf{64B} (1976) 286.

\bibitem{KN}  P.P. Kulish and E.R. Nissimov, JETP Lett. \textbf{24}
  (1976) 220-223.

\bibitem{BK1}  H.M. Babujian and M. Karowski, The Exact Quantum Sine-Gordon
Field Equation and other Non-Perturbative Results, Sfb 288 - preprint 414,
hep-th/9909153, Phys. Lett. \textbf{B 411} (1999) 53-57.

\bibitem{BK}  H. Babujian and M. Karowski, \emph{Exact Form Factors in
Integrable Quantum Field Theories: the Sine-Gordon Model (III)}, in
preparation.

\bibitem{FMS}  A. Fring, G. Mussardo and P. Simonetti, \emph{Nucl. Phys.}%
\textbf{\ B393} (1993) 413.

\bibitem{KM}  A. Koubeck and G. Mussardo, \emph{Phys. Lett. }\textbf{B311}
(1993) 193.

\bibitem{MS}  G. Mussardo and P. Simonetti, \emph{Int. J. Mod. Phys.}\textbf{%
\ A9} (1994) 3307-3338.

\bibitem{K2}  M. Karowski, \emph{The bootstrap program for 1+1 dimensional
field theoretic models with soliton behavior}, in 'Field theoretic methods
in particle physics', ed. W. R\"{u}hl, (Plenum Pub. Co., New York, 1980).

\bibitem{YuZ} V.P.Yurov, Al.B. Zamolodchikov, \emph{Int.J.Mod.Phys.}
\textbf{A 6} (1991) 4557.

\bibitem{Lu2}  S. Lukyanov, \emph{Commun.Math.Phys.} \textbf{ 167} (1995)
183-226.
\end{thebibliography}
\end{document}